\documentstyle[12pt]{article}
\let\ssection=\section
\renewcommand{\section}{\setcounter{equation}{0}\ssection}

\newcommand\mathC{\mkern1mu\raise2.2pt\hbox{$\scriptscriptstyle|$}
        {\mkern-7mu\rm C}}              
\newcommand{\mathR}{{\rm I\! R}}         

\newtheorem{definition}{Definition}[section]
\newtheorem{theorem}{Theorem}[section]

\newcommand\mapdown[1]{\Big\downarrow
                        \rlap{$\vcenter{\hbox{$\scriptstyle#1$}}$}}
\newcommand\mapright[1]{\smash{
        \mathop{\mbox{\large{$\longrightarrow$}}}\limits^{#1}}}
\newcommand\bundle[3]{\begin{array}[t]{c}
        {#1}\\ \mapdown{#2}\\ {#3}\end{array}}
\newcommand\bundlemap[2]{\begin{array}[t]{c}
\mapright{#1}\\
\phantom{\mapdown{}}\\\mapright{#2}\\\end{array}}

\begin{document}
\begin{titlepage}
\hspace{10truecm}Imperial/TP/97--98/28

\begin{center}
{\large\bf A Topos Perspective on the Kochen-Specker
                                            Theorem:\\[6pt]
I. Quantum States as Generalized Valuations}
\end{center}

\vspace{0.8 truecm}
\begin{center}
        C.J.~Isham\footnote{email: c.isham@ic.ac.uk}\\[10pt]
        The Blackett Laboratory\\
        Imperial College of Science, Technology \& Medicine\\
        South Kensington\\
        London SW7 2BZ\\
\end{center}

\begin{center}
and
\end{center}

\begin{center}
        J.~Butterfield\footnote{email:
            jb56@cus.cam.ac.uk;
            jeremy.butterfield@all-souls.oxford.ac.uk}\\[10pt]
            All Souls College\\
            Oxford OX1 4AL
\end{center}

\begin{center}
    May\footnote{Small clarifications added concerning
operators with continuous spectra; September 1998.} 1998
\end{center}

\begin{abstract}
Any attempt to construct a realist interpretation of quantum
theory founders on the Kochen-Specker theorem, which asserts the
impossibility of assigning values to quantum quantities in a way
that preserves functional relations between them.  We construct
a new type of valuation which is defined on all operators, and
which respects an appropriate version of the functional
composition principle. The truth-values assigned to propositions
are (i) contextual; and (ii) multi-valued, where the space of
contexts and the multi-valued logic for each context come
naturally from the topos theory of presheaves.

The first step in our theory is to demonstrate that the
Kochen-Specker theorem is equivalent to the statement that a
certain presheaf defined on the category of self-adjoint
operators has no global elements. We then show how the use of
ideas drawn from the theory of presheaves leads to the
definition of a generalized valuation in quantum theory whose
values are sieves of operators. In particular, we show how each
quantum state leads to such a generalized valuation.

A key ingredient throughout is the idea that, in a situation
where no normal truth-value can be given to a proposition
asserting that the value of a physical quantity $A$ lies in a
subset $\Delta\subset\mathR$, it is nevertheless possible to
ascribe a partial truth-value which is determined by the set of
all coarse-grained propositions that assert that some function
$f(A)$ lies in $f(\Delta)$, and that {\em are\/} true in a
normal sense. The set of all such coarse-grainings forms a sieve
on the category of self-adjoint operators, and is hence
fundamentally related to the theory of presheaves.

\end{abstract}
\end{titlepage}

\section{Introduction}
\label{Sec:Introduction}
\subsection{Preliminary Remarks}
\label{SubSec:IntroPrelRemarks} Anyone who has taught an
introductory course on quantum theory will have encountered the
anguish that can accompany a student's first engagement with the
problematic status of beliefs previously deemed to be
self-evidently true. In particular, it is difficult to remove
the compelling conviction that, at any given time, any physical
quantity must have a value.

In classical physics, there is no problem with this belief since
the underlying mathematical structure is geared precisely to
express it. Specifically, if $\cal S$ is the state space of some
classical system, a physical quantity $A$ is represented by a
real-valued function $\bar A:{\cal S}\rightarrow\mathR$; and
then the value $V_s(A)$ of $A$ in any state $s\in\cal S$ is
simply
\begin{equation}
    V^s(A)=\bar{A}(s).                      \label{Vs(A)=A(s)}
\end{equation}
Thus all physical quantities possess a value in any state.
Furthermore, if $h:\mathR\rightarrow\mathR$ is a real-valued
function, a new physical quantity $h(A)$ can be defined by
requiring the associated function $\overline{h(A)}$ to be
\begin{equation}
    \overline{h(A)}(s):=h(\bar{A}(s))       \label{Def:h(A)}
\end{equation}
for all $s\in\cal S$; {\em i.e.}, $\overline{h(A)}:=h\circ\bar
A:\cal S\rightarrow\mathR$. Thus the physical quantity $h(A)$ is
{\em defined\/} by saying that its value in any state $s$ is the
result of applying the function $h$ to the value of $A$; hence,
by definition, the values of the physical quantities $h(A)$ and
$A$ satisfy the `functional composition principle'
\begin{equation}
    V^s(h(A))=h(V^s(A))                     \label{FUNCT-class}
\end{equation}
for all states $s\in\cal S$.

However, to the distress of angst-ridden students, standard
quantum theory precludes any such naive realist interpretation
of the relation between formalism and physical world. And this
is not just because of some wilfully obdurate philosophical
interpretation of the theory: rather, the obstruction comes from
the mathematical formalism itself, in the guise of the famous
Kochen-Specker theorem which asserts the impossibility of
assigning values to all physical quantities whilst, at the same
time, preserving the functional relations between them
\cite{KS67}.\footnote{As has been emphasized by Brown
\cite{Bro92}, the essential result is already contained in
Bell's seminal first paper on hidden variables \cite{Bel66}.}

In a quantum theory, a physical quantity $A$ is represented by a
self-adjoint operator $\hat A$ on the Hilbert space of the
system, and the first thing one has to decide is whether to
regard a valuation as a function of the physical quantities
themselves, or on the operators that represent them.  From a
mathematical perspective, the latter strategy is preferable, and
we shall therefore define a (global) valuation to be a
real-valued function $V$ on the set of all bounded, self-adjoint
operators, with the properties that : (i) the value $V(\hat A)$
of the physical quantity $A$ represented by the operator $\hat
A$ belongs to the spectrum of $\hat A$ (the so-called `value
rule'); and (ii) the functional composition principle (or FUNC
for short) holds:
\begin{equation}
    V(\hat B)=h(V(\hat A))                  \label{funct-rule}
\end{equation}
for any pair of self-adjoint operators $\hat A$, $\hat B$ such
that $\hat B= h(\hat A)$ for some real-valued function $h$. If
they existed, such valuations could be used to embed the set of
self-adjoint operators in the commutative ring of real-valued
functions on an underlying space $\cal S$ of microstates,
thereby laying the foundations for a hidden-variable
interpretation of quantum theory.

Several important results follow from the definition of a
valuation. For example, if $\hat A_1$ and $\hat A_2$ commute,
there exists an operator $\hat C$ and functions $h_1$ and $h_2$
such that $\hat A_1=h_1(\hat C)$ and $\hat A_2=h_2(\hat C)$; it
then follows from FUNC that
\begin{equation}
    V(\hat A_1+\hat A_2)=V(\hat A_1)+V(\hat A_2)
\end{equation}
and
\begin{equation}
    V(\hat A_1\hat A_2)=V(\hat A_1)V(\hat A_2).\label{VAB}
\end{equation}

The defining equation Eq.\ (\ref{funct-rule}) for a valuation
makes sense whatever the nature of the spectrum $\sigma(\hat A)$
of the operator $\hat A$. However, if $\sigma(\hat A)$ contains
a continuous part, one might doubt the physical meaning of
assigning one of its elements as a value; indeed, in the present
paper, we shall consider valuations in this sense as being
defined only on the subset of operators whose spectrum is purely
discrete.  To handle the more general case, we shall reconceive
a valuation as primarily giving {\em truth-values\/} to {\em
propositions\/} about the values of a physical quantity, rather
than assigning a specific value to the quantity itself.

The propositions concerned are of the type `$A\in\Delta$', which
asserts that the value of the physical quantity $A$ lies in the
Borel subset $\Delta$ of the spectrum $\sigma(\hat A)$ of the
associated operator $\hat A$. Of course, such assertions are
meaningful for both discrete and continuous spectra: which
motivates studying the general mathematical problem of assigning
truth-values to projection operators.

If $\hat P$ is a projection operator, the identity $\hat P=\hat
P^2$ implies that $V(\hat P)=V(\hat P^2)=(V(\hat P))^2$ (from
Eq.\ (\ref{VAB})); and hence, necessarily, $V(\hat P)=0$ or $1$.
Thus $V$ defines a homomorphism from the Boolean algebra $\{\hat
0, \hat 1, \hat P,\neg\hat P\equiv(\hat 1-\hat P)\}$ to the
`false(0)-true(1)' Boolean algebra $\{0,1\}$. More generally, a
valuation $V$ induces a homomorphism
$\chi^V:W\rightarrow\{0,1\}$ where $W$ is any Boolean subalgebra
of the lattice $\cal P$ of projectors on $\cal H$. In
particular,
\begin{equation}
    \hat\alpha\leq\hat\beta\mbox{\ \ implies \ }
    \chi^V(\hat\alpha)\leq\chi^V(\hat\beta) \label{a<b->V(a)<V(b)}
\end{equation}
where `$\hat\alpha\leq\hat\beta$' refers to the partial ordering
in the lattice $\cal P$, and
`$\chi^V(\hat\alpha)\leq\chi^V(\hat\beta)$' is the ordering in
the Boolean algebra $\{0,1\}$. This result has an important
implication for us, to which we shall return shortly.

The Kochen-Specker theorem asserts that no global valuations
exist if the dimension of the Hilbert space $\cal H$ is greater
than two.  The obstructions to the existence of such valuations
typically arise when trying to assign a single value to an
operator $\hat C$ that can be written as $\hat C=g(\hat A)$ and
as $\hat C=h(\hat B)$ with $[\hat A,\,\hat B]\neq 0$.

One response to this result is to note that the theorem does not
preclude the existence of `partial', or `local',
valuations---{\em i.e.}, valuations that are defined only on
some subset of the set of self-adjoint operators; a typical
example would be any complete set of commuting operators on the
Hilbert space. However, if partial valuations are to form part
of a proper interpretative framework, the question immediately
arises as to how the domain of any such valuation is to be
chosen.

The extant interpretations of quantum theory that aspire to use
`beables', rather than `observables', are all concerned in one
way or another with addressing this issue. One well-known
approach is that of Bohm, where certain physical
quantities---for example, the position of a particle---are
declared by fiat to be those that always have a value. In other,
so-called `modal' approaches, the domain of a partial valuation
depends on the quantum state; as, for example, in the works of
van Fraassen \cite{Fra81,Fra91}, Kochen \cite{Koc85}, Healey
\cite{Hea89}, Clifton \cite{Cli95a}, Dieks \cite{Die95}, Vermaas
and Dieks \cite{VD95}, Bacciagaluppi and Hemmo \cite{BH96}, and
Bub \cite{Bub97}.

Inherent in such schemes is a type of `contextuality' in which a
value ascribed to a physical quantity $C$ cannot be part of a
global assignment of values but must, instead, depend on some
{\em context\/} in which $C$ is to be considered. In practice,
contextuality is endemic in any attempt to ascribe properties to
quantities in a quantum theory. For example, as emphasized by
Bell \cite{Bel66}, in the situation where $\hat C=g(\hat
A)=h(\hat B)$, if the value of $C$ is construed counterfactually
as referring to what would be obtained {\em if\/} a measurement
of $A$ or of $B$ is made---and with the value of $C$ then being
{\em defined\/} by applying the relation $C=g(A)$, or $C=h(B)$,
to the result of the measurement---then one can claim that the
actual value obtained depends on whether the value of $C$ is
determined by measuring $A$, or by measuring $B$.

In the programme to be discussed here, the idea of a contextual
valuation will be developed in a different direction from that
of the existing modal interpretations. In particular, rather
than accepting only a limited domain of beables we shall propose
a theory of `generalized' valuations that are defined globally
on {\em all\/} propositions about values of physical quantities.
However, the price of global existence is that any given
proposition may have only a `{\em partial\/}' truth-value.  More
precisely, (i) the truth-value of a proposition `$A\in\Delta$'
belongs to a logical structure that is larger than $\{0,1\}$;
and (ii) these target-logics are context-dependent.

It is clear that the main task is to formulate mathematically
the idea of a contextual, `partial' truth-value in such a way
that the assignment of generalized truth-values is consistent
with an appropriate analogue of the functional composition
principle FUNC. The scheme also has to have some meaningful
physical interpretation; in particular, we want the set of all
possible partial truth-values for any given context to form some
sort of {\em distributive\/} logic, in order to facilitate a
proper semantics for this `neo-realist' view of quantum theory.

\subsection{Generalized Logic in Quantum Physics}
\label{SubSec:GenLogQP} Our central idea is that, although in a
given situation in quantum theory it may not be possible to
declare a particular proposition `$A\in\Delta$' to be true (nor
false), nevertheless there may be (Borel) functions $f$ such
that the associated propositions `$f(A)\in f(\Delta)$' {\em
can\/} be said to be true.  This possibility arises for the
following reason.

Let $W_A$ denote the spectral algebra of the operator $\hat A$
that represents a physical quantity $A$: thus $W_A$ is the
Boolean algebra of projectors $\hat E[A\in\Delta]$ that project
onto the eigenspaces associated with the Borel subsets $\Delta$
of the spectrum $\sigma(\hat A)$ of $\hat A$; physically
speaking, $\hat E[A\in\Delta]$ represents the proposition
`$A\in\Delta$'.  It follows from the spectral theorem that, for
all Borel subsets $J$ of the spectrum of $f(\hat A)$, the
spectral projector $\hat E[f(A)\in J]$ for the operator $f(\hat
A)$ is equal to the spectral projector $\hat E[A\in f^{-1}(J)]$
for $\hat A$. In particular, if $f(\Delta)$ is a Borel subset of
$\sigma(f(\hat A))$ (which is automatically true if the spectrum
of $\hat A$ is discrete; we shall discuss the non-discrete case
later) then, since $\Delta\subseteq f^{-1}(f(\Delta))$, we have
$\hat E[A\in\Delta]\leq \hat E[A\in f^{-1}(f(\Delta))]$; and
hence
\begin{equation}
    \hat E[A\in\Delta]\leq\hat E[f(A)\in f(\Delta)]
                        \label{1EA<EfA}.
\end{equation}

Physically, the inequality in Eq.\ (\ref{1EA<EfA}) reflects the
fact that the proposition `$f(A)\in f(\Delta)$' is generally
weaker than the proposition `$A\in\Delta$' in the sense that the
latter implies the former, but not necessarily vice versa. For
example, the proposition `$f(A)=f(a)$' is weaker than the
original proposition `$A=a$' if the function $f$ is many-to-one
and such that more than one eigenvalue of $\hat A$ is mapped to
the same eigenvalue of $f(\hat A)$. In general, we shall say
that `$f(A)\in f(\Delta)$' is a {\em coarse-graining\/} of
`$A\in\Delta$'.

Now if the proposition `$A\in\Delta$' is evaluated as `true' by,
for example, a partial valuation $V$ of the type mentioned at
the end of Section \ref{SubSec:IntroPrelRemarks}---so that
$V(\hat E[A\in\Delta])=1$---then, from Eq.\
(\ref{a<b->V(a)<V(b)}) and Eq.\ (\ref{1EA<EfA}), it follows that
the weaker proposition `$f(A)\in f(\Delta)$' is also evaluated
as `true'.

This remark provokes the following observation. There may be
situations in which, although the proposition `$A\in\Delta$'
cannot be said to be either true or false, the weaker
proposition `$f(A)\in f(\Delta)$' can be. In particular, if the
latter {\em can\/} be given the value `true' in a total sense,
then---by virtue of the remark above---it is natural to suppose
that any further coarse-graining to give an operator $g(f(\hat
A))$ will yield a proposition `$g(f(A))\in g(f(\Delta))$' that
also is to be evaluated as `true'. Note that there may be more
than one possible choice for the `initial' function $f$, each of
which can then be further coarse-grained in this way.  This
multi-branched picture of coarse-graining is one of the main
justifications for our invocation of the topos-theoretic idea of
a presheaf.

In fact, guided by the remarks above, the procedure we shall
adopt in Section \ref{Sec:PartVal} is first to consider partial
valuations---which assign truth-values $0$ or $1$ in a standard
way, but are defined on less than all the operators---and then
to go on to {\em define\/} the partial truth-value (associated
with each partial valuation $V$) of {\em any\/} proposition
`$A\in \Delta$' to be the set of all operators $\hat B$ of the
form $\hat B=f(\hat A)$ that are in the domain of $V$, and which
are such that the weaker proposition `$f(A)\in f(\Delta)$' is
`totally true'---{\em i.e.}, it is assigned the unit in the
logic of partial truth-values.

We shall then generalize this idea in Section \ref{Sec:GenVal}
where we extract the key properties of these partial
truth-values and use them to formulate a definition of a
`generalized valuation', the semantic interpretation of which is
that the truth-value of a proposition `$A\in\Delta$' is a set of
coarse-grained propositions `$f(A)\in f(\Delta)$' each of which
can be regarded as being totally true.  As we shall show, any
quantum state gives rise to such a generalized valuation.

The key property of such a generalized truth-value is that it is
a {\em sieve\/} in a certain category formed from the
self-adjoint operators on the Hilbert space of the system---and
it is a fundamental property of sieves that they form a {\em
Heyting\/} algebra, and hence have the structure of a
distributive logic; albeit one that is {\em intuitionistic\/},
not classical, in the sense that the logical law of excluded
middle is replaced with the weaker condition
$\alpha\lor\neg\alpha\leq 1$. These sieves are associated with a
certain presheaf---the `spectral presheaf'---that is naturally
associated with any quantum theory: this is how ideas from topos
theory enter our scheme.

This procedure was partly motivated by an earlier paper in which
topos ideas were applied to the consistent histories approach to
quantum theory \cite{Ish97}; in particular, it was shown there
how a topos framework fits naturally with the multi-branched,
coarse-graining operations that play a central role in the
construction of consistent sets of propositions.  Contextuality
arises explicitly there as the need to choose a particular
consistent set of histories; and---in fact---topos-theoretic
ideas can be expected to arise naturally in any physical theory
where contextuality plays a central role.  Presheaves are
particularly important in this respect since they are naturally
associated with contextual, generalized truth-values given by
the so-called `subobject classifier'.

Another motivation for our procedure is more general and
conceptual. In short, it represents a {\em via media\/} between
two extremes in the semantics, or interpretation, of quantum
theory.  For, on the one hand, the Kochen-Specker theorem shows
the impossibility of sustaining any naive realist interpretation
of quantum theory in which propositions about the values of
physical quantities are handled with the simple type of Boolean
logic which is characteristic of, for example, the set of
subsets of a classical state space. And, on the other hand, we
believe that the `logical' structure inherent in the lattice of
projection operators that represent quantum propositions
mathematically is too non-classical---in particular, it is
non-distributive---to fulfill any genuine semantic role. (This
is a well-known viewpoint; for example, see Dummett's
\cite{Dum78} critique of Putnam's proposal to ``read the logic
off Hilbert space'' \cite{Put75}.) Our aim is to find a middle
path between these extremes with the aid of logical structures
that are certainly not just simple Boolean algebras---our logics
are contextual and intuitionistic---but which retain the
semantically crucial property of distributivity.  We hope that
this intermediate position will extend a little our encompassing
of `quantum reality'.

\subsection{Some Expected Properties of Generalized Truth
Values} \label{SubSec:primafacie} To further motivate the
detailed constructions that will be made in this paper it is
helpful at this stage to consider what can be said {\em ab
initio\/} about the assignment of partial truth values.  For
example, presumably the minimum that should be satisfied by the
analogue of FUNC is that if $\hat B=h(\hat A)$, and if the
proposition `$A\in\Delta$' is assigned the value `totally true',
then the proposition `$B\in h(\Delta)$' should also be `totally
true'. As we shall see, this requirement is implemented in a
simple way in the presheaf framework which we employ.

A central problem in handling multi-valued truth-values is to
understand how the internal mathematical operations of the
`target' logic are to be related to the logical structure of the
propositions being evaluated.  More precisely, let $L$ denote
the Boolean algebra of all propositions of the type
`$A\in\Delta$' for some fixed physical quantity $A$---and
suppose we have some assignment of partial truth-values,
$\nu:L\rightarrow T(L)$, where $T(L)$ is the target logic in the
context of $L$.  Then how should the structure of $L$ be
reflected in the properties of $\nu$ and the logical structure
of $T(L)$? For example, is $\nu$ some type of algebraic
homomorphism? The minimum that can be said in this direction
would seem to be the following.

Firstly, the null proposition corresponding to the zero element
$0_L\in L$ should presumably always be valued as totally false;
and hence we expect $\nu(0_L)=0_{T(L)}$ in all contexts.

Secondly, if $\alpha,\beta\in L$ are such that
$\alpha\leq\beta$, then the physical interpretation is that the
proposition $\alpha$ implies the proposition $\beta$; an example
is `$A\in\Delta_1$' and `$A\in\Delta_2$' respectively, with
$\Delta_1\subseteq\Delta_2$. Under these circumstances, the
analogy with Eq.\ (\ref{a<b->V(a)<V(b)}) suggests that the
generalized truth-values should satisfy
$\nu(\alpha)\leq\nu(\beta)$ in the target logic $T(L)$.  In what
follows, we shall refer to this central requirement as the
`monotonicity' condition.

Now, for any $\alpha,\beta\in L$, we have $\alpha\leq
\alpha\lor\beta$ and $\beta\leq\alpha\lor\beta$; hence it
follows from monotonicity that
$\nu(\alpha)\leq\nu(\alpha\lor\beta)$ and
$\nu(\beta)\leq\nu(\alpha\lor\beta)$. This implies that, in the
logic $T(L)$,
\begin{equation}
    \nu(\alpha)\lor\nu(\beta)\leq\nu(\alpha\lor\beta)
                        \label{weak-disj}
\end{equation}
if we assume that the `or' operation in the target logic $T(L)$
behaves as expected, {\em i.e.}, it is the least upper bound for
the partial ordering.

One might wonder if the stronger disjunctive rule
$\nu(\alpha\lor\beta)= \nu(\alpha)\lor\nu(\beta)$ holds but, on
reflection, this is at variance with certain key ideas of
quantum theory.  For example, suppose that $\alpha$ and $\beta$
are the propositions `$A=a_1$' and `$A=a_2$' respectively, with
$a_1\neq a_2$. Then the projection operators that represent
these propositions project onto the eigenstates of $\hat A$
corresponding to the eigenvalues $a_1$ and $a_2$ respectively.
However, in the lattice of projectors, the disjunction of these
operators projects onto the two-dimensional space spanned by
these eigenvectors, which is strictly {\em bigger\/} than the
union of the pair of one-dimensional spaces (which, indeed, is
not a linear subspace at all). Hence a generalized truth-value
$\nu(\alpha\lor\beta)$ of $\alpha\lor\beta$ might be greater (in
the logical sense) than the disjunction of the generalized
truth-values of $\alpha$ and $\beta$ separately.  We shall see
in several concrete examples that this is indeed the case.

Similarly, for any $\alpha,\beta\in L$, we have
$\alpha\land\beta\leq \alpha$ and $\alpha\land\beta\leq\beta$,
so that, by monotonicity, $\nu(\alpha\land\beta)\leq
\nu(\alpha)$ and $\nu(\alpha\land\beta)\leq \nu(\beta)$.
Assuming that the `and' operation, `$\land$', in the target
logic $T(L)$ behaves as expected---{\em i.e.}, is the greatest
lower bound for the partial ordering---it follows that
\begin{equation}
    \nu(\alpha\land\beta)\leq\nu(\alpha)\land\nu(\beta).
            \label{weak-conj}
\end{equation}
Here also, one might wonder if a stronger conjunctive rule
$\nu(\alpha\land\beta)= \nu(\alpha)\land\nu(\beta)$ holds; but
we can see at once that it cannot do so in any scheme in which
`blurred' truth-values occur. For example, suppose once more
that $\alpha$ and $\beta$ are the propositions `$A=a_1$' and
`$A=a_2$' respectively, with $a_1\neq a_2$.  Then, as explained
earlier, our key idea is to assign a partial truth-value to a
proposition like `$A=a$' by finding a `coarse-grained' operator
$\hat B=f(\hat A)$ such that the weaker proposition
`$f(A)=f(a)$' {\em is\/} totally true. One consequence is that,
even though the propositions `$A=a_1$' and `$A=a_2$' are
disjoint---so that $\alpha\land\beta=0$---this does not imply
that $\nu(\alpha)\land\nu(\beta)$ is totally false: all that is
needed is an operator $\hat B=f(\hat A)$ with $f(a_1)=f(a_2)$
and such that `$f(A)=f(a_1)$' is unequivocally true. In this
circumstance, the strict inequality holds in Eq.\
(\ref{weak-conj}).

The monotonicity rule requires supplementing in one respect.
Consider again the propositions `$A=a_1$' and `$A=a_2$' with
$a_1\neq a_2$, and suppose the generalized valuation is such
that $\nu(A=a_1)=1_{T(L)}$---{\em i.e.}, the proposition
`$A=a_1$' is totally true in the logic $T(L)$. Then it seems
natural to require that the disjoint proposition `$A=a_2$'
cannot also be {\em totally\/} true, even though it need not be
totally false either.  However, for the following reason, this
restriction---which we shall refer to as `exclusivity'---cannot
be deduced from the monotonicity condition.

The disjointness condition $\alpha\land\beta=0$ in the Boolean
algebra $L$, implies that $\beta\leq\neg\alpha$; and hence,
using monotonicity,
\begin{equation}
    \nu(\beta)\leq \nu(\neg\alpha).     \label{mubnega}
\end{equation}
Now, {\em if\/} we assumed that $\nu:L\rightarrow T(L)$ commutes
with the negation operation, in the sense that
\begin{equation}
    \nu(\neg\alpha)=\neg\nu(\alpha),        \label{muneg=negmu}
\end{equation}
then Eq.\ (\ref{mubnega}) plus the hypothesis
$\nu(\alpha)=1_{T(L)}$, would imply that
$\nu(\beta)\leq\neg\nu(\alpha)=\neg 1_{T(L)}=0_{T(L)}$; hence
$\nu(\beta)=0_{T(L)}$, which certainly satisfies exclusivity.
However, it turns out that the equality Eq.\ (\ref{muneg=negmu})
is precisely what {\em cannot\/} be assumed in our theory since,
as we shall see later, the target logic for the generalized
truth-values is a Heyting algebra, and the negation operation in
an intuitionistic logic of this type behaves differently from
that in a Boolean algebra.  As a result, the exclusivity
condition cannot be derived from monotonicity, and it must
therefore be added as an extra requirement.

Putting together all these remarks, we arrive at the following
tentative, minimal list of algebraic properties that we expect
to be satisfied by a generalized valuation $\nu:L\rightarrow
T(L)$ of a Boolean logic $L$:
\begin{eqnarray}
    &\mbox{Null condition:\quad} &\nu(0_L)=0_{T(L)}\label{NullC}\\
    &\mbox{Monotonicity:\quad}\; &\alpha\leq\beta
\mbox{\ implies\ } \nu(\alpha)\leq\nu(\beta) \label{MonoC}\\
&\mbox{Exclusivity:\qquad}\; &\mbox{If $\alpha\land\beta=0_L$
and $\nu(\alpha)=1_{T(L)}$, then $\nu(\beta)<1_{T(L)}$}
                                \label{ExclC}
\end{eqnarray}
As we shall see, the examples of generalized valuations in
quantum theory given in this paper satisfy these requirements.
Another condition that we might want to add is
\begin{equation}
    \mbox{Unit condition:\quad} \nu(1_L)=1_{T(L)}
                    \hspace{6cm}    \label{UnitC}
\end{equation}
which, as we shall see, is also satisfied by the valuations
associated with quantum states. On the other hand, it can be
violated by the generalized valuations, which we mentioned in
Section \ref{SubSec:GenLogQP} that are associated with partial
valuations (mentioned in Section \ref{SubSec:GenLogQP}). We
shall see this explicitly in Section \ref{AlgPropnuV}.

\subsection{Prospectus}

The plan of the paper is as follows.  In Section
\ref{KSintopos}, we show how the Kochen-Specker theorem can be
viewed as asserting the non-existence of global sections of
certain presheaves that are naturally associated with any
quantum theory.  A key ingredient here is the idea that the set
of all bounded, self-adjoint operators forms an appropriate
category on which to form presheaves, as does the set of all
Boolean subalgebras of projectors. Readers unfamiliar with topos
theory may find it helpful to read the Appendix before embarking
on this section.

By rewriting the Kochen-Specker theorem in terms of presheaves,
several ways of generalizing the idea of a valuation present
themselves. In this paper we pursue {\em one\/} particular
scheme: to motivate the definition we finally arrive at, we show
in Section \ref{Sec:PartVal} how a partial valuation (of the
type used in the extant modal interpretations of quantum theory)
gives rise to a generalized valuation whose truth-values lie in
the Heyting algebra of sieves on an object in the category of
self-adjoint operators. By these means, we arrive naturally at
contextualized, multi-valued truth-value assignments.

Then, in Section \ref{Sec:GenVal} we use these results to
motivate the formal definition of a generalized valuation, and
we show how any state in a quantum theory gives rise to one
such. In Section \ref{Sec:BoolSubContext}, we extend these ideas
to the case where the space of contexts is taken as the category
of all Boolean subalgebras of projectors, rather than the
category of self-adjoint operators.

This paper is intended to be the first in a series devoted to an
extensive analysis of the possible uses of topos ideas in
quantum theory. Our main aim in the present paper is to present
the basic mathematical tools and some general ideas about using
quantum states to produce generalized valuations, but this
leaves much work to be done: in particular, an analysis of the
philosophical implications of generalized truth-values will be
given in a future paper, as will the way in which similar ideas
can arise in classical physics \cite{BI98a}.  For this reason,
the present paper concludes with a short summary of what has
been achieved so far, and a list of some of the more significant
topics for further research.

\section{The Kochen-Specker Theorem in the Language of Topos
Theory} \label{KSintopos}
\subsection{The Categories of Boolean Algebras and
Self-adjoint Operators} A key step in formulating the
Kochen-Specker theorem in the language of topos theory is the
construction of several categories that will form the domains of
the presheaf functors we shall be using later. Readers
unfamiliar with topos theory may find it helpful to read the
Appendix first. This contains a short introduction to the
relevant parts of topos theory, particularly the theory of
presheaves and the associated use of sieves as generalized
truth-values.

We start with the set $\cal W$ of all Boolean subalgebras of the
lattice ${\cal P}({\cal H})$ of projection operators on the
Hilbert space $\cal H$ of the quantum system. This forms a poset
under subalgebra inclusion, $W_2\subseteq W_1$.  As with any
poset, $\cal W$ can be regarded as a category in which (i) the
objects are defined to be the elements $W\in\cal W$ of the
poset; and (ii) a morphism is defined to exist from $W_2$ to
$W_1$ if $W_2\subseteq W_1$; we shall write this morphism as
$i_{W_2W_1}:W_2\rightarrow W_1$. Thus there is at most one
morphism between any two objects.

The next step is to introduce the set $\cal O$ of all bounded,
self-adjoint operators on the Hilbert space $\cal H$. First,
recall that any such operator $\hat A$ has the spectral
representation\footnote{As usual, the expression in Eq.\
(\ref{spectral-rep}) is shorthand for the equation
$\langle\psi,\hat A\phi\rangle=\int \lambda\, d\langle\psi, \hat
E_\lambda \phi\rangle$ for all $\psi,\phi\in\cal H$, whose right
hand side is to be interpreted as a Stieltjes integral. A
similar remark applies to the integrals in Eq.\ (\ref{Def:EAD})
and Eq.\ (\ref{Def:f(hatA)}).}
\begin{equation}
    \hat A=\int_{\sigma(\hat A)} \lambda\, d\hat E^A_\lambda
                                \label{spectral-rep}
\end{equation}
where $\sigma(\hat A)\subset\mathR$ is the spectrum of $\hat A$,
and $\{\hat E^A_\lambda\mid \lambda\in\sigma(\hat A)\}$ is the
spectral family of $\hat A$. The spectral projection operators
$\hat E[A\in\Delta]$ are determined by the spectral family
according to
\begin{equation}
    \hat E[A\in\Delta]=\int_\Delta d\hat E^A_{\lambda}
                                    \label{Def:EAD}
\end{equation}
where $\Delta$ is any Borel subset of the spectrum of $\hat A$.
In particular, if $a$ belongs to the discrete spectrum of $\hat
A$, the projector onto the eigenspace with eigenvalue $a$ is
\begin{equation}
        \hat E[A=a]:=\hat E[A\in\{a\}].
\end{equation}

    Then, if $f:\mathR\rightarrow\mathR$ is any
bounded Borel function, the operator $f(\hat A)$ is defined by
\begin{equation}
    f(\hat A):=\int_{\sigma(\hat A)}f(\lambda)\, d\hat
                E^A_\lambda.            \label{Def:f(hatA)}
\end{equation}
Note that if functions $f$ and $g$ exist such that $\hat
B=f(\hat A)$ and $\hat B =g(\hat A)$, this does not imply that
$f$ and $g$ are equal: in the discrete case it means only that
their restrictions to $\sigma(\hat A)$ are equal; more
generally, measure-theoretic issues arise, and we shall define
two bounded Borel functions $f,g:\sigma(\hat
A)\rightarrow\mathR$ to be {\em equivalent\/} if $f(\hat
A)=g(\hat A)$.

    We are now ready to turn $\cal O$ into a category. We define
the objects to be the elements of $\cal O$, and we say that
there is a `morphism' from $\hat B$ to $\hat A$ if there exists
a Borel function (more precisely, an equivalence class of Borel
functions) $f:\sigma(\hat A)\rightarrow \mathR$ such that $\hat
B=f(\hat A)$. As implied above, any such function on
$\sigma(\hat A)$ is unique (up to the equivalence relation), and
hence there is at most one morphism between any two operators;
if such exists---{\em i.e.}, if $\hat B=f(\hat A)$, for some
$f:\sigma(\hat A)\rightarrow\mathR$---the corresponding morphism
in the category $\cal O$ will be denoted $f_{\cal O}: \hat
B\rightarrow\hat A$. Note that we could make the corresponding
definitions for any subset of $\cal O$ that it is closed under
the action of constructing functions of its members. In what
follows, we shall be especially concerned with the category
${\cal O}_d$ of all bounded self-adjoint operators whose spectra
are discrete.

The categories $\cal W$ and $\cal O$ are closely
related\footnote{Another, closely related, category has as its
objects the abelian subalgebras of the algebra of bounded,
self-adjoint operators. The fact that this can be regarded as a
category was mentioned in the original paper of Kochen and
Specker \cite{KS67}.} via a certain covariant functor ${\bf
W}:{\cal O}\rightarrow\cal W$:
\begin{definition}\label{Defn:functorW}
The {\em spectral algebra functor\/} is the covariant functor
${\bf W}:{\cal O}\rightarrow\cal W$ defined as follows:
\begin{itemize}
\item On objects:  ${\bf W}(\hat A):=W_A$, where
$W_A$ is the spectral algebra of the operator $\hat A$ ({\em
i.e.}, the collection of all projectors onto the subspaces of
$\cal H$ associated with Borel subsets of $\sigma(\hat A)$).

\item On morphisms: If $f_{\cal O}:\hat B\rightarrow\hat A$,
then ${\bf W}(f_{\cal O}):W_B\rightarrow W_A$ is defined as the
subset inclusion $i_{W_BW_A}:W_B\rightarrow W_A$.
\end{itemize}
\end{definition}
In defining the map ${\bf W}(f_{\cal O}):W_B\rightarrow W_A$ we
have exploited the fact that the spectral algebra for $\hat
B=f(\hat A)$ is naturally embedded in the spectral algebra for
$\hat A$ according to the result $\hat E[f(A)\in J]=\hat E[A\in
f^{-1}(J)]$, for all Borel subsets $J\subseteq\sigma(\hat B)$.
Rigorously speaking, we could write $i_{W_{f(A)}W}(\hat
E[f(A)\in J])=\hat E[A\in f^{-1}(J)]$.

Note that we have defined $f_{\cal O}$ to be a morphism from
$\hat B$ to $\hat A$---rather than from $\hat A$ to $\hat
B$---so as to ensure that the categories $\cal O$ and $\cal W$
match up in this way. One consequence of this choice is the
reversal of arrows in the equation
\begin{equation}
        f_{\cal O}\circ g_{\cal O}=(g\circ f)_{\cal O}
\end{equation}
where the left hand side denotes composition in the category
$\cal O$, and the right hand side denotes normal composition of
functions, so that if $\hat B=f(\hat A)$ and $\hat C=g(\hat B)$,
the functional relation $\hat C= g(f(\hat A))\equiv g\circ
f(\hat A)$ translates to the morphism $f_{\cal O}\circ g_{\cal
O}:\hat C\rightarrow\hat A$ in the category $\cal O$.

It should be noted that pairs of operators $\hat A \neq\hat B$
exist such that $\hat B=f(\hat A)$ and $\hat A=g(\hat B)$ for
suitable functions $f$ and $g$. In the category $\cal O$, these
relations become $f_{\cal O}:\hat B\rightarrow \hat A$ and
$g_{\cal O}:\hat A\rightarrow \hat B$ with
\begin{equation}
    g_{\cal O}\circ f_{\cal O}=\mbox{id}_B;\ \
    f_{\cal O}\circ g_{\cal O}=\mbox{id}_A. \label{iso-objects}
\end{equation}
One consequence of the existence of such pairs is that $\cal O$
is only a pre-ordered space since it lacks the antisymmetry
property\footnote{A pre-ordered set $X$ is said to have the {\em
antisymmetry\/} property if $x\preceq y$ and $y\preceq x$
implies $x=y$.} of a true poset (which $\cal W$ is). However, it
follows from Eq.\ (\ref{iso-objects}) that two such operators
are {\em isomorphic\/} objects in the category $\cal O$, and it
is therefore possible to construct a new category $[\cal O]$
whose objects are the equivalence classes of operators, where
two operators are regarded as being equivalent if they are
isomorphic as objects in $\cal O$.  Finally, we note that if
$\hat A$ and $\hat B$ are related as in Eq.\ (\ref{iso-objects})
then they have the same spectral algebras; {\em i.e.},
$W_A=W_B$, and hence $[{\cal O]}$ is closely related to the
category $\cal W$.

\subsection{The Spectral Presheaf on ${\cal O}_d$ and the
Kochen-Specker Theorem} A central step in developing our use of
topos theory is the observation that the spectra of the
self-adjoint operators on a Hilbert space can be used to form a
presheaf on the category ${\cal O}_d$ of self-adjoint operators
whose spectra are discrete. Specifically:
\begin{definition}\label{Defn:spectral-presheaf}
The {\em spectral presheaf} on ${\cal O}_d$ is the contravariant
functor ${\bf\Sigma}:{\cal O}_d \rightarrow {\rm Set}$ defined
as follows:
\begin{enumerate}
    \item On objects: ${\bf\Sigma}(\hat A):=\sigma(\hat
A)$---the spectrum of the self-adjoint operator $\hat A$.

    \item On morphisms: If $f_{{\cal O}_d}:\hat B\rightarrow \hat
A$, so that $\hat B=f(\hat A)$, then ${\bf\Sigma}(f_{{\cal
O}_d}):\sigma(\hat A)\rightarrow \sigma (\hat B)$ is defined by
${\bf\Sigma}(f_{{\cal O}_d})(\lambda):=f(\lambda)$ for all
$\lambda\in \sigma(\hat A)$.

\end{enumerate}
\end{definition}
Note that ${\bf\Sigma}(f_{{\cal O}_d})$ is well-defined since,
if $\lambda\in\sigma(\hat A)$, then $f(\lambda)$ is indeed an
element of the spectrum of $\hat B$; indeed, for these
discrete-spectrum operators we have $\sigma(f(\hat
A))=f(\sigma(\hat A))$.

It is straightforward to see that ${\bf\Sigma}$ is a genuine
functor.  It clearly respects domains and codomains of a
morphism in ${\cal O}_d$ in the desired way, and
${\bf\Sigma}({\rm id}_A)=\mbox{id}_{\sigma(A)}$. The key step is
to show that ${\bf\Sigma}(f_{{\cal O}_d}\circ g_{{\cal
O}_d})={\bf\Sigma}(g_{{\cal O}_d})\circ{\bf\Sigma}(f_{{\cal
O}_d})$.  So, suppose that $f_{{\cal O}_d}:\hat B\rightarrow
\hat A$ and $g_{{\cal O}_d}:\hat C\rightarrow\hat B$, so that
$\hat B=f(\hat A)$ and $\hat C=g(\hat B)$. Then $f_{{\cal
O}_d}\circ g_{{\cal O}_d}:\hat C\rightarrow\hat A$ with $\hat
C=g(f(\hat A))=g\circ f(\hat A)$.  Hence, for all
$\lambda\in\sigma(\hat A)$, we have
\begin{equation}
{\bf\Sigma}(f_{{\cal O}_d}\circ g_{{\cal O}_d})(\lambda)
=g(f(\lambda)) ={\bf\Sigma}(g_{{\cal O}_d})(f(\lambda))
={\bf\Sigma}(g_{{\cal O}_d})({\bf\Sigma}(f_{{\cal
O}_d})(\lambda)) ={\bf\Sigma}(g_{{\cal O}_d})\circ
            {\bf\Sigma}(f_{{\cal O}_d})(\lambda)
\end{equation}
so that
\begin{equation}
    {\bf\Sigma}(f_{{\cal O}_d}\circ g_{{\cal O}_d})=
        {\bf\Sigma}(g_{{\cal O}_d})\circ
            {\bf\Sigma}(f_{{\cal O}_d})
\end{equation}
as required. Thus ${\bf\Sigma}$ is a contravariant functor from
${\cal O}_d$ to ${\rm Set}$, and hence a presheaf on ${\cal
O}_d$.

    The key remark now is the following. As discussed in the
Appendix, a {\em global section\/}, or {\em global element\/},
of a contravariant functor ${\bf X}:{\cal C}\rightarrow{\rm
Set}$ is defined to be a function $\gamma$ that assigns to each
object $A$ in the category $\cal C$ an element $\gamma_A\in{\bf
X}(A)$ in such a way that if $f:B\rightarrow A$ then ${\bf
X}(f)(\gamma_A)=\gamma_B$, as in Eq.\ (\ref{Def:global}).

In the case of the spectral functor ${\bf\Sigma}:{\cal O}_d^{\rm
op}\rightarrow {\rm Set}$, a global section/element is therefore
a function $\gamma$ that assigns to each self-adjoint operator
$\hat A$ with a purely discrete spectrum, a real number
$\gamma_A\in \sigma(\hat A)$ such that if $\hat B=f(\hat A)$
then $f(\gamma_A)=\gamma_B$. But this is precisely the condition
FUNC in Eq.\ (\ref{funct-rule}) for a valuation!  Thus, for
operators with a discrete spectrum, the Kochen-Specker theorem
is equivalent to the statement that, if $\dim{\cal H}>2$, there
are no global sections of the spectral presheaf
${\bf\Sigma}:{\cal O}_d^{\rm op}\rightarrow {\rm Set}$.

The situation for operators whose spectra contains continuous
parts is more complex since it is no longer necessarily true
that $\sigma(f(\hat A))=f(\sigma(\hat A)$. Indeed, the most that
can be proved in general\footnote{For details see page 900 of
\cite{DS64}.} is that
\begin{equation}
    \sigma(f(\hat A))=\bigcap_{\Delta}\{\overline {f(\Delta)}
    \mid \hat E[A\in\Delta]=\hat 1\}    \label{sf(A)<}
\end{equation}
where $\overline{f(\Delta)}$ is the topological closure of
$f(\Delta)\subset\mathR$, and $\Delta$ denotes Borel subsets of
$\mathR$. The idea of the spectral presheaf can be extended to
this case by using a more sophisticated approach that involves
the spectral theorem for commutative von Neumann algebras.
However, we shall not develop this particular approach further
in the present paper because of the problematic physical meaning
of assigning an exact value to a quantity whose range of values
is continuous.  Of much greater relevance is the assignment of
truth-false values to propositions of the type `$A\in\Delta$',
as discussed in Section \ref{Sec:Introduction} and in the
original Kochen-Specker paper: as we shall see shortly in
Section \ref{SubSec:KSdualpresheaves}, the relevant presheaf in
this case can be defined for the category $\cal O$ of all
bounded self-adjoint operators on the Hilbert space of the
quantum system.

Note that, in the form above, the Kochen-Specker theorem looks
remarkably like the theorem in fibre-bundle theory which says
that there are no global cross-sections of a non-trivial
principal bundle \cite{Sin78}. Thus, {\em cum grano salis\/},
one might be tempted to say that the Kochen-Specker theorem in
quantum theory is analogous to the `Gribov effect' in Yang-Mills
gauge theories (which arises from the non-triviality of the
gauge bundle)!

More seriously, the non-triviality of a principal fibre bundle
is related to the existence of certain non-vanishing cohomology
classes that arise as obstructions to the step-wise construction
of a cross-section on the simplices of a locally-trivializing
triangulation of the base manifold. It would be intriguing to
see if the non-existence of global valuations in the quantum
theory can be related to the non-vanishing of some topos-based
cohomology structure. If so, this would open an perspective on
the Kochen-Specker theorem that would be extremely interesting;
not least because most of the existing literature on the theorem
is concerned with finding concrete counter-examples to the
existence of a global valuation rather than studying the
phenomenon in a general sense.

However, from our immediate perspective the most important
reason for presenting this topos-theoretic restatement of the
Kochen-Specker theorem is that, as we shall see, it suggests
specific ways of implementing the idea of constructing
generalized valuations; particularly in regard to using the
contextual logic that forms the heart of the theory of
presheaves.

\subsection{The Kochen-Specker Theorem in Terms of the Dual
Presheaves on $\cal W$ and on $\cal O$. }
\label{SubSec:KSdualpresheaves}

\paragraph*{1. The Dual Presheaf on $\cal W$:}

The Kochen-Specker theorem is usually stated in terms of the
features of a valuation on the Boolean subalgebras of the
lattice ${\cal P}({\cal H})$ of projectors on the Hilbert space
$\cal H$. This is useful for handling operators whose spectra
contain continuous parts; it is also the starting point for most
constructions of explicit counter-examples to the existence of
global valuations.  For these reasons, it is very useful to
restate, and extend, the results above using the category $\cal
W$ rather than ${\cal O}_d$. This will enable us in Section
\ref{SubSec:KSdualpresheaves}.2 to state the Kochen-Specker
theorem in terms of the category $\cal O$ of all bounded
self-adjoint operators.

Once again we start with the definition of an appropriate
presheaf; this time on the category $\cal W$.

\begin{definition}
\label{Defn:dual-presheaf-W} The {\em dual presheaf\/} on $\cal
W$ is the contravariant functor ${\bf D}:{\cal W}\rightarrow{\rm
Set}$ defined as follows:
\begin{enumerate}
\item On objects: ${\bf D}(W)$ is the {\em dual\/} of $W$;
thus it is the set ${\rm Hom}(W,\{0,1\})$ of all homomorphisms
from the Boolean algebra $W$ to the Boolean algebra $\{0,1\}$.

\item On morphisms: If $i_{W_2W_1}:W_2\rightarrow W_1$ then
${\bf D}(i_{W_2W_1}): {\bf D}(W_1)\rightarrow {\bf D}(W_2)$ is
defined by ${\bf D}(i_{W_2W_1})(\chi):=\chi|_{W_2}$ where
$\chi|_{W_2}$ denotes the restriction of $\chi\in {\bf D} (W_1)$
to the subalgebra $W_2\subseteq W_1$.
\end{enumerate}
\end{definition}

    A global section of the functor ${\bf D}:{\cal W}\rightarrow
{\rm Set}$ is then a function $\gamma$ that associates to each
$W\in\cal W$ an element $\gamma_W$ of the dual of $W$ such that
if $i_{W_2W_1}: W_2\rightarrow W_1$ then
$\gamma_{W_1}|_{W_2}=\gamma_{W_2}$; thus, for all $\hat\alpha\in
W_2$,
\begin{equation}
\gamma_{W_2}(\hat\alpha)=\gamma_{W_1}((i_{W_2W_1}(\hat\alpha)).
\end{equation}

Since each projection operator belongs to at least one Boolean
algebra (for example, the algebra $\{\hat 0,\hat
1,\hat\alpha,\neg\hat\alpha\}$) it follows that a global section
of ${\bf D}:{\cal W}^{\rm op}\rightarrow{\rm Set}$ associates to
each projection operator $\hat\alpha$ a number $V(\hat\alpha)$
which is either $0$ or $1$, and is such that, if
$\hat\alpha\land\hat\beta=\hat 0$, then
$V(\hat\alpha\lor\hat\beta)=V(\hat\alpha)+V(\hat\beta)$. These
are precisely the types of valuation considered in the
discussions of the Kochen-Specker theorem that focus on the
construction of specific counter-examples. Thus an alternative
way of expressing the Kochen-Specker theorem is that, if
$\dim{\cal H}>2$, the dual presheaf ${\bf D}:{\cal W}^{\rm
op}\rightarrow{\rm Set}$ has no global sections.

\paragraph*{2. The Dual Presheaf on $\cal O$:}

The covariant functor ${\bf W}:{\cal O}\rightarrow {\cal W}$ of
Definition \ref{Defn:functorW} and the contravariant functor
${\bf D}:{\cal W}\rightarrow{\rm Set}$, can be composed to give
a contravariant functor ${\bf D}\circ{\bf W}:{\cal O}\rightarrow
{\rm Set}$, which we shall call the {\em dual presheaf\/} on
$\cal O$. It has the following properties:
\begin{enumerate}
    \item On objects: ${\bf D}\circ{\bf W}(\hat A)$ is the dual
of the spectral Boolean algebra $W_A$; thus it is the set ${\rm
Hom}(W_A,\{0,1\})$ of all homomorphisms from $W_A$ to the
Boolean algebra $\{0,1\}$.

    \item On morphisms: If $f_{\cal O}:\hat B\rightarrow\hat A$
then ${\bf D}\circ{\bf W}(f_{\cal O}):D(W_A)\rightarrow D(W_B)$
is defined by ${\bf D}\circ{\bf W}(f_{\cal O})(\chi):=
\chi|_{W_{f(A)}}$ where $\chi|_{W_{f(A)}}$ denotes the
restriction of $\chi\in D(W_A)$ to the subalgebra
$W_{f(A)}\subseteq W_A$.
\end{enumerate}

Note that a global section $\gamma$ of the presheaf ${\bf
D}\circ{\bf W}:{\cal O}\rightarrow {\rm Set}$ would correspond
to a consistent association of each physical quantity $A$ with
an element $\gamma_A\in{\rm Hom}(W_A,\{0,1\})$, and hence with a
statement of which propositions of the form `$A\in\Delta$' are
true, and which are false. The non-existence of such global
sections is perhaps the most physically transparent statement of
the Kochen-Specker theorem in the language of presheaves.

Finally, we note that, as might be expected, there is a close
relationship between the spectral presheaf ${\bf\Sigma}$ on
${\cal O}_d$ and the corresponding dual presheaf ${\bf
D}\circ{\bf W}$ on ${\cal O}_d$.  Specifically, there is a
natural transformation $T:{\bf \Sigma}\rightarrow {\bf
D}\circ{\bf W}$ between these presheaves, whose component
$T_A:{\bf \Sigma}(\hat A)\rightarrow {\bf D}\circ{\bf W}(\hat
A)$ at each stage $\hat A\in{\cal O}_d$,
\begin{equation}
    T_A:\sigma(\hat A)\rightarrow {\rm Hom}(W_A,\{0,1\})
\end{equation}
is defined by (where $\lambda\in\sigma(\hat A)$)
\begin{equation}
    T_A(\lambda)(\hat E[A\in\Delta]):=\left\{
                \begin{array}{ll}
                    1 & \mbox{if $\lambda\in\Delta$}\\
                    0 & \mbox{otherwise}
                \end{array}
                \right.
\end{equation}
for all projection operators $\hat E[A\in\Delta]\in W_A$.

\section{From Partial Valuations to Generalized Valuations}
\label{Sec:PartVal}
\subsection{Some Implications of the Presheaf Version of the
Kochen-Specker Theorem}

We are now ready to begin the presentation of our theory of
generalized valuations.  From a pedagogical perspective, this
could be done in several ways.  One possibility would be to
start with the formal definition and then to exhibit some
physically relevant examples. However, although the definition
of a generalized valuation is partly motivated by the
conclusions of our earlier discussion in Section
\ref{SubSec:primafacie}, one of the central components---the
presheaf analogue of the functional composition principle
FUNC---is best justified by seeing how it arises in a particular
case. Therefore, we shall devote this section to a fairly
extensive discussion of a concrete example of a particular class
of generalized valuation that will serve to illustrate the ideas
that lie behind our later, more abstract, constructions in
Sections \ref{Sec:GenVal} and \ref{Sec:BoolSubContext}.

As we have seen, the Kochen-Specker theorem asserts that, if
$\dim{\cal H} > 2$, there do not exist valuations that are
globally defined in the sense that FUNC is satisfied for all
pairs of operators $\hat A$, $\hat B$ (with discrete spectra) in
the Hilbert space with $\hat B=f(\hat A)$ for some $f$; or, in
the language of topos theory, the spectral presheaf
${\bf\Sigma}:{{\cal O}_d}^{\rm op}\rightarrow {\rm Set}$ has no
global sections. More generally, the theorem asserts that there
are no global sections of the dual presheaf ${\bf D}\circ{\bf
W}$ on $\cal O$; and hence there is no consistent way of
assigning the values true or false to propositions of the type
`$A\in\Delta$' for all bounded physical quantities $A$.

Rewriting the Kochen-Specker theorem in the language of topos
theory suggests several ways in which the idea of a valuation
might be generalized so that globally-defined entities {\em
do\/} exist.  For example, one possibility is to embed the
spectral presheaf $\bf \Sigma$ in a larger presheaf that does
have global elements.  The existence of at least one such
presheaf follows from some general considerations in topos
theory\footnote{The existence of injective resolutions of a
presheaf.}: in the present case, a relevant example is the
presheaf on $\cal O$ whose objects are {\em subsets\/} of
$\sigma(\hat A)$ at each stage of truth $\hat A$. A global
section of this presheaf would comprise a consistent assignment
of a {\em range\/} of values for each physical quantity. This
option sounds physically plausible, and is something to which we
may return in a later paper.

Another possibility is to replace the dual presheaf ${\bf
D}\circ{\bf W}$ on $\cal O$ with a presheaf $\bf H$ in which
${\bf H}(\hat A)$ is defined to be the set of homomorphisms from
$W_A$ into some larger algebra than the $\{0,1\}$ used by ${\bf
D}\circ{\bf W}$, thus building in the idea of multi-valued truth
in a rather direct way. Of course, guided by our remarks in the
Introduction, this target logic could itself depend on the stage
of truth $\hat A$ ({\em i.e.}, it could be contextual), and it
is not clear that we would want to use genuine homomorphisms;
for example, if the target algebra was an intuitionistic logic,
then the negation operation would behave differently from that
in $W_A$, as was mentioned briefly in the Introduction in the
context of the (incorrect!) equation Eq.\ (\ref{muneg=negmu}).
We shall see an example of this type of structure in Section
\ref{Sec:BoolSubContext} in the form of the `valuation presheaf'
of Definition \ref{Defn:valuation-presheaf}.

However, in this Section, we will take our departure from the
property of presheaves that even if a global section/element
does not exist, typically there will be plenty of {\em local \/}
sections (just as there are in a non-trivial principal bundle),
which are defined to be morphisms of a subobject of the terminal
object into the presheaf. In the case of the spectral presheaf
${\bf\Sigma}$, any such local element corresponds to what we
shall call a `partial' valuation, and the main thrust of this
section of the paper is to show how each such locally-defined
normal valuation (`normal' in the sense that assigned values lie
in the minimal Boolean algebra $\{0,1\}$) gives rise to a {\em
globally\/} defined `generalized' valuation with truth-values in
the Heyting algebra of sieves on $\cal O$.  This also allows
comparison to be made with the various modal approaches to the
interpretation of quantum theory, all of which use local
valuations of one type or another; however, we shall not pursue
that comparison in this paper.

\subsection{The Idea of a Partial Valuation}
\label{SubSec:part-val} The precise definition of a `partial
valuation' is that it is a local section of the spectral
presheaf $\bf\Sigma$ on the category ${\cal O}_d$ of bounded
self-adjoint operators with discrete spectra. This translates
into the following explicit set of properties:

\begin{definition}
A {\em partial valuation\/} on the set of bounded, self-adjoint
operators with discrete spectra is a map $V:{\rm dom\,}
V\rightarrow\mathR$ defined on a subset ${\rm dom\,}V$ of such
operators (called the {\em domain\/} of $V$) such that:
\begin{enumerate}
\item If $\hat A\in{\rm dom\,}V$, then $V(\hat A)\in\sigma
(\hat A)$.

\item If $\hat A\in{\rm dom\,}V$ and $\hat B=f(\hat A)$ then
(i) $\hat B\in{\rm dom\,}V$; and (ii) $V(\hat B)=f(V(\hat A))$.
\end{enumerate}
\end{definition}
One consequence of this definition is that if $\hat A$ belongs
to the domain of $V$, then so do all its spectral projectors.
This is because any such projector $\hat E[A\in\Delta]$ can be
written as
\begin{equation}
        \hat E[A\in\Delta]=\chi_\Delta(\hat A)
\end{equation}
where $\chi_\Delta:\sigma(\hat A)\rightarrow\mathR$ is the
characteristic function of $\Delta\subseteq\sigma(\hat A)$. It
follows that
\begin{equation}
    V(\hat E[A\in\Delta])=\chi_\Delta(V(\hat A))
            = \left\{ \begin{array}{ll}
                    1 &\mbox{if $V(\hat A)\in\Delta$;}\\[3pt]
                    0 & \mbox{otherwise.}
                        \end{array}
                \right.\label{V[AinD]}
\end{equation}

    Note that, provided ${\rm dom\,}V\neq\emptyset$, real
multiples of the unit operator $\hat 1$ belong to the domain of
any partial valuation $V$. This is because if $\hat A$ is any
operator in ${\rm dom\,}V$, then $r\hat 1=c_r(\hat A)$ where
$c_r:\sigma(\hat A)\rightarrow\mathR$, $r\in\mathR$, is the
constant map $c_r(a):=r$ for all $a\in\sigma(\hat A)$. This also
shows that $V(\hat 1)=1$.

    The definition of a partial valuation is not empty since
there clearly exists a `trivial' example $V_0$ whose domain is
defined as ${\rm dom\,}V_0:=\{r\hat 1\mid r\in\mathR\}$, and
with $V_0(r\hat 1):=r$.  However, non-trivial partial valuations
also exist. For example, we have the following definition:
\begin{definition}
Let $\hat M$ be any bounded, self-adjoint operator with a purely
discrete spectrum, and let $m\in\sigma(\hat M)$ be one its
eigenvalues.  Then the associated partial valuation $V^{M,m}$ is
defined as follows:
\begin{enumerate}
\item {The domain of $V^{M,m}$ is
\begin{equation}
    {\rm dom\,}V^{M,m}:=\downarrow\!\hat M:=\{f_{{\cal O}_d}:\hat
A\rightarrow\hat M\}= \{\hat A\mid \exists f {\rm\ s.t.\ } \hat
A=f(\hat M)\}, \end{equation} where the last equality holds
since there is at most one morphism between two objects in
${\cal O}_d$.  }

\item{If $\hat A\in{\rm dom\,}V^{M,m}$ with $\hat A=f(\hat M)$,
then the value of $V^{M,m}(\hat A)$ is
    \begin{equation}
        V^{M,m}(\hat A):=f(m).  \label{Def:Vm}
    \end{equation}
    }
\end{enumerate}
\end{definition}
It is straightforward to check that this satisfies the
requirements for a partial valuation.

Note that, generally speaking, a partial valuation of this type
can be extended `upwards' in the sense that if there is a
morphism $h_{{\cal O}_d}:\hat M\rightarrow\hat N$, so that $\hat
M=h(\hat N)$, then $V^{M,m}$ can be extended to $\hat N$ by
defining $V^{M,m}(\hat N)$ to be any eigenvalue $n$ of $\hat N$
such that $h(n)=m$ (there must be at least one such eigenvalue
since $\sigma(\hat M)=h(\sigma(\hat N))$). Therefore, one might
as well suppose in the first place that $\hat M$ is a {\em
maximal\/} operator\footnote{In the present context, we could
define an operator $\hat M$ to be {\em maximal\/} if, for any
operator $\hat N$ and function $h:\sigma(\hat
N)\rightarrow\mathR$ such that $\hat M=h(\hat N)$, there exists
$g:\sigma(\hat M)\rightarrow\mathR$ such that $\hat N=g(\hat
M)$; {\em i.e.}, $h_{\cal O}:\hat M\rightarrow \hat N$ implies
that $\hat M$ and $\hat N$ are isomorphic objects in the
category $\cal O$.}.

The domain of a valuation $V^{M,m}$ forms a commutative set of
operators; however, there is no reason in general why the domain
of a partial valuation should be commutative. For example, the
use of a non-abelian domain forms an integral part of the modal
interpretation of Clifton and Bub \cite{Cli95a,BC96,Bub97}.

    We note that a partial valuation $V$ gives a simple
`false-true' assignment to propositions of the type
`$A\in\Delta$' provided that $A$ lies in the domain of $V$;
specifically:
\begin{equation}
    V(A\in\Delta):=\left\{\begin{array}{ll}
            \mbox{`true' if  $V(A)\in\Delta$};\\[2pt]
            \mbox{`false' otherwise}.
                            \end{array}
                    \right.
\end{equation}
Thus the proposition `$A\in\Delta$' is true if $A$ lies in the
domain of $V$, and if the value of $A$ lies in the range
$\Delta$; it is false, if $A$ lies in the domain of $V$ and the
value of $A$ does not lie in $\Delta$. If $A$ is {\em not\/} in
the domain of $V$, no truth-value at all is assigned to
propositions about the value of $A$. Of course, these
assignments are consistent with the assignment in Eq.\
(\ref{V[AinD]}) of a $0-1$ value to the projection operator
$\hat E[A\in\Delta]$.

\subsection{The Construction of a Generalized Valuation from a
Partial Valuation}
\paragraph*{1. The Basic Idea:}

Let $V$ be any partial valuation, and consider a proposition of
the form `$A=a$', where $a$ is an eigenvalue of $\hat A$ and
where $\hat A$ does {\em not\/} lie in the domain of $V$. The
implication of the Kochen-Specker theorem is that it may not be
possible to extend the domain of $V$ to include $\hat A$. If
this is indeed the case, then the proposition `$A=a$' cannot be
given a value true or false in a way that is consistent with the
values already given by $V$ to the operators in its domain.

However, consider a proposition of the form `$f(A)=f(a)$'. As
was emphasized earlier, this will generally be weaker than
`$A=a$'; both in a conceptual sense---knowing that the quantity
$f(A)$ has the value $f(a)$ gives only limited information on
the value of $A$ itself (it could be any number $b$ such that
$f(b)=f(a)$)---and in the mathematical sense that, in the
lattice of projection operators, ({\em cf.}, Eq.\
(\ref{1EA<EfA})),
\begin{equation}
        \hat E[A=a]\leq \hat E[f(A)=f(a)]
                \label{E[A=a]<=FfA}
\end{equation}
where $\hat E[A=a]$ projects onto the eigenspace of $\hat A$
with eigenvalue $a$, and $\hat E[f(A)=f(a)]$ projects onto the
eigenspace of $f(\hat A)$ with eigenvalue $f(a)$. More precisely
\begin{equation}
    \hat E[f(A)=f(a)]=\sum_{b\in\sigma(A), f(b)=f(a)}
    \hat E[A=b]\quad = \quad \hat E[A\in f^{-1}(f(\{a\}))].
                \label{Ef(A)=sum}
\end{equation}
In other words, $\hat E[f(A)=f(a)]$ is the sum of the
(orthogonal) set of those projectors in the spectral
decomposition of $\hat A$ whose corresponding eigenvalues are
mapped into the number $f(a)$ by the function $f:\sigma(\hat
A)\rightarrow\mathR$.

The key remark is then the following. It is possible that, for
one (or more) function $f$, (i) the coarse-grained operator
$f(\hat A)$ {\em does\/} lies in the domain of $V$ ({\em i.e.},
at least part of the spectral algebra of $\hat A$ lies in ${\rm
dom\,}V$); and (ii) $V(f(\hat A))=f(a)$.  Under these
circumstances, we can assign a {\em true\/} value to the {\em
weaker\/} proposition `$f(A)=f(a)$', and thereby assign a {\em
partial\/} truth-value to the original proposition
`$A\in\Delta$'. We note that if $g:\sigma(f(\hat
A))\rightarrow\mathR$ then $V(g(f(\hat A)))=g(f(a))$---{\em
i.e.}, $V(g\circ f(\hat A))=g\circ f(a)$---and hence if the
function $f$ satisfies the above conditions, so does $g\circ f$
for any $g$; in other words, the set of such functions
determines a {\em sieve\/} on $\hat A$ in the category ${\cal
O}_d$.

Motivated by these remarks, we propose the following definition
of a {\em generalized valuation associated\/} with a partial
valuation.

\begin{definition}\label{Defn:partval-genval}
Given a partial valuation $V$ on the set of bounded self-adjoint
operators with discrete spectra, the {\em associated generalized
valuation\/} is defined on a proposition `$A=a$' as
\begin{equation}
\nu^V(A=a):=\left\{\begin{array}{ll} \{f_{{\cal O}_d}:\hat
B\rightarrow\hat A\mid \hat B\in {\rm
            dom\,}V,\; V(\hat B)=f(a)\} & \mbox{if
$a\in\sigma(\hat A)$;}\\[5pt] \emptyset & \mbox{otherwise.}
\end{array} \label{Def:nVA=a} \right.
\end{equation}
\end{definition}

A crucial consequence of this definition is that, as indicated
above, $\nu^V(A=a)$ is a {\em sieve\/} on $\hat A$ in the
category ${\cal O}_d$.  Indeed, suppose $f_{{\cal O}_d}:\hat
B\rightarrow\hat A$ belongs to $\nu^V(A=a)$, and consider any
morphism $g_{{\cal O}_d}:\hat C\rightarrow\hat B$. Then, since
$\hat B\in{\rm dom\,}V$, and $\hat C=g(\hat B)$, the definition
of a partial valuation shows that (i) $\hat C\in{\rm dom\,}V$,
and (ii) $V(\hat C)=g(V(\hat B))$. However, $g(V(\hat
B))=g(f(a))=g\circ f(a)$; and hence $V(\hat C)=g\circ f(a)$.
Thus $f_{{\cal O}_d}\circ g_{{\cal O}_d}:\hat C\rightarrow\hat
A$ is in the set $\nu^V(A=a)$, which is therefore a sieve.

Thus the partial truth-value $\nu^V(A=a)$ of the proposition
`$A=a$' is defined to be the sieve on $\hat A$ of
coarse-grainings $f(\hat A)$ of $\hat A$, at which the
proposition `$f(A)= f(a)$' is `totally' true according to the
partial valuation $V$.

\paragraph*{2. The Origin of Contextuality:}

The fact that $\nu^V(A=a)$ is a sieve is of considerable
importance since it shows that the target space of the valuation
$\nu^V(A=a)$ is a genuine mathematical logic: namely, the
Heyting algebra ${\bf\Omega}(\hat A)$ of sieves on the object
$\hat A$ in the category ${\cal O}_d$.

In more general terms, the sieve-like nature of the generalized
valuation gives strong support to our claim that topos theory is
the appropriate mathematical framework in which to develop these
ideas. This is particularly so in regard to the presheaf idea of
`contextual' logic.  From the defining property of a generalized
valuation in Eq.\ (\ref{Def:nVA=a}), it is clear that if the
propositions `$A=a$' and `$C=c$' happen to correspond to the
{\em same\/} projection operator $\hat P$, so that $\hat
E[A=a]=\hat E[C=c]=\hat P$, this does {\em not\/} mean that
$\nu^V(A=a)$ is equal to $\nu^V(C=c)$; indeed, the former is a
sieve on $\hat A$, whilst the latter is a sieve on $\hat C$.
Furthermore, if the projection operator $\hat P$ is thought of
as representing some physical quantity $P$ directly, then the
proposition `$P=1$' can also be assigned a partial truth-value
$\nu^V(P=1)$; which, as a sieve on $\hat P$, is different from
both $\nu^V(A=a)$ and $\nu^V(C=c)$.

The situation can be summarized by saying that if we think of
ourselves as assigning partial truth-values to projection
operators, then the actual value assigned to any specific
projector $\hat P$ will depend on the {\em context\/}
chosen---{\em i.e.}, we have to choose a particular self-adjoint
operator $\hat A$ from the set of all operators whose associated
sets of spectral projectors include $\hat P$; hence each context
corresponds to a `stage of truth' for the presheaf.

Thus we see that, in the notation `$\nu^V(A=a)$', the argument
`$A=a$' serves two purposes: (i) it specifies the associated
projection operator $\hat E[A=a]$; and (ii) it indicates the
context ({\em i.e.,} $\hat A$) in which a partial truth-value is
to be ascribed to this projector. This manifest contextuality is
one of the crucial features that distinguishes our scheme from a
naive one in which one tries simply to assign to each projector
the value $1$ or $0$---an attempt that immediately falls foul of
the Kochen-Specker theorem.

If desired, this situation can be reflected in the notation by
rewriting $\nu^V(A=a)$ as $\nu^V_A(\hat P)$ to emphasize that
the former can be construed as the partial truth-value assigned
to the projection operator $\hat P$ ($=\hat E[A=a]$) in the
context/stage of truth of the self-adjoint operator $\hat A$.
Notice that, as is characteristic of presheaf logic, the Heyting
algebra to which $\nu^V_A(\hat P)$ belongs itself depends on the
context $\hat A$; namely, it is the algebra ${\bf\Omega}(\hat
A)$ of sieves on $\hat A$.

\paragraph*{3. Extending to Propositions `$A\in\Delta$':}

The construction above of a generalized valuation can be
extended in an obvious way to include more general propositions
of the form `$A\in\Delta$', where $\Delta$ is any
Borel\footnote{Note that any subset of the spectrum of an
operator in ${\cal O}_d$ is Borel, and hence the qualification
is unnecessary. However, we shall leave in references to `Borel'
subsets as this {\em is\/} of importance for operators whose
spectra is not just discrete.} subset of the spectrum
$\sigma(\hat A)$. Note that the set of these propositions is
naturally equipped with the logical structure of the Boolean
algebra of all Borel subsets of $\sigma(\hat A)$; in the quantum
theory, this algebra is represented by the spectral algebra
$W_A$ of projectors onto the eigenspaces associated with these
Borel subsets.

Specifically, we define:

\begin{definition}\label{Defn:partval-genval-Delta}
Given a partial valuation $V$, the associated {\em generalized
valuation\/} is defined on a proposition `$A\in\Delta$' as
\begin{equation}
\nu^V(A\in\Delta):=\{f_{{\cal O}_d}:\hat B\rightarrow \hat A
\mid \hat B\in{\rm dom\,}V,\; V(\hat B)\in
 f(\Delta)\}            \label{nVinDelta}
\end{equation}
\end{definition}
It is a straightforward exercise to show that the right hand
side is a sieve on $\hat A$ in the category ${\cal O}_d$.

\paragraph*{4. `Totally true' and `Totally false':}

This is a convenient point at which to give a precise meaning to
the concepts `totally true' and `totally false' that have been
employed up to now in a rather heuristic way. These concepts,
too, are contextual in nature.

The formal definition is as follows:
\begin{definition}
\
\begin{enumerate}
\item The proposition `$A\in\Delta$' is {\em totally true\/} at
the stage of truth $\hat A$ if
    \begin{equation}
        \nu^V(A\in\Delta)={\rm true}_A:=\;
\downarrow\!\!\hat A= \{f_{{\cal O}_d}:\hat B\rightarrow\hat
A\}.
                        \label{Def:totallytrue}
\end{equation}

\item The proposition `$A\in\Delta$' is {\em totally false\/} at
the stage of truth $\hat A$ if
\begin{equation}
        \nu^V(A\in\Delta)={\rm false}_A:=\emptyset.
                        \label{Def:totallyfalse}
\end{equation}
\end{enumerate}
\end{definition}
Thus a proposition is totally true in the context $\hat A$ if
its partial truth-value is equal to the principal sieve on $\hat
A$, which is the unit element in the Heyting algebra
${\bf\Omega}(\hat A)$; the proposition is totally false if it is
equal to the empty sieve, which is the zero element in
${\bf\Omega}(\hat A)$.

Note that if $\nu^V(A=a)=\downarrow\!\!\hat A$, then, in
particular, the identity morphism ${\rm id}_A:\hat
A\rightarrow\hat A$ belongs to the sieve $\nu^V(A=a)$. According
to the Definition \ref{Defn:partval-genval} this means that (i)
$\hat A\in{\rm dom\,}V$, and (ii) $V(\hat A)=a$. Conversely, if
$\hat A\in{\rm dom\,}V$ and $V(\hat A)=a$ then
$\nu^V(A=a)=\downarrow\!\!\hat A$. Thus the proposition `$A=a$'
is totally true if, and only if, $\nu^V(A=a)={\rm true}_A$.
Hence the notion of total truth of the proposition `$A=a$'
captures precisely the idea that the quantity $A$ does indeed
have a value, and that value is $a$.

More generally, $\nu^V(A\in\Delta)={\rm true}_A$ if, and only
if, $A$ lies in the domain of $V$, and the value of $A$ assigned
by $V$ lies in the subset $\Delta\subseteq\sigma(\hat A)$.

\paragraph*{5. Possible Modification of ${\cal O}_d$ to Remove
Minimal Truth-Values:}

As things stand, if $\Delta\neq\emptyset$, the proposition
`$A\in\Delta$' is never totally false since, as mentioned in
Section \ref{SubSec:part-val}, real multiples of the unit
operator $\hat 1$ belongs to the domain of any partial
valuation, and so $c_{r\,{{\cal O}_d}}:r\hat 1\rightarrow \hat
A$ with $V(r\hat 1)=r=c_r(a)$ for all $a\in\sigma(\hat A)$, is
bound to be in $\nu^V(A\in\Delta)$ if $\Delta\neq\emptyset$.
Thus the morphism $c_{r\,{{\cal O}_d}}:r\hat 1\rightarrow \hat
A$ always belongs to the sieve $\nu^V(A\in\Delta)$ provided only
that $\Delta$ is not the empty set.

If $\nu(A\in\Delta)=\{c_{r\,{{\cal O}_d}}:r\hat 1\rightarrow
A\mid r\in\mathR\}$ then we will say that the proposition
`$A\in\Delta$' is {\em minimally true\/}; that is, it really
provides no interesting information about the value of $A$.  If
desired, the existence of such minimal truth-values can be
removed by the simple expedient of replacing the category ${\cal
O}_d$ with the category ${\cal O}_{d*}$, which is defined to be
${\cal O}_d$ minus (i) the objects $r\hat 1$, $r\in\mathR$, and
(ii) all morphisms that have these objects as domains. Clearly
there is a precise analogue of this construction for the
category $\cal O$ of all bounded, self-adjoint operators on
$\cal H$. The analogous modification of the category $\cal W$
consists in removing the trivial Boolean algebra $\{0,1\}$ as a
possible context/stage of truth; we shall denote the resulting
category by ${\cal W}_*$.

Whether or not one wants to make the change from $\cal O$ to
${\cal O}_*$ is not totally clear, and for the moment we prefer
to keep the two options open as two slightly different schemes.
Most of the material that follows is valid irrespective of
whether $\cal O$ or ${\cal O}_*$ is used; where there is a
significant difference, we shall point it out.

\paragraph*{6. The Analogue of FUNC:}

Let us turn now to the crucial question of the analogue of the
functional composition condition FUNC; in particular, we must
check that if the proposition `$A\in\Delta$' is given the value
`totally true' then, in an appropriate sense, this is also the
case for the proposition `$h(A)\in h(\Delta)$' for any bounded
Borel function $h:\sigma(\hat A)\rightarrow\mathR$. The
following theorem provides the key to seeing that this is so.

\begin{theorem}
If $h_{{\cal O}_d}:\hat C\rightarrow \hat A$, so that $\hat
C=h(\hat A)$, then
\begin{equation}
        \nu^V(C\in h(\Delta))=h_{{\cal O}_d}^*(\nu^V(A\in\Delta))
                        \label{FUNCTpartval}
\end{equation}
where the pull-back $h_{{\cal O}_d}^*(S)$ of
$S\in{\bf\Omega}(\hat A)$ by $h_{{\cal O}_d}:\hat C\rightarrow
\hat A$ is the sieve on $\hat C$ defined as ({\em cf.}, Eq.\
(\ref{Def:Om(f)}))
\begin{equation}
h_{{\cal O}_d}^*(S):=\{k_{{\cal O}_d}:\hat D\rightarrow \hat
C\mid h_{{\cal O}_d}\circ k_{{\cal O}_d}\in S\}.
\end{equation}
\end{theorem}

\noindent{\bf Proof}\smallskip

\noindent We have
\begin{equation}
\nu^V(C\in h(\Delta)):= \{k_{{\cal O}_d}:\hat D\rightarrow\hat
C\mid \hat D\in {\rm dom\,}V,\; V(\hat D)\in k(h(\Delta))\}
\label{FUNCT-sieves1}
\end{equation}
and so, since $\hat C=h(\hat A)$, if $k_{{\cal
O}_d}\in\nu^V(C\in h(\Delta))$ then $h_{{\cal O}_d}\circ
k_{{\cal O}_d}:\hat D\rightarrow\hat A$ with $\hat D\in{\rm
dom\,}V$ and $V(\hat D)\in k\circ h(\Delta)$; hence $h_{{\cal
O}_d}\circ k_{{\cal O}_d}\in\nu^V(A\in\Delta)$, so that
$k_{{\cal O}_d}\in h^*_{{\cal O}_d}(\nu^V(A\in\Delta))$. Thus
$\nu^V(C\in h(\Delta))\subseteq h^*_{{\cal
O}_d}(\nu^V(A\in\Delta))$.

Conversely, let $k_{{\cal O}_d}:\hat D\rightarrow\hat C$ belong
to $h^*_{{\cal O}_d}(\nu^V(A\in\Delta))$; thus $h_{{\cal
O}_d}\circ k_{{\cal O}_d}\in \nu^V(A\in\Delta)$. Then $\hat
D\in{\rm dom\,}V$, and $V(\hat D)\in k(h(\Delta))$, and so
$k_{{\cal O}_d}\in\nu^V(C\in h(\Delta))$.  Hence $h^*_{{\cal
O}_d}(\nu^V(A\in\Delta))\subseteq \nu^V(C\in h(\Delta))$. \hfill
{\bf Q.E.D.}

In particular, suppose that $\nu^V(A\in \Delta)$ has the value
`totally true', {\em i.e.}, it is equal to the unit
$1_A=\downarrow\!\!\hat A$ (or `${\rm true}_A$') of the Heyting
algebra ${\bf\Omega}(\hat A)$ of sieves on $\hat A$. Then
\begin{equation}
h^*_{{\cal O}_d}(\nu^V(A\in\Delta))=
    h^*_{{\cal O}_d}(\downarrow\!\!\hat A)=
        \downarrow\!\!\hat C        \label{h*nuVA=1C}
\end{equation}
and so, by Eq.\ (\ref{FUNCTpartval}), we get $\nu^V(C\in
h(\Delta))=\downarrow\!\!C=1_C$; hence the proposition `$C\in
h(\Delta)$' has the value `totally true' in the Heyting algebra
of sieves on $\hat C$.

In summary: if the proposition `$A\in\Delta$' is totally true at
the stage of truth $\hat A$, then the weaker proposition
`$h(A)\in h(\Delta)$' is also totally true at the stage of truth
$h(\hat A)$. This result is precisely the type of thing we
wanted, and justifies our taking Eq.\ (\ref{FUNCTpartval}) to be
the presheaf analogue of the functional composition rule.

Furthermore,  the pull-back of a sieve by a morphism that is
itself a member of the sieve, is the principal sieve (see the
discussion around Eq.\ (\ref{f*S}) in the Appendix). Thus Eq.\
(\ref{FUNCTpartval}) implies that the partial truth-value of a
proposition `$A\in\Delta$' is the set of coarse-grainings of
$\hat A$ which are such that the associated propositions are
totally true at their own `stages of truth'.

\subsection{Algebraic Properties of the Generalized
Valuation $\nu^V$} \label{AlgPropnuV} Let us consider now the
extent to which the generalized valuation Eq.\ (\ref{nVinDelta})
satisfies the conditions listed in
Eqs.(\ref{NullC}--\ref{ExclC}) in the Introduction. We shall
also consider explicitly the possibility that the generalized
valuation might satisfy strong disjunctive or conjunctive
conditions.

\paragraph*{1. The Null Proposition Condition:}

 The null proposition regarding the value of the
physical quantity $A$ is `$A\in\emptyset$', and then
$\nu^V(A\in\emptyset):=\{f_{{\cal O}_d}:\hat B\rightarrow\hat
A\mid \hat B\in{\rm dom\,}V,\; V(\hat B)\in f(\emptyset)\}$. But
$f(\emptyset)=\emptyset$, and hence
$\nu^V(A\in\emptyset)=\emptyset$, which is the zero element of
the Heyting algebra ${\bf\Omega}(\hat A)$. Hence, as required,
$\nu^V(0)=0_A$; or, to indicate the context in a more precise
way,
\begin{equation}
    \nu^V_A(\hat 0)=0_A.                        \label{nuV0=0}
\end{equation}

\paragraph*{2. The Monotonicity Condition:}

To check monotonicity we consider propositions `$A\in\Delta_1$'
and `$A\in\Delta_2$' with $\Delta_1\subseteq\Delta_2$, which is
equivalent to the propositional relation `$A\in\Delta_1 \leq
A\in\Delta_2$'.

Then if $f_{{\cal O}_d}:\hat B\rightarrow \hat A$ belongs to
$\nu^V(A\in\Delta_1)$, we have $\hat B\in {\rm dom\,}V$ and
$V(\hat B)\in f(\Delta_1)$. However, $\Delta_1\subseteq\Delta_2$
implies $f(\Delta_1)\subseteq f(\Delta_2)$; and hence $V(\hat
B)\in f(\Delta_2)$. Thus $f_{{\cal O}_d}$ also belongs to
$\nu^V(A\in\Delta_2)$. This proves the monotonicity condition
\begin{equation}
\mbox{`$A\in\Delta_1 \leq A\in\Delta_2$' implies }
        \nu^V(A\in\Delta_1)\leq\nu^V(A\in\Delta_2).
                    \label{mono-nuV}
\end{equation}

\paragraph*{2.1\ A Strong Disjunctive Condition:}

As noted in Section  \ref{SubSec:primafacie}, the monotonicity
condition implies the weak disjunctive and conjunctive
conditions
\begin{equation}
    \nu^V(A\in\Delta_1)\lor\nu^V(A\in\Delta_2)\leq
\nu^V(A\in\Delta_1\lor A\in\Delta_2)    \label{weak-disj-nuV}
\end{equation}
and
\begin{equation}
\nu^V(A\in\Delta_1\land A\in\Delta_2)\leq
    \nu^V(A\in\Delta_1)\land\nu^V(A\in\Delta_2)
                                        \label{weak-conj-nuV}
\end{equation}
respectively.

However, it turns out that $\nu^V$ satisfies a strong
disjunctive condition in which the inequality in Eq.
(\ref{weak-disj-nuV}) is replaced by an equality.

    To see this, consider propositions `$A\in\Delta_1$' and
`$A\in\Delta_2$', so that `$A\in\Delta_1\lor A\in\Delta_2$' is
the equivalent to the proposition `$A\in\Delta_1\cup\Delta_2$',
({\em i.e.}, the logical `$\lor$' operation is taken in the
Boolean algebra of propositions about the value of $A$ lying in
Borel subsets of $\sigma(\hat A)$).  Then
\begin{equation}
\nu^V(A\in\Delta_1\lor A\in\Delta_2):=
    \{f_{{\cal O}_d}:\hat B\rightarrow\hat A\mid
        \hat B\in{\rm dom\,}V,\; V(\hat B)\in
            f(\Delta_1\cup \Delta_2)\}
\end{equation}
which, since $f(\Delta_1\cup\Delta_2)=f(\Delta_1)\cup
f(\Delta_2)$, gives
\begin{equation}
\nu^V(A\in\Delta_1\lor A\in\Delta_2)= \{f_{{\cal O}_d}:\hat
B\rightarrow\hat A\mid
        \hat B\in{\rm dom\,}V,\; V(\hat B)\in f(\Delta_1),
            \mbox{\ or\ } V(\hat B)\in f(\Delta_2)\}.
\end{equation}
However, the right hand side of this expression is just
$\nu^V(A\in\Delta_1)\cup \nu^V(A\in\Delta_2)$. Thus we see that
\begin{equation}
    \nu^V(A\in\Delta_1\lor A\in\Delta_2)=
        \nu^V(A\in\Delta_1)\lor \nu^V(A\in\Delta_2)
                    \label{nuVconj}
\end{equation}
where the `$\lor$'-operation on the right hand side is taken in
the Heyting algebra ${\bf\Omega}(\hat A)$, and where we recall
from Eq.\ (\ref{Def:S1lorS2}) that if $S_1$ and $S_2$ are sieves
on the same object, then $S_1\lor S_2:=S_1\cup S_2$.  Thus, the
generalized valuation $\nu^V$ satisfies a disjunctive condition
in the strong sense that the equality holds. As we shall see
later , this is not the case for other types of generalized
valuation (see the discussion around Eqs.\
(\ref{GV1dum}--\ref{a1a2}) in Section
\ref{SubSec:Qu-statemixed-gen-val}).

\paragraph*{2.2\ No Strong Conjunctive Condition:}

One might wonder if there is not a strong version of the
conjunctive condition too, in which the inequality in Eq.\
(\ref{weak-conj-nuV})---which comes purely from
monotonicity---is replaced by an equality.

To check this, we note that the conjunction `$A\in\Delta_1\land
A\in\Delta_2$' $=$ `$A\in\Delta_1\cap\Delta_2$', receives the
truth-value
\begin{equation}
\nu^V(A\in\Delta_1\land A\in\Delta_2):= \{f_{{\cal O}_d}:\hat
B\rightarrow\hat A\mid \hat B \in{\rm dom\,}V,\; V(\hat B)\in
f(\Delta_1\cap\Delta_2)\}
                            \label{nuvand}
\end{equation}
whereas
\begin{eqnarray}
\lefteqn{\nu^V(A\in\Delta_1)\land\nu^V(A\in\Delta_2)=
\nu^V(A\in\Delta_1)\cap\nu^V(A\in\Delta_2):=}
                        \hspace{1.0cm}\nonumber\\
&&\{f_{{\cal O}_d}:\hat B\rightarrow\hat A\mid \hat B\in{\rm
dom\,}V,\; V(\hat B)\in f(\Delta_1)\mbox{\ and \ }V(\hat B)\in
f(\Delta_2)\}
                            \label{nuvand2}
\end{eqnarray}
where we have used the definition in Eq.\ (\ref{Def:S1landS2})
that if $S_1$ and $S_2$ are sieves on the same object, then
$S_1\land S_2:=S_1\cap S_2$.

However $f(\Delta_1\cap\Delta_2)\subseteq f(\Delta_1)\cap
f(\Delta_2)$, and the equality may not hold if $f$ is
many-to-one.  Thus the most that can be deduced from Eqs.\
(\ref{nuvand}--\ref{nuvand2}) is that
$\nu^V(A\in\Delta_1\cap\Delta_2)\subseteq
\nu^V(A\in\Delta_1)\cap\nu^V(A\in\Delta_2)$, which gives only
the inequality
\begin{equation}
\nu^V(A\in\Delta_1\land A\in\Delta_2)\leq \nu^V(A\in\Delta_1)
\land \nu^V(A\in\Delta_2)
\end{equation}
that could have been derived directly from the monotonicity
result in Eq.\ (\ref{mono-nuV}).

As anticipated in the Introduction (the discussion in Section
\ref{SubSec:primafacie}), there are good reasons for expecting
the strict equality not to hold. For example, consider the
propositions `$A\in\{a_1\}$' and `$A\in\{a_2\}$' with $a_1\neq
a_2$. Then
\begin{equation}
    \nu^V(A\in\{a_1\}\cap\{a_2\})=\nu^V(A\in\emptyset)=\emptyset
\end{equation}
whereas
\begin{equation}
        \nu^V(A\in\{a_1\})\land \nu^V(A\in\{a_2\}) =
    \{f_{{\cal O}_d}:\hat B\rightarrow A\mid \hat B\in{\rm dom\,}V,\;
V(\hat B)=f(a_1)=f(a_2)\}
\end{equation}
and there is no reason for this to be the empty set, or even to
be just minimally true: all that is necessary is that there is
some nontrivial function $f:\sigma(\hat A)\rightarrow\mathR$
such that $f(\hat A)\in{\rm dom\,}V$ and $f(a_1)=f(a_2)$. Thus,
in this special `topos' sense, a physical quantity can have more
than one partial value at once!

\paragraph*{3. The Exclusivity Condition:}

It is necessary to check the exclusivity condition since this
cannot be derived directly from the monotonicity result in Eq.\
(\ref{mono-nuV}).

So, suppose that $\nu^V(A\in\Delta_1)={\rm
true}_A=\downarrow\!\hat A$, and that $\Delta_2$ is such that
$\Delta_1\cap\Delta_2=\emptyset$. Then, from the definition of
$\nu^V$, it follows that ${\rm id}_A$ belongs to the sieve
$\nu^V(A\in\Delta_1)$, and hence $\hat A\in{\rm dom\,}V$ and
$V(\hat A)\in\Delta_1$. Therefore, since
$\Delta_1\cap\Delta_2=\emptyset$, we have $V(\hat
A)\notin\Delta_2$, and hence ${\rm id}_A$ is not a member of the
sieve $\nu^V(A\in\Delta_2)$. This does not mean that
$\nu^V(A\in\Delta_2)$ is equal to ${\rm false}_A$
($=\emptyset$), but it does make it strictly less than ${\rm
true}_A$. Thus we have shown that if $A\in\Delta_1$ and
$A\in\Delta_2$ are disjoint propositions, and if
$\nu^V(A\in\Delta)={\rm true}_A$, then $\nu^A(A\in\Delta_2)<{\rm
true}_A$; hence exclusivity is satisfied.

\paragraph*{4. No Unit Proposition Condition:}

The unit proposition in the Boolean algebra of propositions
about $A$ is simply `$A\in\sigma(\hat A)$', and {\em a priori\/}
one might expect that this is always given the value
`$\mbox{true}_A$', so that there is a unit analogue of the null
condition Eq.\ (\ref{nuV0=0}).  We shall refer to this as the
`unit proposition condition', and state it formally as:

\noindent{\em Unit Proposition Condition}: For all stages of
truth $\hat A$
\begin{equation}
    \nu(A\in\sigma(\hat A))=\mbox{true}_A \label{unit-prop-cond}
\end{equation}
or, in the alternative notation for valuations on projection
operators,
\begin{equation}
    \nu_A(\hat 1)=\mbox{true}_A.    \label{unit-prop-condP}
\end{equation}

However, in fact, this is not necessarily satisfied by the
generalized valuation $\nu^V$. Indeed, from the definition of
$\nu^V$ we see at once that
\begin{eqnarray}
\nu^V(A\in\sigma(\hat A))&:=&\{f_{{\cal O}_d}:\hat B
\rightarrow\hat A
    \mid \hat B\in\mbox{dom\,}V,\; V(\hat B)\in f(\sigma(\hat
A))\}   \nonumber\\
    &=&\{f_{{\cal O}_d}:\hat B\rightarrow\hat A\mid
                \hat B\in\mbox{dom\,}V\}\label{gladys}
\end{eqnarray}
where the last equality holds since, for these discrete-spectra
operators, $f(\sigma(\hat A))=\sigma(\hat B)$, which means that
$V(\hat B)$ is always an element of the set $f(\sigma(\hat A))$.
Thus
\begin{equation}
        \nu^V(A\in\sigma(\hat A))=\mbox{dom\,}V\cap
            \downarrow\!\!\hat A
\end{equation}
which could well be a proper subset of the sieve
$\mbox{true}_A:=\downarrow\!\!\hat A$. Thus, in this situation,
even the proposition `$A$ has {\em some\/} value' is not totally
true!  Rather, the partial truth-value of this proposition is a
measure of the `proximity' of the observable $A$ to the domain
of the partial valuation.  Borrowing a standard piece of
nomenclature from topos theory, one could say that the physical
quantity $A$ only `partially exists' in this situation. As we
shall see in Section \ref{SubSec:Qu-state-gen-val}, the unit
proposition condition {\em is\/} satisfied by the generalized
valuation associated with a quantum state.

\section{Generalized Valuations and Quantum States}
\label{Sec:GenVal} Motivated by Definition
\ref{Defn:partval-genval} as an example of a sieve-valued
generalized valuation, and by the properties of these
valuations, we turn now in Section \ref{SubSec:DefGenVal} to the
formal definition of a generalized valuation that is not based
on the existence of any partial valuation. In Section
\ref{SubSec:topos-int-gen-val} we discuss the precise way in
which this fits into a topos framework; finally in Section
\ref{SubSec:Qu-state-gen-val} we show how any quantum state
gives rise to a generalized valuation.

\subsection{The Definition of a Generalized Valuation}
\label{SubSec:DefGenVal} Since we wish to apply these methods to
the category $\cal O$ of all bounded, self-adjoint operators,
the first step is to give meaning to the projector $\hat
E[f(A)\in f(\Delta)]$ in those cases in which $f(\Delta)$ is not
a Borel subset of $\sigma(f(\hat A))$. The main ingredient is
the following theorem (which is also used in Section
\ref{SubSec:defncoarse-graining}):

\begin{theorem}\label{Theorem:inf-defn}
If $\Delta$ is a Borel subset of $\sigma(\hat A)$, and if
$f:\sigma(\hat A)\rightarrow\mathR$ is a Borel function such
that $f(\Delta)$ is a Borel subset of $\sigma(f(\hat A))$, then
if $W_{f(A)}$ is viewed as a subalgebra of the Boolean algebra
$W_A$ we have
\begin{equation}
\hat E[f(A)\in f(\Delta)] =\inf \{\hat Q\in W_{f(A)}\subseteq
W_A
    \mid \hat E[A\in\Delta]\leq \hat Q\}
                        \label{Theorem:inf}
\end{equation}
where the infinum of projectors is taken in the (complete)
lattice structure of $\cal P$.
\end{theorem}

\smallskip\noindent{\bf Proof}

\noindent Let $\hat I:=\inf \{\hat Q\in W_{f(A)}\subseteq W_A
\mid \hat E[A\in\Delta]\leq \hat Q\}$; then, since
$E[A\in\Delta]\leq \hat E[f(A)\in f(\Delta)]$, we clearly have
$\hat I\leq \hat E[f(A)\in f(\Delta)]$.

Conversely, suppose $\hat Q\in W_{f(A)}\subseteq W_A$ is such
that $\hat E[A\in\Delta]\leq \hat Q$. There is some Borel subset
$K\subseteq\sigma(f(\hat A))$ such that $\hat Q=\hat E[f(A)\in
K]$, and so $\hat E[A\in\Delta]\leq \hat E[f(A)\in K]$. However,
$\hat E[f(A)\in K]=\hat E[A\in f^{-1}(K)]$; and hence the
inequality reads $\hat E[A\in\Delta]\leq \hat E[A\in
f^{-1}(K)]$, which implies $\Delta\subseteq f^{-1}(K)$ (up to
sets of spectral-measure zero), and hence that
$f(\Delta)\subseteq f(f^{-1}(K))\subseteq J$. In turn, this
implies that $\hat Q=\hat E[f(A)\in K]\geq \hat E[f(A)\in
f(\Delta)]$.  In summary: $\hat E[A\in\Delta]\leq \hat Q$
implies that $\hat E[f(A)\in f(\Delta)] \leq \hat Q$, and hence
$\hat E[f(A)\in f(\Delta)] \leq \hat I$. Thus $\hat E[f(A)\in
f(\Delta)] = \hat I$.  \hfill{\bf Q.E.D.}

The key idea now is to use the right hand side of Eq.\
(\ref{Theorem:inf}) as the {\em definition\/} of the symbol
$\hat E[f(A)\in f(\Delta)]$ in those cases in which $f(\Delta)$
is not a Borel subset of $\sigma(f(\hat A))$. In this context,
we note that since $W_{f(A)}$ is a complete sublattice of $\cal
P$, the right hand side of Eq.\ (\ref{Theorem:inf}) is always of
the form $E[f(A)\in J]$ for {\em some\/} Borel subset $J$ of
$\sigma(f(\hat A))$. Note also that, considered as a definition
of $\hat E[f(A)\in f(\Delta)]$, the expression Eq.\
(\ref{Theorem:inf}) can be usefully rewritten as
\begin{eqnarray}
 \hat E[f(A)\in f(\Delta)]&:=&\inf_{K\subseteq \sigma(f(\hat A))}
        \{\hat E[f(A)\in K]\mid \hat E[A\in\Delta]\leq
                    \hat E[f(A)\in K]\}\label{Def:f(A)inD}\\
    &=&\inf_{K\subseteq \sigma(f(\hat A))}
        \{\hat E[f(A)\in K]\mid \hat E[A\in\Delta]\leq
                    \hat E[A\in f^{-1}(K)]\}\\
    &=&\inf_{K\subseteq \sigma(f(\hat A))}
        \{\hat E[f(A)\in K]\mid \Delta\subseteq f^{-1}(J)\}
\end{eqnarray}
where the {\em infinum\/} is taken over all Borel subsets $J$ of
$\sigma(f(\hat A))$. From now on we shall use Eq.\
(\ref{Def:f(A)inD}) as the definition of $\hat E[f(A)\in
f(\Delta)]$ for the category of operators $\cal O$.

    Equipped with this idea, we can give the definition of a
generalized valuation on propositions about the values of any
physical quantity represented by a bounded self-adjoint operator
$\hat A$ in $\cal O$:

\begin{definition}\label{Defn:gen-val-O}
A {\em generalized valuation\/} on the propositions in a quantum
theory is a map $\nu$ that associates to each proposition of the
form `$A\in\Delta$' (where $\Delta$ is a Borel subset of
$\sigma(\hat A)$) a sieve $\nu(A\in\Delta)$ on $\hat A$ in $\cal
O$. These sieves must satisfy the following properties:
\end{definition}
\noindent {\em (i) Functional composition}:
\begin{eqnarray}
\lefteqn{\mbox{\ For any Borel function } h:\sigma(\hat A)
\rightarrow\mathR \mbox{ we have }} \hspace{3cm}\nonumber\\[3pt]
    &&\nu(h(A)\in h(\Delta))=h_{\cal O}^*(\nu(A\in\Delta)).
\hspace{2cm} \ \label{FC-gen}
\end{eqnarray}

\noindent {\em (ii) Null proposition condition}:
\begin{equation}
    \nu(A\in\emptyset)=0_A          \label{Null-gen}
\end{equation}

\noindent {\em (iii) Monotonicity}:
\begin{equation}
\mbox{If }\Delta_1\subseteq\Delta_2\mbox{ then }
                \nu(A\in\Delta_1)\leq\nu(A\in\Delta_2).
                                    \label{Mono-gen}
\end{equation}

\noindent {\em (iv) Exclusivity}:
\begin{equation}
    \mbox{If $\Delta_1\cap\Delta_2=\emptyset$ and
$\nu(A\in\Delta_1)= {\rm true}_A$, then $\nu(A\in\Delta_2)< {\rm
true}_A$}. \label{Excl-gen}
\end{equation}

\noindent We may also wish to add the `unit proposition
condition':

\smallskip
\noindent {\em (v) Unit proposition condition}:

\begin{equation}
    \nu(A\in\sigma(\hat A))=\mbox{true}_A.
                            \label{unit-prop-cond2}
\end{equation}

    Note that this definition of a generalized valuation makes
sense for operators whose spectra contains continuous parts, as
well as for those whose spectra is purely discrete. However, in
order to give meaning to the proposition `$h(A)\in h(\Delta)$'
in Eq.\ (\ref{FC-gen}) if $h(\Delta)$ is not a Borel subset of
$\sigma(h(\hat A))$, it is more appropriate to think of a
generalized valuation as being defined on the projectors $\hat
E[A\in\Delta]$, rather than on the more abstract propositions
`$A\in\Delta$' themselves; for this enables the definition in
Eq.\ (\ref{Def:f(A)inD}) to be used.

The physical interpretation of a generalized valuation is
motivated by the special case of the valuations $\nu^V$
discussed in the last section.  Namely, the partial truth-value
$\nu(A\in\Delta)$ of the proposition `$A\in\Delta$' is a sieve
of coarse-grainings $f(\hat A)$ of $\hat A$, at which each
associated proposition `$f(A)\in f(\Delta)$' is totally
true---reflecting the fact that if $f_{\cal O}:\hat
B\rightarrow\hat A$ belongs to the sieve $\nu(A\in\Delta)$ on
$\hat A$ then, by the definition of a sieve, the pull-back
$f_{\cal O}^*\nu(A\in\Delta)$ to $f(\hat A)$ of this sieve is
necessarily the principal sieve on $f(\hat A)$ (see Eq.\
(\ref{f*S})). In general terms, we can say that the `size' of
the sieve $\nu(A\in\Delta)$ determines the degree of the partial
truth of the proposition `$A\in\Delta$'.

We note that, as in the earlier discussion of the generalized
valuation $\nu^V$, the phrase `$A\in\Delta$' in
$\nu(A\in\Delta)$ performs the dual function of specifying (i)
the projection operator whose partial truth-value is to be
given; and (ii) the context---the operator $\hat A$---in which
this valuation takes place.

    As in the previous section, this contextuality can be made
more explicit by shifting the emphasis to think of valuations as
being defined on projection operators in the explicit context of
a specific physical quantity. Then, the truth-value associated
with a specific $\hat P\in\cal P$ depends on the context of a
particular self-adjoint operator $\hat A$ whose set of spectral
projectors $W_A$ includes $\hat P$.  In this manifestly
contextual form, the definition of a generalized valuation would
read as follows:

\begin{definition}\label{Defn:gen-val-W}
A {\em generalized valuation\/} on the lattice of projection
operators $\cal P$ in a quantum theory is a collection of maps
$\nu_A:W_A\rightarrow {\bf\Omega}(\hat A)$, one for each `stage
of truth' $\hat A$ in the category $\cal O$, with the following
properties:
\end{definition}
\noindent {\em (i) Functional composition}:
\begin{eqnarray}
\lefteqn{\mbox{\ For any Borel function } h:\sigma(\hat A)
\rightarrow\mathR,} \hspace{3cm}\nonumber   \\[3pt]
    &&\nu_{h(A)}(\hat E[h(A)\in h(\Delta)])=
h_{\cal O}^*(\nu_A(\hat E[A\in\Delta]).  \hspace{2cm} \
                                            \label{FC-genP}
\end{eqnarray}

\noindent {\em (ii) Null proposition condition}:
\begin{equation}
        \nu_A(\hat 0)=0_A        \label{Null-genP}
\end{equation}

\noindent {\em (iii) Monotonicity}:
\begin{equation}
    \mbox{If }\hat\alpha,\hat\beta\in W_A\mbox{ with }
    \hat\alpha\leq\hat\beta,\mbox{ then }
    \nu_A(\hat\alpha)\leq\nu_A(\hat\beta). \label{Mono-genP}
\end{equation}

\noindent {\em (iv) Exclusivity}:
\begin{equation}
    \mbox{If $\hat\alpha,\hat\beta\in W_A$ with
$\hat\alpha\hat\beta=\hat 0$ and $\nu_A(\hat\alpha)= {\rm
true}_A$, then $\nu_A(\hat\beta)< {\rm true}_A$}.
\label{Excl-genP}
\end{equation}

\noindent We may wish to supplement this list with:

\smallskip\noindent
{\em (v) Unit proposition condition}:
\begin{equation}
    \nu_A(\hat 1)=\mbox{true}_A.    \label{unit-prop-condP2}
\end{equation}

Note that, in writing Eq.\ (\ref{FC-genP}) we have employed the
specific `coarse-graining' function from the Boolean algebra
$W_A$ to the Boolean algebra $W_{h(A)}$, defined by the map
\begin{equation}
    \hat E[A\in\Delta]\mapsto \hat E[h(A)\in h(\Delta)]
                                            \label{CG}
\end{equation}
where, if necessary, the right hand side is to be understood in
the sense of Eq.\ (\ref{Def:f(A)inD}). In Section
\ref{SubSec:defncoarse-graining} we shall consider a more
general way of understanding this operation.

\subsection{The Topos Interpretation of Generalized Valuations}
\label{SubSec:topos-int-gen-val}
\paragraph*{1. The Coarse-graining Presheaf:}

From what has been said so far it should be clear that ideas of
topos theory lie at the heart of our constructions. However, the
only explicit feature that has appeared so far is our use of
sieves as truth-values, and we wish now to explain more fully
how our ideas fit in with the theory of presheaves.

A key ingredient in exhibiting the underlying topos framework of
generalized valuations is a certain presheaf on $\cal O$ that
incorporates our central idea of operator coarse-graining. This
is contained in the following definition.

\begin{definition}
The {\em coarse-graining presheaf\/} over $\cal O$ is the
covariant functor ${\bf G}:{{\cal O}^{\rm op}}\rightarrow{\rm
Set}$ defined as follows.
\begin{enumerate}
\item {\em On objects in $\cal O$:} ${\bf G}(\hat A):=
W_A$, where $W_A$ is the spectral algebra of $\hat A$.

\item {\em On morphisms in $\cal O$:} If $f_{\cal O}:\hat B
\rightarrow\hat A$ ({\em i.e.}, $\hat B=f(\hat A)$), then ${\bf
G}(f_{\cal O}): W_A\rightarrow W_B$ is defined as
\begin{equation}
        {\bf G}(f_{\cal O})(\hat E[A\in\Delta]):=
        \hat E[f(A)\in f(\Delta)]       \label{Def:G(fO)}
\end{equation}
where, if necessary, the right hand side is to be understood in
the sense of Eq.\ (\ref{Def:f(A)inD}).
\end{enumerate}
\end{definition}
Note that ${\bf G}(f_{\cal O}): W_A\rightarrow W_{f(A)}$ is just
the coarse-graining operation considered above in Eq.\
(\ref{CG}).

The main step in proving that ${\bf G}$ is a contravariant
functor from $\cal O$ to ${\rm Set}$ is to show that if $f_{\cal
O}:\hat B\rightarrow\hat A$ and $g_{\cal O}:\hat
C\rightarrow\hat B$, then ${\bf G}(f_{\cal O}\circ g_{\cal O})
={\bf G}(g_{\cal O})\circ {\bf G}(f_{\cal O})$, as in Eq.\
(\ref{Def:confunct}). However, ${\bf G}(f_{\cal O})(\hat
E[A\in\Delta]):= \hat E[f(A)\in f(\Delta)]$, and therefore, if
$f(\Delta)$ and $g(f(\Delta))$ are Borel subsets of the
appropriate spectra, then
\begin{equation}
    {\bf G}(g_{\cal O})(\hat E[f(A)\in f(\Delta)]):=
\hat E[g(f(A))\in g(f(\Delta))]
\end{equation}
while
\begin{equation}
    {\bf G}(f_{\cal O}\circ g_{\cal O})(\hat E[A\in\Delta]):=
\hat E[g(f(A))\in g(f(\Delta))].
\end{equation}
Hence ${\bf G}(f_{\cal O}\circ g_{\cal O}) ={\bf G}(g_{\cal
O})\circ {\bf G}(f_{\cal O})$, as desired. If $f(\Delta)$ or
$g(f(\Delta))$ are not Borel subsets then the result follows
(using the definition in Eq.\ (\ref{Def:f(A)inD})) as a special
case of the result stated after Definition
\ref{Defn:can-coarse-grain}.

\paragraph*{2. The Natural Transformation Between ${\bf G}$
and ${\bf\Omega}$:}

A key technical result in revealing the topos content of our
constructions is the following.

\begin{theorem}\label{genval=NT}
To each generalized valuation $\nu$ on $\cal P$ there
corresponds a natural transformation $N^\nu$ between the
contravariant functors ${\bf G}$ and ${\bf\Omega}$, in which, at
each stage of truth $\hat A$, the component $N^\nu_A:{\bf
G}(\hat A)\rightarrow {\bf\Omega}(\hat A)$ is defined by
\begin{equation}
        N^\nu_A(\hat P):=\nu_A(\hat P)
\end{equation}
for all $\hat P\in W_A={\bf G}(\hat A)$.
\end{theorem}

\noindent{\bf Proof}\smallskip

\noindent We recall that the subobject classifier ${\bf\Omega}$
in the topos ${\rm Set}^{{\cal O}^{\rm op}}$ is defined (i) on
objects by ${\bf\Omega}(\hat A):=\{S\mid S\mbox{\ is a sieve on
$\hat A$ in $\cal O$}\}$; and (ii) on morphisms $f_{\cal O}:\hat
B\rightarrow\hat A$ by ${\bf\Omega}(f_{\cal O}):{\bf\Omega}(\hat
A)\rightarrow{\bf\Omega}(\hat B)$ where ${\bf\Omega}(f_{\cal
O})(S):=f_{\cal O}^*(S)$ for all sieves $S\in{\bf\Omega}(\hat
A)$.

As discussed in Section \ref{SubSec:presheaves-gen-cat}, a
natural transformation $N$ between the contravariant functors
${\bf G}$ and ${\bf\Omega}$ is defined to be a family of
functions $N_A:{\bf G}(\hat A)\rightarrow{\bf\Omega}(\hat
A)$---one for each stage of truth $\hat A$---such that, if
$f_{\cal O}:\hat B\rightarrow\hat A$, the composite map ${\bf
G}(\hat A)\stackrel{N_A}\longrightarrow{\bf\Omega}(\hat
A)\stackrel{{\bf\Omega}(f_{\cal
O})}\longrightarrow{\bf\Omega}(\hat B)$ is equal to ${\bf
G}(\hat A)\stackrel{{\bf G}(f_{\cal O})}\longrightarrow {\bf
G}(\hat B)\stackrel{N_B}\longrightarrow {\bf\Omega}(\hat B)$
({\em cf.} the commutative diagram in Eq.\ (\ref{cdNT})).

In our case, if $\nu$ is a generalized valuation, the associated
natural transformation $N^\nu$ is defined at stage $\hat A$ on
${\bf G}(\hat A):=W_A$ by $N^\nu_A(\hat
E[A\in\Delta]):=\nu_A(\hat E[A\in\Delta])\equiv\nu(A\in\Delta)$.
Then
\begin{equation}
{\bf\Omega}(f_{\cal O})\circ N^\nu_A(\hat E[A\in\Delta])=
{\bf\Omega}(f_{\cal O})(N^\nu_A(\hat E[A\in\Delta]))=f_{\cal
O}^*(\nu_A(\hat E[A\in\Delta]))             \label{ON}
\end{equation}
while
\begin{eqnarray}
N^\nu_B\circ {\bf G}(f_{\cal O})(\hat E[A\in\Delta]) & =&
N^\nu_B({\bf G}(f_{\cal O})(\hat E[A\in\Delta])) \nonumber\\ &=&
N^\nu_B(\hat E[f(A)\in f(\Delta)])=
        \nu_{f(A)}(\hat E[f(A)\in f(\Delta)])       \label{NG}
\end{eqnarray}
However, the functional composition principle Eq.\
(\ref{FC-genP}) says that the right hand sides of Eq.( \ref{ON})
and Eq.\ (\ref{NG}) are identical, which shows that
${\bf\Omega}(f_{\cal O})\circ N^\nu_A(\hat
E[A\in\Delta])=N^\nu_B\circ{\bf G} (f_{\cal O})(\hat
E[A\in\Delta])$. Hence $N^\nu$ is a natural transformation
between the contravariant functors ${\bf G}$ and ${\bf\Omega}$.
\hfill {\bf Q.E.D.}

Note that, in the language of Definition \ref{Defn:gen-val-O},
the components of the natural transformation are $N^\nu_A(\hat
E[A\in\Delta]):=\nu(A\in\Delta)$.

The next three subsections bring out some of the implicit `topos
content' of Theorem \ref{genval=NT}.

\paragraph*{3. Another Perspective on the Coarse-Graining
Presheaf:}

There is another way of looking at the coarse-graining presheaf
which may help to clarify its place in the theory; at least in
the case of operators with purely discrete spectra. Associated
with the spectral presheaf ${\bf\Sigma}:{{\cal O}_d}^{\rm
op}\rightarrow{\rm Set}$ of Definition
\ref{Defn:spectral-presheaf} there is another covariant functor
$B{\bf\Sigma}:{{\cal O}_d}^{\rm op}\rightarrow{\rm Set}$,
defined as follows:
\begin{enumerate}
    \item{\em On objects: $B{\bf\Sigma}(\hat A):=B(\sigma(\hat
A))$---the Boolean algebra of Borel subsets of the spectrum of
$\hat A$}.

    \item{\em On morphisms: If $f_{{\cal O}_d}:\hat
B\rightarrow\hat A$, so that $\hat B=f(\hat A)$, then
$B{\bf\Sigma}(f_{{\cal O}_d}):B(\sigma(\hat A))\rightarrow
B(\sigma(\hat B))$ is defined by
\begin{equation}
    B{\bf \Sigma}(f_{{\cal O}_d})(\Delta):=f(\Delta)
                            \label{Def:BSigmaf}
\end{equation}
for all Borel subsets $\Delta\subseteq\sigma(\hat A)$. }
\end{enumerate}

Note that the spectral Boolean algebra $W_A$ is isomorphic to
the Boolean algebra $B(\sigma(\hat A))$ by the map that
associates the projector $\hat E[A\in\Delta]\in W_A$ with the
Borel subset $\Delta\in B(\sigma(\hat A))$. From equations Eq.\
(\ref{Def:G(fO)}) and Eq.\ (\ref{Def:BSigmaf}), it is clear
therefore that the coarse-graining presheaf $\bf G$ is
essentially the same thing as the `power-object'
$B\bf\Sigma$.\footnote{With any object $X$ in a topos, there is
associated another object $PX:=\Omega^X$, known as the `power
object', which is the topos analogue of the power set of a set
(the set of all subsets of the set). In our case, $B\bf\Sigma$
is the subobject of the power object $P\Sigma$ obtained by
requiring the elements of $B{\bf\Sigma}(\hat A)$ to be {\em
Borel\/} subsets of ${\bf\Sigma}(\hat A):=\sigma(\hat A)$
only---rather than arbitrary subsets---at each stage $\hat A$.
Thus the `coarse-graining' presheaf is closely related to the
power object $P\bf\Sigma$.}

\paragraph*{4. Generalized Valuations as Subobjects of
${\bf G}$:}

We recall that, in a topos of presheaves such as ${\rm
Set}^{{\cal O}^{\rm op}}$, a morphism between a pair of functors
({\em i.e.}, a pair of objects in the topos) is defined to be a
natural transformation between them. Therefore, Theorem
\ref{genval=NT} implies that to each generalized valuation $\nu$
there corresponds a morphism $N^\nu:{{\bf
G}}\rightarrow{\bf\Omega}$ between the coarse-graining object
${\bf G}$ and the subobject classifier ${\bf\Omega}$ in the
topos ${\rm Set}^{{\cal O}^{\rm op}}$.  However, precisely
because ${\bf\Omega}$ {\em is\/} the subobject classifier in
this topos, morphisms ${\bf G}\rightarrow{\bf\Omega}$ are in
one-to-one correspondence with subobjects of ${\bf G}$ (see the
end of Section \ref{SubSec:presheaves-gen-cat}; especially
equations Eq.\ (\ref{Def:chiKA}) and Eq.\ (\ref{Def:KchiA})).
Thus, we conclude that to every generalized valuation there
corresponds a subobject of the coarse-graining object ${\bf G}$;
or, equivalently, of the power object $B\bf\Sigma$.

Conversely, of course, we could turn this around and {\em
define\/} a generalized valuation to be any subobject of ${\bf
G}$, or $B\bf\Sigma$, that is subject to the conditions Eqs.\
(\ref{Null-genP}--\ref{Excl-genP}), or to the equivalent set
Eqs.\ (\ref{Null-gen}--\ref{Excl-gen}).

One important consequence of looking at a generalized valuation
as a certain type of morphism from $\bf G$ to $\bf \Omega$,
comes from the fact that, in any topos, the collection of all
subobjects of a given object has the structure of a Heyting
algebra.  This is of considerable interest to us since it raises
the possibility that the subset of subobjects that satisfy our
extra conditions Eqs.\ (\ref{Null-genP}--\ref{Excl-genP})---{\em
i.e.}, the set of generalized valuations---may inherit some, or
all, of this logical structure.  This could be expected to play
an important role in exploring the physical implications of
these valuations. We shall return in a later paper to discussing
the structure of the space of all generalized valuations.

\paragraph*{5. Generalized Valuations as Global Sections of a
Presheaf:}

We note in passing that there is a bijection between morphisms
from $\bf G$ to $\bf\Omega$, and global elements of the
`exponential' object ${\bf \Omega}^{\bf G}$ which, roughly
speaking, is the topos analogue of the set $Y^X$ of all maps
from $X$ to $Y$ in normal set theory. Thus a generalized
valuation {\em does\/} turn out to be a global section of a
certain presheaf on $\cal O$, but it is the presheaf ${\bf
\Omega}^{\bf G}$, not the simple dual presheaf ${\bf D}\circ{\bf
W}$ to which the Kochen-Specker `no-go' theorem applies.

\paragraph*{6. The Generalized Valuation of a Physical Quantity:}

Definition \ref{Defn:gen-val-O} gives generalized truth-values
to propositions of the form `$A\in\Delta$' but this leaves open
the question whether in the case of operators with purely
discrete spectra, there is some corresponding concept of a
`generalized value' for the physical quantity $A$.

Clearly this cannot generally be a single real number, unless
all the propositions `$A=a$', $a\in\sigma(\hat A)$, (where $\hat
A$ has a purely discrete spectrum) are evaluated as ${\rm
false}_A$ except for one, `$A=a_0$' say, which is evaluated as
${\rm true}_A$; in this case one can say that the value of $A$
is $a_0$. More generally, however, the quantity $A$ has to be
given some sort of `smeared' value, corresponding to the
collection of propositions `$A=a$' that are not evaluated as
totally false. In fact, this suggests that, given a generalized
valuation $\nu$, we might try {\em defining\/} the `value' of
the physical quantity $A$ as $V^\nu(A):=\{\langle
a,\nu(A=a)\rangle\mid a\in\sigma(\hat A)\}$, so that we assign
to $A$ the collection of the eigenvalues of $\hat A$ `weighted'
with the generalized valuations of the associated propositions.

    With this preliminary definition, $V^\nu(\hat A)$ is a subset
of $\sigma(\hat A)\times{\bf\Omega}(\hat A)$, and is hence a
relation between $\sigma(\hat A)$ and ${\bf\Omega}(\hat A)$.
However, since each $a\in\sigma(\hat A)$ is associated with a
{\em unique\/} element $\nu(A=a)\in{\bf\Omega}(\hat A)$, this
relation defines a function from $\sigma(\hat A)$ to
${\bf\Omega}(\hat A)$, and thus we arrive at the idea that
$V^\nu(A)$ should be such a function. However, this holds at
each stage of truth $\hat A$ and, it transpires, these fit
together nicely to give a morphism between the presheaves
$\bf\Sigma$ and $\bf\Omega$ in the category ${\rm Set}^{{\cal
O}^{\rm op}}$. More precisely, we have the following theorem:

\begin{theorem}\label{Theorem:gen-valNT}
To each generalized valuation $\nu$ in the sense of Definition
\ref{Defn:gen-val-O} applied to the category ${\cal O}_d$, there
is associated a natural transformation
$V^\nu:{\bf\Sigma}\rightarrow{\bf\Omega}$ for which, at each
stage of truth $\hat A$, the component $V^\nu_A:{\bf\Sigma}(\hat
A)\rightarrow {\bf\Omega}(\hat A)$ is defined by
\begin{equation}
        V^\nu_A(a):=\nu(A=a).       \label{Def:VnuNT}
\end{equation}
\end{theorem}

\noindent{\bf Proof}\smallskip

\noindent To see that this is a natural transformation we have
to show that, if $f_{{\cal O}_d}:\hat B\rightarrow\hat A$, the
composite map ${\bf\Sigma}(\hat
A)\stackrel{V^\nu_A}\longrightarrow {\bf\Omega}(\hat
A)\stackrel{{\bf\Omega}(f_{{\cal O}_d})}\longrightarrow
{\bf\Omega}(\hat B)$ is equal to ${\bf\Sigma}(\hat
A)\stackrel{{\bf\Sigma}(f_{{\cal O}_d})}\longrightarrow
{\bf\Sigma}(\hat B)\stackrel {V^\nu_B}\longrightarrow
{\bf\Omega}(\hat A)$ ({\em cf.}, the commutative diagram in Eq.\
(\ref{cdNT})).

    It is a straightforward task to prove this directly, but in
fact this is not necessary since the theorem can be derived at
once from the earlier result in Theorem \ref{genval=NT} that
$N^\nu$ is a natural transformation from $\bf G$ (or
$B{\bf\Sigma}$) to $\bf\Omega$.  The main step is to note the
existence of a natural transformation\footnote{The notation
reflects that fact that $\{\}_{\bf\Sigma}:{\bf\Sigma}\rightarrow
B{\bf\Sigma}$ is a topos analogue of the map $X\rightarrow PX$,
$x\mapsto\{x\}$, in standard set theory.}
$\{\}_{\bf\Sigma}:{\bf\Sigma}\rightarrow B{\bf\Sigma}$ whose
components $\{\}_{\bf\Sigma}{}_A$ that map ${\bf\Sigma}(\hat
A)=\sigma(\hat A)$ to $B{\bf\Sigma}(\hat A)=B(\sigma(\hat A))$
are
\begin{equation}
        \{\}_{\bf\Sigma}{}_A(a):=\{a\}.
\end{equation}
This is well-defined since $\{a\}$ is a Borel subset of the
(discrete) spectrum $\sigma(\hat A)$ of $\hat A$; that it
satisfies the requirements for a natural transformation is
obvious. Then, identifying the coarse-graining presheaf $\bf G$
with $B\bf\Sigma$, we see that $V^\nu_A:{\bf\Sigma}(\hat
A)\rightarrow {\bf\Omega}(\hat A)$ can be written as
$V^\nu_A=(N^\nu\circ\{\}_{\bf\Sigma})_A$ for all stages $\hat
A$. Thus
\begin{equation}
    V^\nu=N^\nu\circ\{\}_{\bf\Sigma} \label{Vnu=GS}
\end{equation}
which, as a composition of natural transformations, is itself a
natural transformation. \hfill {\bf Q.E.D.}

    In particular, it follows that each generalized valuation
defines a subobject of the spectral presheaf $\bf\Sigma$ ({\em
cf.} the remarks in Subsection 4. above, or Eq.\
(\ref{Def:KchiA}) for the general definition of the subobject
associated with a morphism into $\bf\Omega$). Note that the
exclusivity condition means that, in the map
$V^\nu_A:{\bf\Sigma}(\hat A)\rightarrow{\bf\Omega}(\hat A)$, at
most one element in ${\bf\Sigma}(\hat A)=\sigma(\hat A)$ is
assigned the value `totally true' (${\rm true}_A$). Thus, the
subobject of $\bf\Sigma$ defined by $V^\nu$ has the property
that the associated subset of each ${\bf\Sigma}(\hat A)$ is
either a singleton or it is empty. In fact, it defines a partial
section of the presheaf $\bf\Sigma$, and hence a partial
valuation in the sense of Section \ref{Sec:PartVal} ({\em i.e.},
a number-valued valuation with a limited domain)---which we will
also denote $V^\nu$---with
\begin{equation}
    {\rm dom\,}V^\nu:=\{\hat A\mid \exists a\in\sigma(\hat A),
\mbox{ s.t. }V_A^\nu(a)={\rm true}_A\} \label{Def:domVnu}
\end{equation}
and with the value of any operator $\hat A$ in this domain being
defined as the associated real number $a\in\sigma(\hat A)$.

    In Definition \ref{Defn:partval-genval} we showed how to go
from a partial valuation/section to a generalized valuation;
here we have shown how each generalized valuation leads back to
a partial valuation. We note in passing that the chain
\begin{equation}
    \mbox{partial valuation }\rightarrow
    \mbox{generalized valuation}\rightarrow
    \mbox{partial valuation}
\end{equation}
takes any given partial valuation back to itself.  However we do
not necessarily return to the starting point if we begin the
`chain' with a generalized valuation; {\em i.e.},
\begin{equation}
    \mbox{generalized valuation}\rightarrow
    \mbox{partial valuation }\rightarrow
    \mbox{generalized valuation}.   \label{gen-part-gen}
\end{equation}
We shall see an explicit example of this in Section
\ref{SubSec:GVPVGV}.

\subsection{The Generalized Valuation Associated with a Quantum
State Vector} \label{SubSec:Qu-state-gen-val} We shall now show
that any quantum state gives rise to an associated generalized
valuation.

Let us start by considering the extent to which a vector
$\psi\in\cal H$ can be regarded as assigning a value to a
physical quantity $A$ represented by a self-adjoint operator
$\hat A$ (whose spectrum may not necessarily be purely
discrete). In the standard interpretation of quantum theory, one
makes only the minimal claim that a physical quantity $A$
possesses a value $a$ when the state $\psi$ is an eigenstate of
$\hat A$ with eigenvalue $a$; {\em i.e.}, $\hat A\psi=a\psi$.

However, the ideas we have been developing in this paper suggest
that even when $\psi$ is {\em not\/} an eigenvector of $\hat A$,
it may still be possible to give a {\em partial\/} truth-value
to the proposition `$A=a$'.  Indeed, in the light of our earlier
discussion, it is natural to reflect on the possibility that
some function $f(\hat A)$ of $\hat A$ may have $\psi$ as an
eigenvector, even though $\hat A$ itself does not.  Thus we are
led to define, for each state $\psi\in\cal H$, an associated
generalized valuation $\nu^\psi$ on propositions `$A=a$' as
\begin{equation}
\nu^\psi(A=a):=\{f_{\cal O}:\hat B\rightarrow\hat A\mid \hat
B\psi=f(a)\psi\}.       \label{Def:nupsi}
\end{equation}

The condition $\hat B\psi=f(a)\psi$ is equivalent to $\hat
E[B=f(a)]\psi=\psi$, and this suggests an obvious extension to
include propositions of the form `$A\in\Delta$':

\begin{definition}
The generalized valuation $\nu^\psi$ associated with a vector
$\psi\in\cal H$ is
\begin{equation}
    \nu^\psi(A\in\Delta):=\{f_{\cal O}:\hat B\rightarrow\hat A
\mid \hat E[B\in f(\Delta)]\psi=\psi\}  \label{Def:nupsiDelta}
\end{equation}
where $\Delta$ is a Borel subset of the spectrum $\sigma(\hat
A)$ of $\hat A$. If necessary, the right hand side of Eq.\
(\ref{Def:nupsiDelta}) is to be understood in the sense of Eq.\
(\ref{Def:f(A)inD}).
\end{definition}

Note that if $\psi$ is actually an eigenstate of $\hat A$ with
eigenvalue $a$, then $\nu^\psi(A\in\Delta)={\rm true}_A$, if
$a\in\Delta$. This is a good illustration of the general
rule-of-thumb that if a proposition is evaluated as `totally
true', this is equivalent to saying that it is true in the
normal sense; {\em i.e.} in the sense of simple two-valued
logic.

At this point we could check explicitly that the right hand side
of Eq.\  (\ref{Def:nupsi}) is a sieve, and that $\nu^\psi$
possesses the extra properties Eqs.\
(\ref{FC-gen}--\ref{Excl-gen}) required for a generalized
valuation.  However, we shall first give a few simple examples,
and then press on to give a substantial extension of the
definition to include generalized valuations associated with
density matrices, and then prove all the needed results for
that.

\paragraph*{1. An Example with Spin-$1/2$:}

We take a two-dimensional spin system with
$\psi:={1\over\sqrt2}\left({1\atop 1}\right)$---which is an
eigenstate of $\hat S_x$---and consider the generalized
evaluation of the propositions `$S_z={1\over 2}$' and
`$S_z=-{1\over 2}$' (we choose units in which $\hbar=1$).

The physical quantity $S_z$ is represented by the matrix
${1\over 2}\left({1\atop 0}\;{0\atop -1}\right)$, and the only
functions of this for which $\psi$ is an eigenvector are $r(\hat
S_z)^2$, $r\in\mathR$. Thus, if we use the category ${\cal O}$,
the definition Eq.\ (\ref{Def:nupsi}) of the generalized
valuation $\nu^\psi$, says that both the propositions
`$S_z={1\over 2}$' and `$S_z=-{1\over 2}$' are only minimally
true. If we use the category ${\cal O}_*$ (so that multiples of
the unit operator are excluded as stages of truth) then
\begin{equation}
        \nu^\psi(S_z=\textstyle{{1\over 2}})=\emptyset;\quad
        \nu^\psi(S_z=-{1\over 2})=\emptyset.
                        \label{Szisnull}
\end{equation}
Hence we see that, in this particular example, the physical
quantity $S_x$---which, unequivocally, has the value ${1/2}$ in
the state $\psi$---is sufficiently `far' from $S_z$ that
propositions assigning a definite value to the latter cannot be
evaluated as anything other than (i) totally false, if ${\cal
O}_*$ is used as the category of contexts; or (ii) minimally
true, if $\cal O$ is used.

On the other hand, the spectral projector corresponding to the
proposition $S_z\in\{-1/2,1/2\}=\sigma(\hat S_z)$ is the unit
operator $\hat 1$, and hence
\begin{equation}
    \nu^\psi(S_z\in\{-1/2,1/2\})={\rm true}_{S_z}.
\end{equation}
This result might be construed as asserting that the quantity
$S_z$ `exists', even if it is not possible to assign a
non-trivial truth-value to a proposition that asserts it has any
specific value.  As we shall see shortly, this unit proposition
condition (defined earlier in Eq.\ (\ref{unit-prop-cond})) is
always satisfied by a generalized valuation produced by a
quantum state.

We note in passing that the result in Eq.\ (\ref{Szisnull})
means that this particular type of generalized valuation cannot
be used by itself to construct a stochastic hidden variable
theory. More precisely, the example shows that, given a
valuation $\nu^\psi$ generated by a normalised state
$\psi\in\cal H$, one cannot expect to find a measure $\mu_A$ on
the space of sieves on $\hat A$, such that
$\mu_A[\nu^\psi(A\in\Delta)]$ is equal to the quantum-mechanical
value $\langle\psi,\hat E[A\in\Delta]\psi\rangle$ for the
probability that a measurement of $A$ will yield a result lying
in $\Delta$. Thus, in the example, we have ${\rm
Prob}(S_z=1/2;\psi)=1/2$ and ${\rm Prob}(S_z=-1/2;\psi)=1/2$,
whereas the generalized truth-values of the propositions
`$S_z={1\over 2}$' and `$S_z=-{1\over 2}$' are both null (or
minimal).

\paragraph*{2. An Example with Spin-$1$:}
We shall now consider an example where a non-trivial generalized
valuation is obtained. This involves a spin-$1$ system where the
physical quantities $S_x$ and $S_z$ are represented by the
matrices
\begin{equation}
 \hat S_x={1\over\sqrt2}\pmatrix{0&1&0\cr
                                    1&0&1\cr
                                    0&1&0};\quad
 \hat S_z={1\over\sqrt2}\pmatrix{1&0&0\cr
                                    0&0&0\cr
                                    0&0&-1}
\end{equation}
respectively.

Let the quantum state $\psi$ be $(0,1,0)$---which is an
eigenstate of $\hat S_z$ with eigenvalue $0$---and consider the
propositions `$S_x=1$' and `$S_x=-1$'.  Since $\psi$ is not an
eigenstate of $\hat S_x$, neither of these propositions is
totally true at stage $\hat S_x$. On the other hand,
\begin{equation}
    \hat S_x^2={1\over 2}\pmatrix{1&0&1\cr
                                        0&2&0\cr
                                        1&0&1}
\end{equation}
and we see that $\hat S_x^2\psi=\psi$. Furthermore, taking the
square of $\hat S_x^2$ gives just a multiple of itself, and
taking the cube of $\hat S_x$ gives just a multiple of $\hat
S_x$; hence all functions of $\hat S_x$ are of the form $t\hat
1+k\hat S_x+r\hat S_x^2$. Note that the real numbers $t,k,r$
have to be such that $k$ and $r$ are not both zero if we use the
category ${\cal O}_*$, since that excludes multiples of $\hat 1$
as possible contexts/stages of truth.

It is easy to check that $\psi$ is an eigenstate of $t\hat
1+k\hat S_x+r\hat S_x^2$ if, and only if, $k=0$; hence, if
$s_{t,r\, {\cal O}}:t\hat 1+r\hat S_x^2\rightarrow \hat S_x$
denotes the morphism in $\cal O$ that corresponds to the
function $s_{t,r}:\sigma(\hat S_x)\rightarrow\mathR$ defined by
$s_{t,r}(\lambda):=t+r\lambda^2$, we see that
\begin{equation}
    \nu^\psi(S_x=1)=
\{s_{t,r\, {\cal O}}:t\hat 1+r\hat S_x^2\rightarrow \hat S_x\mid
        t,r\in \mathR\}\label{nuSx=1}
\end{equation}
and
\begin{equation}
    \nu^\psi(S_x=-1)=
\{s_{t,r\,\cal O}:t\hat 1+r\hat S_x^2\rightarrow \hat S_x\mid
        t,r\in\mathR\} \label{nuSx=-1}.
\end{equation}
The conclusion is that the propositions `$S_x=1$' and `$S_x=-1$'
are both assigned a non-trivial partial truth-value: namely the
sieve $\{s_{t,r\,{\cal O}}:t\hat 1+r\hat S_x^2\rightarrow \hat
S_x\mid t,r\in\mathR\}$; if ${\cal O}_*$ is used, then the value
$r=0$ is excluded.

On the other hand, we note that the proposition
`$S_x\in\{-1,1\}$' is represented by the projector $\hat
E[S_x=-1]+\hat E[S_x=+1]$, and also
\begin{equation}
\psi:=\pmatrix{0\cr
             1\cr
             0} = {1\over 2\sqrt2}\pmatrix{ 1\cr
                                            \sqrt2\cr
                                            1}
                - {1\over 2\sqrt2}\pmatrix{1\cr
                                            -\sqrt2\cr
                                            1}
\end{equation}
where the column vectors on the right hand side are eigenvectors
of $\hat S_x$ with eigenvalues $+1$ and $-1$ respectively.  It
follows that $\hat E[S_x\in\{-1,1\}]\psi=\psi$, and hence
\begin{equation}
    \nu^\psi(S_x\in\{-1,1\}])={\rm true}_{S_x}
                    \label{Sx-11=true}
\end{equation}
whereas, as shown by Eqs.\ (\ref{nuSx=1}--\ref{nuSx=-1}),
\begin{equation}
    \nu^\psi(S_x=1)\lor\nu^\psi(S_x=-1)=
\{s_{t,r\, {\cal O}}:t\hat 1+r\hat S_x^2\rightarrow \hat S_x\mid
    t,r\in \mathR\} <{\rm true\/}_{S_x}.
                    \label{sx1-<true}
\end{equation}
This failure of a strong disjunctive condition is typical of the
generalized valuations produced by quantum states, and we shall
return to this feature shortly. As emphasized in the
Introduction, it can be regarded as a fundamental consequence of
the superposition principle of quantum theory.

\subsection{The Generalized Valuation Associated with a
Density Matrix} \label{SubSec:Qu-statemixed-gen-val} We shall
now show that it is possible to associate a generalized
valuation to each density matrix state $\rho$ in the quantum
theory. To this end, we note that the previous definition Eq.\
(\ref{Def:nupsiDelta}) for $\nu^\psi$ can be re-expressed as
\begin{equation}
    \nu^\psi(A\in\Delta)=\{f_{\cal O}:\hat B\rightarrow\hat A
\mid \langle\psi, \hat E[B\in f(\Delta)]\psi\rangle
=\langle\psi, \psi\rangle\}
\end{equation}
or, in more physical terms,
\begin{equation}
    \nu^\psi(A\in\Delta)=\{f_{\cal O}:\hat B\rightarrow\hat A
\mid {\rm Prob}(B\in f(\Delta);\psi)=1\}
\end{equation}
where ${\rm Prob}(B\in f(\Delta);\psi)$ denotes the usual
quantum mechanical probability that the result of a measurement
of $B$ will lie in $f(\Delta)\subseteq\sigma(\hat
B)\subset\mathR$, given that the quantum state is $\psi$.

This way of expressing $\nu^\psi$ clarifies a little the
physical meaning of the generalized valuation---it is the set of
coarse-grainings $f(\hat A)$ of $\hat A$ such that the
probability that $f(A)$ lies in $f(\Delta)$ is $1$; something
that is construed in the standard interpretation as equivalent
to saying that $f(A)$ actually  has a value in $f(\Delta)$. It
also suggests the following definition for a generalized
valuation associated with any density matrix: \label{Def:nurho}

\begin{definition}
The {\em generalized valuation\/} $\nu^\rho$ associated with a
density matrix $\rho$ is
\begin{eqnarray}
    \nu^\rho(A\in\Delta)&:=&\{f_{\cal O}:\hat B\rightarrow \hat A
\mid {\rm Prob}(B\in f(\Delta);\rho)=1\}\nonumber \\[2pt]
                &\,=&\{f_{\cal O}:\hat B\rightarrow \hat A
\mid {\rm tr}(\rho\,\hat E[B\in f(\Delta)])=1\}.
                            \label{Def:nurhoDelta}
\end{eqnarray}
If necessary, the right hand side of Eq.\ (\ref{Def:nurhoDelta})
is to be understood in the sense of Eq.\ (\ref{Def:f(A)inD}).
\end{definition}

This class of generalized valuation is clearly of considerable
physical interest, and therefore it is important to check that
the necessary conditions are satisfied (of course, this will
include as a special case the generalized valuations $\nu^\psi$,
$\psi\in\cal H$.)

First we show that $\nu^\rho(A\in\Delta)$ is a sieve on $\hat A$
in $\cal O$. Thus, suppose $f_{\cal O}\in\nu^\rho(A\in\Delta)$,
and let $h_{\cal O}:\hat C\rightarrow\hat B$. Then, in the
lattice $\cal P$ of projection operators, $\hat E[B\in
f(\Delta)]\leq \hat E[C\in h(f(\Delta))]$; and hence ${\rm
tr}(\rho\,\hat E[B\in f(\Delta)])\leq {\rm tr}(\rho\hat E[C\in
h(f(\Delta))]$. In particular, since $f_{\cal O}\in
\nu^\rho(A\in\Delta)$, we have ${\rm tr}(\rho\,\hat E[B\in
f(\Delta)]) =1$, and hence ${\rm tr}(\rho\hat E[C\in
h(f(\Delta))])=1$ (since ${\rm tr}(\rho \hat P)\leq 1$ for all
projection operators $\hat P$). Thus $h_{\cal O}\in
\nu^\rho(A\in\Delta)$, which proves that $\nu^\rho(A\in\Delta)$
is a sieve on $\hat A$.

\paragraph*{1. The Functional Composition Rule:}

Next we must show that the functional composition rule is
satisfied. If $k_{\cal O}:\hat C\rightarrow \hat A$, then
\begin{eqnarray}
k^*_{\cal O}(\nu^\rho(A\in\Delta))&:=& \{j_{\cal O}:\hat
D\rightarrow \hat C\mid k_{\cal O}\circ j_{\cal
O}\in\nu^\rho(A\in\Delta)\}                 \nonumber\\ &\,=&
\{j_{\cal O}:\hat D\rightarrow\hat C\mid {\rm tr}(\rho\,\hat
E[D\in j(k(\Delta))])=1\}
\end{eqnarray}
whereas
\begin{equation}
    \nu^\rho(k(A)\in k(\Delta)):=\{h_{\cal O}:\hat D\rightarrow
k(\hat A)\mid {\rm tr}(\rho\,\hat E[D\in h(k(\Delta))])=1\}.
\end{equation}
Thus $k^*_{\cal O}(\nu^\rho(A\in\Delta))=\nu^\rho(k(A)\in
k(\Delta))$, as required.

We shall now consider the extent to which the object $\nu^\rho$
defined in Eq.\ (\ref{Def:nurhoDelta}) satisfies the remaining
conditions Eqs.\ (\ref{Null-gen}--\ref{Excl-gen}) in the formal
definition of a generalized valuation.

\paragraph*{2. The Null Proposition Condition:}

To check this, we note that $\nu^\rho(A\in\emptyset):=\{f_{\cal
O}:\hat B\rightarrow\hat A\mid {\rm tr}(\rho\,\hat E[B\in
f(\emptyset)])=1\}$. But this is the empty set since $\hat
E[B\in f(\emptyset)]=\hat 0$. Hence the null proposition
condition is satisfied.

\paragraph*{3. The Monotonicity Condition:}

Suppose $f_{\cal O}\in\nu^\rho(A\in\Delta_1)$ where
\begin{equation}
    \nu^\rho(A\in\Delta_1):=\{f_{\cal O}:\hat B\rightarrow\hat A
    \mid {\rm tr}(\rho\,\hat E[B\in f(\Delta_1)])=1\}.
\end{equation}
If $\Delta_1\subseteq\Delta_2$, then $f(\Delta_1)\subseteq
f(\Delta_2)$; and in the lattice of projection operators we then
have
\begin{equation}
    \hat E[B\in f(\Delta_1)])\leq \hat E[B\in f(\Delta_2)]).
\end{equation}
But then ${\rm tr}(\rho\,\hat E[B\in f(\Delta_1)])=1$ implies
that ${\rm tr}(\rho\,\hat E[B\in f(\Delta_2)])=1$ (since ${\rm
tr}(\rho\hat P)\leq 1$ for all projection operators $\hat P$).
Thus $f_{\cal O}$ also belongs to $\nu(A\in\Delta_2)$, which
means that
$\nu^\rho(A\in\Delta_1)\subseteq\nu^\rho(A\in\Delta_2)$.
However, in the Heyting algebra of sieves on $\hat A$, the
partial ordering operations is just subset inclusion; hence we
have shown that
\begin{equation}
 \Delta_1\subseteq\Delta_2 \mbox{ implies }
\nu^\rho(A\in\Delta_1)\leq \nu^\rho(A\in\Delta_2),
            \label{proof-mono}
\end{equation}
as required.

\paragraph*{3.1  No Strong Disjunctive Condition:}

From the monotonicity result in Eq.\ (\ref{proof-mono}) one can
immediately derive the weak disjunctive condition
\begin{equation}
    \nu^\rho(A\in\Delta_1)\cup\nu^\rho(A\in\Delta_2)\leq
        \nu^\rho(A\in\Delta_1\cup\Delta_2).     \label{GV1dum}
\end{equation}
However, in Section \ref{SubSec:primafacie} we remarked, in
rather general terms, that the existence of the quantum
superposition principle suggests that the reverse inequality may
not hold in Eq.\ (\ref{GV1dum}). To see this explicitly,
consider the special case when $\rho$ comes from a state vector
$\psi$, and let $\Delta_1:=\{a_1\}$, $\Delta_2:=\{a_2\}$ with
$a_1\neq a_2$---{\em i.e.}, we are considering the propositions
`$A=a_1$' and `$A=a_2$'. Then
\begin{equation}
\nu^\psi(A\in\{a_1\})\cup\nu^\psi(A\in\{a_2\})
    =\{f_{\cal O}:\hat B \rightarrow\hat A\mid
\hat B\psi=f(a_1)\psi\mbox{\ or\ } \hat B\psi=f(a_2)\psi\}
                                    \label{a1ora2}
\end{equation}
whereas
\begin{equation}
\nu^\psi(A\in\{a_1,a_2\}):=\{f_{\cal O}:\hat B\rightarrow\hat A
\mid \hat E[B\in f(\{a_1,a_2\})]\psi=\psi\}. \label{a1a2}
\end{equation}
Now suppose $f:\sigma(\hat A)\rightarrow\mathR$ is such that
$f(a_1)\neq f(a_2)$. Then satisfaction of the condition in Eq.\
(\ref{a1a2}) requires only that $\psi$ lies in the direct sum of
the eigenspaces of the operator $\hat B:=f(\hat A)$ that are
associated with the eigenvalues $f(a_1)$ and $f(a_2)$; in
particular, if $\psi$ is a non-trivial linear superposition of
these eigenstates of $\hat B$, then $f_{\cal O}: \hat
B\rightarrow\hat A$ will belong to the sieve
$\nu^\psi(A\in\{a_1,a_2\})$, but it will not belong to
$\nu^\psi(A\in\{a_1\})\cup\nu^\psi(A\in\{a_2\})$. Thus, there is
a strict inequality in Eq.\ (\ref{GV1dum}); an explicit example
is Eqs.(\ref{Sx-11=true}--\ref{sx1-<true}) in the spin-$1$ model
discussed above, with $f$ chosen to be the identity map on $\hat
S_x$. This should be contrasted with the generalized valuation
$\nu^V$ that satisfies the strong disjunctive condition Eq.\
(\ref{nuVconj}).

\paragraph*{3.2 No Strong Conjunctive Condition:}

We can also confirm the absence of any strong conjunctive
condition. Indeed, using the same pair of propositions as above,
we have `$A\in\{a_1\}\land A\in\{a_2\}$' $=$
`$A\in\{a_1\}\cap\{a_2\}$' $=$ `$A\in\emptyset$'; and hence
\begin{equation}
    \nu^\rho(A\in\{a_1\}\land A\in\{a_2\})=\emptyset=0_A.
\end{equation}
On the other hand
\begin{eqnarray}
\lefteqn{\nu^\rho(A\in\{a_1\})\cap\nu^\rho(A\in\{a_2\})=
\{f_{\cal O}:\hat B\rightarrow\hat A\mid {\rm tr}(\rho\,\hat
E[B\in f(\{a_1\})])=1\mbox{\ and\ }} \hspace{8cm} \nonumber\\
&&{\rm tr}(\rho\,\hat E[B\in f(\{a_2\})])=1\}. \label{horace}
\end{eqnarray}
Then, if we chose $f$ such that $f(a_1)=f(a_2)$ it is perfectly
possible for the right hand side of Eq.\ (\ref{horace}) to be
non-trivial. Thus, in general, there is no strong conjunctive
condition.

\paragraph*{4. The Exclusivity Condition:}

Finally, we must check the exclusivity condition. Thus suppose
$\Delta_1\cap\Delta_2=\emptyset$ and
$\nu^\rho(A\in\Delta_1)={\rm true}_A$; then, in particular,
${\rm tr}(\rho\,\hat E[A\in \Delta_1])=1$. Now define the real
number $k:={\rm tr}(\rho\,\hat E[A\in\Delta_2])$; this satisfies
$0\leq k\leq 1$. Then, since $\Delta_1\cap\Delta_2=\emptyset$,
the projectors $\hat E[A\in\Delta_1]$ and $\hat E[A\in\Delta_2]$
are orthogonal, and therefore $\hat
E[A\in\Delta_1\cup\Delta_2]=\hat E[A\in\Delta_1]+\hat
E[A\in\Delta_2]$. Thus ${\rm tr}(\rho\,\hat E[A\in
\Delta_1\cup\Delta_2])=1+k$. However, since ${\rm tr}(\rho\hat
P)\leq 1$ for all projection operators $\hat P$, and $k\geq 0$,
we deduce that $k=0$, {\em i.e.}, ${\rm tr}(\rho\,\hat
E[A\in\Delta_2])=0$. This means that
$\nu^\rho(A\in\Delta_2)<{\rm true}_A$; which proves exclusivity.

\paragraph*{5. The Unit Proposition Condition:}

We recall that, in the case of the generalized valuation
$\nu^V$, the unit proposition $A\in\sigma(\hat A)$ is not
necessarily given the truth-value $\mbox{true}_A$ but instead
satisfies the equation Eq.\ (\ref{gladys}).

The situation for $\nu^\rho$ is as follows. We have
\begin{equation}
\nu^\rho(A\in\sigma(\hat A)):= \{f_{\cal O}:\hat B\rightarrow
\hat A\mid \mbox{tr}(\rho\,
    \hat E[B\in f(\sigma(\hat A))])=1\}.        \label{nurAinR}
\end{equation}
But, according to the definition in Eq.\ (\ref{Def:f(A)inD}),
$\hat E[B\in f(\sigma(\hat A))]=\hat 1$; and thus, for these
types of generalized valuation, we do have
\begin{equation}
    \nu^\rho(A\in\sigma(\hat A))=\mbox{true}_A
\end{equation}
or, equivalently,
\begin{equation}
        \nu^\rho_A(\hat 1)=\mbox{true}_A
\end{equation}
for all contexts $\hat A$.

\paragraph*{The Negation Operation:}

We have not made any use so far of the negation operation in the
Heyting algebra of sieves, which is defined in general in Eq.\
(\ref{Def:negS}). In the case of the sieve
$\nu^\rho(A\in\Delta)$, this gives
\begin{eqnarray}
\neg\nu^\rho(A\in\Delta)&:=&\{f_{\cal O}:\hat B\rightarrow\hat A
        \mid \forall g_{\cal O}:\hat C\rightarrow\hat B,\;
    f_{\cal O}\circ g_{\cal O}\not\in \nu^\rho(A\in\Delta)\}
                                            \nonumber\\
        &\,=&\{f_{\cal O}:\hat B\rightarrow\hat A
        \mid \forall g_{\cal O}:\hat C\rightarrow\hat B,\;
            {\rm tr}(\rho\hat E[g(f(A))\in g(f(\Delta))])<1\}.
                        \qquad          \label{negnurho}
\end{eqnarray}
This is one point at which there is a real difference between
using $\cal O$ and ${\cal O}_*$ as the category of contexts. In
the former case, we are allowed the unit operator $\hat 1$ as an
allowed stage of truth, and then the choice of $g$ as the
constant map $c_{1,{\cal O}}:\hat 1\rightarrow\hat B$ gives the
spectral projector $\hat E[g(f(A)\in g(f(\Delta))=\hat
E[1\in\{1\}] =\hat 1$, for which ${\rm tr}(\rho\hat E)=1$. Thus
the right hand side of Eq.\ (\ref{negnurho}) would always be the
empty set, since this particular $g$ would exist and violate the
strict inequality.\footnote{In fact, this is true of presheaves
defined over {\em any\/} category $\cal C$ that has an initial
object;{\em i.e.}, an object $I$ such that there is a morphism
from $I$ to every object in the category.}

Thus, the negation operation is essentially trivial if the
category $\cal O$ is used, and this might suggest employing
${\cal O}_*$ instead. On the other hand, if we do keep the unit
operator as a possible stage of truth, then the definition of
$\nu^\rho$ shows that the set of operators appearing as the
domains of morphisms in the sieve $\nu^\rho(A\in\Delta)$ form an
{\em abelian algebra\/} of operators. This is an attractive
feature, and might suggest that using $\cal O$ has certain
advantages too.  Note that the spin-$1$ example discussed
earlier shows this effect very clearly: the set of operators
$\{t\hat 1+r\hat S_x^2\mid t,r\in\mathR\}$ that appear in the
right hand sides of Eq.\ (\ref{nuSx=1}) and Eq.\ (\ref{nuSx=-1})
are both abelian subalgebras, but cease to be so if the value
$r=0$ is excluded---as would be the case if ${\cal O}_*$ is used
as the category of contexts.

\paragraph*{A Generalization of the Valuations $\nu^\rho$:}
Finally, we note in passing that there exists a one-parameter
family of extensions of our valuations $\nu^\rho$. Namely, we
define
\begin{eqnarray}
    \nu^{r,\rho}(A\in\Delta)&:=&\{f_{\cal O}:\hat B\rightarrow
    \hat A
\mid {\rm Prob}(B\in f(\Delta);\rho)\geq r\}
                         \label{Def:nurhor}     \\[2pt]
        &\,=&\{f_{\cal O}:\hat B\rightarrow \hat A
\mid {\rm tr}(\rho\,\hat E[B\in f(\Delta)])\geq r\}\nonumber
\end{eqnarray}
where $r$ is a real parameter satisfying $0<r\leq 1$. It is
straightforward to show that for all real numbers $r$ in this
range, $\nu^{r,\rho}$ satisfies all our defining conditions for
a generalized valuation, with the exception of exclusivity.
Exclusivity is also satisfied if the parameter $r$ lies in the
range ${1\over 2}\leq r\leq 1$. This is an intriguing class of
generalized valuation, because it seems to promise a topos
perspective on the probabilistic statements of quantum theory.

\subsection{From Generalized Valuation to Partial Valuation, and
Back Again}\label{SubSec:GVPVGV}

As mentioned in the context of Theorem \ref{Theorem:gen-valNT}:
in the case of operators with a purely discrete spectrum, each
generalized valuation $\nu$ on propositions leads to the
valuation $V^\nu:{\bf\Sigma}\rightarrow\Omega$ on physical
quantities, as defined in Eq.\ (\ref{Def:VnuNT}). In particular,
for the generalized valuation $\nu^\rho$ associated with a
density matrix $\rho$, we have
\begin{equation}
    V^{\nu^\rho}_A(a)=\{f_{{\cal O}_d}:\hat B\rightarrow\hat A
\mid {\rm tr}(\rho \hat E[B=f(a)])=1\}.  \label{Vnurho}
\end{equation}

Now consider the generalized valuation $\nu^\psi$ associated
with a quantum state $\psi$ (as in Eq.\ (\ref{Def:nupsiDelta}))
in a situation where all the operators concerned have a discrete
spectrum only. The corresponding generalized valuation Eq.\
(\ref{Vnurho}) on physical quantities gives rise to a partial
valuation, which we shall denote $V^\psi$, whose domain is
defined as in Eq.\ (\ref{Def:domVnu}); thus
\begin{equation}
    \mbox{dom\,}V^\psi=\{\hat B\mid \hat B\psi=b\psi
    \mbox{ for some }b\}
\end{equation}
and, of course, if $\hat B$ belongs to the domain of $V^\psi$,
then $V^\psi(\hat B):=b$.

We can now apply Definition \ref{Defn:partval-genval-Delta} to
the partial valuation $V^\psi$ to get an associated generalized
valuation $\nu^{V^\psi}$ with
\begin{equation}
\nu^{V^\psi}(A\in\Delta):=\{f_{{\cal O}_d}:\hat B\rightarrow\hat
A
        \mid \exists b, \hat B\psi=b\psi \mbox{ and } b\in
f(\Delta)\},\label{Def:nuVpsi}
\end{equation}
which should be contrasted with the original definition of
$\nu^\psi$:
\begin{equation}
    \nu^\psi(A\in\Delta):=\{f_{{\cal O}_d}:\hat B\rightarrow\hat A
\mid \hat E[B\in f(\Delta)]\psi=\psi\}. \label{Def:nupsiDelta2}
\end{equation}

We point we wish to emphasize is that the generalized valuations
in Eq.\ (\ref{Def:nuVpsi}) and Eq.\ (\ref{Def:nupsiDelta2})
assign the same truth-values to propositions of the type
`$A=a$', but they differ in the way they treat more general
propositions `$A\in\Delta$'.

Thus, the definition of $\nu^{V^\psi}$ in Eq.\
(\ref{Def:nuVpsi}) shows that $f_{{\cal O}_d}:\hat
B\rightarrow\hat A$ belongs to the sieve
$\nu^{V\psi}(A\in\Delta)$ if, and only if, (i) $\psi$ is an
eigenvector of $\hat B=f(\hat A)$; and (ii) the corresponding
eigenvalue $b$ belongs to $f(\Delta)$---in other words, the
coarse-grained operator $f(\hat A)$ {\em has\/} a value in the
state $\psi$, and this value lies in $f(\Delta)$. On the other
hand, for $f_{{\cal O}_d}:\hat B\rightarrow\hat A$ to belong to
the sieve $\nu^\psi(A\in\Delta)$ requires only that $\psi$ is
some {\em linear combination\/} of such eigenstates of $f(\hat
A)$. In particular, this proves our earlier remark that the
chain in Eq.\ (\ref{gen-part-gen}) is not the identity
transformation on generalized valuations.

\subsection{The Generalized Valuation Produced by a
Projection Operator}

    There is apparently another way of constructing generalized
valuations using the mathematical ingredients of quantum theory.
To see this, we note that the defining condition $\hat E[B\in
f(\Delta)]\psi=\psi$ for $\nu^\psi(A\in\Delta)$ (see equation
Eq.\ (\ref{Def:nupsiDelta})) can be written as
\begin{equation}
\hat E[B\in f(\Delta)]|\psi\rangle\langle\psi|=
|\psi\rangle\langle\psi|\hat E[B\in f(\Delta)] =
|\psi\rangle\langle\psi|            \label{Epsi=psiE}
\end{equation}
where $|\psi\rangle\langle\psi|$ denotes the projector onto the
vector $\psi$. The expression Eq.\ (\ref{Epsi=psiE}) suggests an
immediate generalization to
\begin{equation}
\hat E[B\in f(\Delta)]\,\hat P = \hat P\,\hat E[B\in f(\Delta)]
                            = \hat P
\end{equation}
where $\hat P$ is now an arbitrary projection operator. In turn,
this condition is equivalent to the relation $\hat P\leq \hat
E[B\in f(\Delta)]$ in the lattice of projection operators. Hence
we are led to the following definition:
\begin{definition}
The {\em generalized valuation\/} $\nu^P$ associated with a
projection operator $\hat P$ is
\begin{equation}
\nu^P(A\in\Delta):=\{f_{\cal O}:\hat B\rightarrow\hat A\mid
                    \hat P\leq \hat E[B\in f(\Delta)]\}.
\end{equation}
\end{definition}
It is relatively straightforward to show that the necessary
conditions for a generalized valuation are satisfied; and, for
reasons of space, we shall not go into the details here. In
fact, if $\hat P$ is a finite projector ({\em i.e.}, its range
is a finite-dimensional subspace of the Hilbert space $\cal H$)
then $\nu^P$ is just a special case of the density-matrix
construction given above:
\begin{theorem}
If $\hat P$ is a projector such that $\dim\,\hat P=n<\infty$,
then, for all propositions `$A\in\Delta$',
\begin{equation}
    \nu^P(A\in\Delta)=\nu^{\rho^P}(A\in\Delta)
\end{equation}
where $\rho^P:={1\over n}\hat P$ is the density matrix given by
the projection operator $\hat P$.
\end{theorem}
\smallskip\noindent{\bf Proof}

\smallskip\noindent
If $\hat P\leq\hat E[B\in f(\Delta)]$ we have $\hat P\hat E[B\in
f(\Delta)]=\hat P$, and hence $\mbox{tr}(\rho^P\,\hat E[B\in
f(\Delta)])=\mbox{tr}\,{1\over n}\hat P=1$.  Thus
$\nu^P(A\in\Delta)\subseteq\nu^{\rho^P}(A\in\Delta)$.

Conversely, suppose $\hat B$ is such that
$\mbox{tr}(\rho^P\,\hat E[B\in f(\Delta)])=1$. Then
$\mbox{tr}\,{1\over n}\hat P=1=\mbox{tr}\,({1\over n}\hat P\hat
E[B\in f(\Delta)])$, which implies at once that $\hat P\leq\hat
E[B\in f(\Delta)]$. Therefore $\nu^{\rho^P}(A\in\Delta)\subseteq
\nu^P(A\in\Delta)$.  Hence $\nu^{\rho^P}(A\in\Delta)=
\nu^P(A\in\Delta)$. \hfill {\bf Q.E.D.}

    Thus nothing new is gained by introducing the valuations
$\nu^P$ on a finite-dimensional Hilbert space. However, if $\cal
H$ has an infinite dimension, then $\nu^P$ does give a new type
of valuation provided that the projection operator $\hat P$ has
an infinite range.

\section{Using the Set of Boolean Sub-Algebras as the
 Space of Contexts}\label{Sec:BoolSubContext}
\subsection{Preliminary Definitions}
We remarked earlier on the existence of a number of isomorphic
pairs of objects in the category $\cal O$. This occurs whenever
 operators $\hat A$ and $\hat B$ are related by
$\hat B=f(\hat A)$ and $\hat A=g(\hat B)$ for some functions
$f:\sigma(\hat A)\rightarrow\mathR$ and $g:\sigma(\hat
B)\rightarrow\mathR$.

From a physical perspective, if we know the value of one member
of such a pair of physical quantities, then we automatically
know the value of the other, and vice versa. In this sense, the
quantities are `physically equivalent' and, in some
circumstances, it is natural therefore to concentrate on the
equivalence classes, rather than on the individual quantities
themselves. In particular---since the spectral Boolean algebras
$W_A$ and $W_B$ of such pairs of operators are isomorphic---a
unique Boolean algebra can be associated with each equivalence
class of physical quantities.

Viewed mathematically, this suggests moving towards a formalism
in which the space of contexts, or stages of truth, is the
category $\cal W$ of all Boolean subalgebras of the projection
lattice, rather than the category $\cal O$ of self-adjoint
operators. Actually, we could have started {\em ab initio\/}
with $\cal W$ as the space of contexts, but we elected to use
$\cal O$ instead since the physical motivation for some of the
mathematical constructions is more transparent in this case; in
particular, this is true of the coarse-graining operation.
However, as we shall see in Sections \ref{Subsec:MotivationCGA}
and \ref{SubSec:defncoarse-graining}, the use of $\cal W$ also
suggests generalizations of the idea of coarse-graining which do
not arise in such a natural way if the category $\cal O$ is
used. Another significant reason for studying the use of $\cal
W$ is that most of the discussion extends at once to the general
quantum logic situation in which all that is said of the basic
mathematical structure of a quantum theory is that the
propositions are represented by elements in an orthomodular,
orthocomplemented lattice; however, we do not take up this
generalization here.

We start by constructing several important presheaf objects in
the topos ${\rm Set}^{{\cal W}^{\rm op}}$. The dual presheaf
${\bf D}:{\cal W}^{\rm op}\rightarrow {\rm Set}$ on $\cal W$ was
introduced in Definition \ref{Defn:dual-presheaf-W}, with ${\bf
D}(W)$ defined to be the dual of the Boolean algebra $W$; {\em
i.e.}, the set of homomorphisms from $W$ to the Boolean algebra
$\{0,1\}$. In our case, we are interested in a generalization of
this situation in which the `homomorphisms' from $W$ takes their
values in the Heyting algebra ${\bf\Omega}(W)$ of sieves on $W$
in the category $\cal W$ rather than in $\{0,1\}$.  Furthermore,
we must satisfy the algebraic conditions that specify a
generalized valuation.  To formalize these ideas we start with
the following definition.

\begin{definition}
A {\em valuation\/} of a Boolean algebra $B$ in a Heyting
algebra $H$ is a map $\phi:B\rightarrow H$ such that the
following conditions are satisfied:
\begin{eqnarray}
&&{\rm Null\ proposition\ condition:}\quad \phi(0_B)=0_H    \\
&&{\rm Monotonicity:}\quad \alpha\leq\beta\mbox{ implies }
                    \phi(\alpha)\leq\phi(\beta)             \\
&&{\rm Exclusivity:}\quad   \mbox{ If }\alpha\land\beta=0_B
\mbox{ and }\phi(\alpha)=1_H, \mbox{ then }\phi(\beta)<1.
\end{eqnarray}

The set of all valuations from $B$ to $H$ will be denoted
\rm{Val}$(B,H)$.
\end{definition}
These have been chosen to be the analogues of the conditions
that we have used a number of times already; and, as before, we
may also want to add the `Unit condition':
\begin{equation}
    \mbox{Unit proposition condition:}\quad \phi(1_B)=1_H
                    \hspace{5cm}
\end{equation}
In the case when $B$ is a Boolean subalgebra $W\in \cal W$, and
$H$ is ${\bf\Omega}(W)$, the elements of ${\rm
Val}(W,{\bf\Omega}(W))$ will be referred to as `{\em local
valuations\/}'.

We can now define a natural generalization of the dual presheaf
$\bf D$ on $\cal W$ (see Definition \ref{Defn:dual-presheaf-W})
in which the standard dual of a Boolean algebra is replaced with
an ${\bf\Omega}(W)$-valued valuation.

\begin{definition}\label{Defn:valuation-presheaf}
The {\em valuation presheaf\/} of $\cal W$ is the contravariant
functor ${\bf V}:{\cal W}\rightarrow{\rm Set}$ defined as
follows:
\begin{enumerate}
\item  {\em On objects in $\cal W$:}\ ${\bf V}(W):=
{\rm Val}(W,{\bf\Omega}(W))$, the set of local valuations on
$W$.

\item {\em On morphisms in $\cal W$:}\ If
$i_{W_2W_1}:W_2\rightarrow W_1$ ({\em i.e.}, $W_2\subseteq
W_1$), then ${\bf V}(i_{W_2W_1}):{\rm
Val}(W_1,{\bf\Omega}(W_1))\rightarrow {\rm
Val}(W_2,{\bf\Omega}(W_2))$ is defined by
\begin{equation}
     [{\bf V}(i_{W_2W_1})(\phi)](\hat\alpha):=
        i_{W_2W_1}^*(\phi(i_{W_2W_1}(\hat\alpha)))
\end{equation}
where $\phi\in{\rm Val}(W_1,{\bf\Omega}(W_1))$ and
$\hat\alpha\in W_2$, and where, in the poset category $\cal W$,
we have $i_{W_2W_1}^*(S)=\downarrow\!W_2\cap S$ for all
$S\in{\bf\Omega}(W_1)$ ({\em cf.}, Eq.\ (\ref{Def:Omqp})).
\end{enumerate}
\end{definition}

It is interesting consider global elements of $\bf V$ for two
reasons: (i) in order to compare with the dual presheaf $\bf D$,
for which---as we saw in Section
\ref{SubSec:KSdualpresheaves}---global elements are ruled out by
the Kochen-Specker theorem; and (ii) in order to make a contrast
with the definition of a generalized valuation in Section
\ref{Sec:GenVal}.

A global element $\gamma$ of the valuation presheaf corresponds
to a family of local valuations
$\gamma_W\in\mbox{Val}(W,{\bf\Omega}(W))$, $W\in\cal W$, such
that, if $W_2\subseteq W_1$ then, for all $\hat\alpha\in W_2$,
\begin{equation}
\gamma_{W_2}(\hat\alpha)=[{\bf V}(i_{W_2W_1})
(\gamma_{W_1})](\hat\alpha) =
        i_{W_2W_1}^*\{\gamma_{W_1}(i_{W_2W_1}(\hat\alpha))\}.
\label{matching-condition}
\end{equation}

In order to see the potential application for such global
elements, it is instructive to study these equations in the
special case where $W_1=W_A$ and $W_2=W_{h(A)}$ for some
function $h:\sigma(\hat A)\rightarrow\mathR$. Thus, suppose that
$\hat\alpha$ is the projection operator $\hat E[h(A)\in\Lambda]$
for some Borel subset $\Lambda\subseteq\sigma(h(\hat A))$.  Then
\begin{equation}
    i_{W_{h(A)}W_A}(\hat E[h(A)\in\Lambda])=
            \hat E[A\in h^{-1}(\Lambda)]
\end{equation}
and hence the matching condition in Eq.\
(\ref{matching-condition}) reads
\begin{equation}
\gamma_{W_{h(A)}}(\hat E[h(A)\in\Lambda])=  i_{W_{h(A)}W_A}^*\{
        \gamma_{W_A}(\hat E[A\in h^{-1}(\Lambda)])\}.
\end{equation}
In particular,
\begin{equation}
\gamma_{W_{h(A)}}(\hat E[h(A)\in h(\Delta)])=
i_{W_{h(A)}W_A}^*\{\gamma_{W_A}(\hat E[A\in
h^{-1}(h(\Delta))])\}.
                                            \label{match1}
\end{equation}

The corresponding matching equation in Section \ref{Sec:GenVal}
for the case of a generalized valuation on $\cal O$ was (Eq.\
(\ref{FC-gen}))
\begin{equation}
    \nu(h(A)\in h(\Delta))=h_{\cal O}^*\{\nu(A\in\Delta)\}
\end{equation}
or, in explicit contextual form,
\begin{equation}
    \nu_{h(A)}(\hat E[h(A)\in h(\Delta)])=
    h_{\cal O}^*\{\nu_A(\hat E[A\in\Delta])\}.\label{match2}
\end{equation}
Here it is important to contrast equations Eq.\ (\ref{match1})
and Eq.\ (\ref{match2}). In Eq.\ (\ref{match2}) the truth-value
of the proposition `$h(A)\in h(\Delta)$' in the context $h(\hat
A)$ is equated with the pull-back of the truth value of the {\em
finer\/} proposition `$A\in\Delta$' at stage $\hat A$; whereas
in Eq.\ (\ref{match1}) it is equated with the pull-back of the
valuation of the proposition `$A\in h^{-1}(h(\Delta))$'.

However, in the lattice of projectors, the projectors $\hat
E[A\in h^{-1}(h(\Delta))]$ and $\hat E[h(A)\in h(\Delta)]$ are
equal: so one is not pulling back a valuation of a finer
proposition.  Indeed, this equality is reflected in Eq.\
(\ref{matching-condition}) which guarantees the consistency of
(i) the sieve valuation of a given projector $\hat\alpha\in
W_2$, in the context $W_2$, with (ii) the sieve valuation of
$\hat\alpha$ if $W_2$ is embedded in the larger Boolean algebra
$W_1$ and the valuation is then taken in the context of $W_1$.

To sum up: the equality of $\hat E[A\in h^{-1}(h(\Delta))]$ and
$\hat E[h(A)\in h(\Delta)]$ means that if we were to define a
generalized valuation to be a global section of the valuation
presheaf $\bf V$, this would {\em not\/} be equivalent to our
earlier definition \ref{Defn:gen-val-O}, or
\ref{Defn:gen-val-W}, using the category $\cal O$.

Global sections of $\bf V$ could possibly be used to develop
another topos semantics for quantum theory---certainly, we would
not wish to claim that the approach adopted in the present paper
is necessarily the {\em only\/} one.  The first step would be to
show that global sections of $\bf V$ actually exist; preferably
by finding concrete examples in analogy to, for example, the
quantum-state induced general valuations $\nu^\rho$ discussed
earlier.

We may return in a later paper to the possible use of $\bf V$ in
the semantics of quantum theory.  But for the remainder of this
section we shall concentrate on showing how the analogue of the
coarse-graining operation---which played a central role in our
definition of a generalized valuation on $\cal O$---can be
introduced into the mathematical framework based on $\cal W$.

\subsection{The Motivation for the Coarse-Graining Axioms}
\label{Subsec:MotivationCGA} Motivated by what we did using the
category $\cal O$, we wish to define a coarse-graining operation
from $W_1$ to $W_2$ where $W_1$ and $W_2$ are Boolean
subalgebras of projectors with $W_2\subseteq W_1$. This is
intended to play an analogous role to that of the
coarse-graining functor ${\bf G}:{\cal O}^{\rm op}\rightarrow
{\rm Set}$, where the map ${\bf G}(f_{\cal O}): W_A\rightarrow
W_B$, with $\hat B=f(\hat A)$, was defined in Eq.\
(\ref{Def:G(fO)}) to map the projector $\hat E[A\in\Delta]$ to
$\hat E[f(A)\in f(\Delta)]$.

The procedure we shall follow is to extract certain key
properties of the coarse-graining process in $\cal O$ in this
Section, and then in Section \ref{SubSec:defncoarse-graining}
use these  as the basis for an axiomatization of an analogous
procedure for $\cal W$.

\paragraph*{1. Coarse Graining:}

The first step is to express more precisely the coarse-graining
property itself. We start by recalling that, in the lattice of
projection operators,
\begin{equation}
    \hat E[A\in\Delta]\leq \hat E[f(A)\in f(\Delta)].
                            \label{EA<EfA}
\end{equation}
However, if we wish to think of the operators on the left and
right hand sides of Eq.\ (\ref{EA<EfA}) as elements of the
Boolean subalgebras $W_A$ and $W_{f(A)}$ respectively, then it
pays to be pedantic by rewriting Eq.\ (\ref{EA<EfA}) as
\begin{equation}
\hat E[A\in\Delta]\leq i_{W_{f(A)}W_A}(\hat E[f(A)\in
f(\Delta)])
\end{equation}
where $i_{W_{f(A)}W_A}:W_{f(A)}\rightarrow W_A$ is the embedding
of the Boolean algebra $W_{f(A)}$ in $W_A$. In this sense, the
precise statement of the coarse-graining property is
\begin{equation}
\hat E[A\in\Delta]\leq i_{W_{f(A)}W_A}({\bf G}(f_{\cal O}) (\hat
E[A\in\Delta]))
\end{equation}
where the partial ordering `$\leq$' takes place in the Boolean
algebra $W_A$. The analogue of this expression will play a key
role in what follows.

\paragraph*{2. The Retraction Property:}

Considered as an element of $W_A$, the spectral projector $\hat
E[f(A)\in J]$ is $E[A\in f^{-1}(J)]$; more precisely,
\begin{equation}
    i_{W_{f(A)}W_A}(\hat E[f(A)\in J]) =
                \hat E[A\in f^{-1}(J))],
\end{equation}
and hence
\begin{equation}
{\bf G}(f_{\cal O})\circ i_{W_{f(A)}W_A}
            (\hat E[f(A)\in J]) =
    \hat E[f(A)\in f(f^{-1}(J))].   \label{Gi0}
\end{equation}
However, using the definition in Eq.\ (\ref{Def:f(A)inD}), it is
easy to show that the right hand side of Eq.\ (\ref{Gi0}) is
equal to $\hat E[f(A)\in J]$. Hence Eq.\ (\ref{Gi0}) becomes,
for all Borel subsets $J\subseteq\sigma(f(\hat A)))$,
\begin{equation}
{\bf G}(f_{\cal O})\circ i_{W_{f(A)}W_A}
            (\hat E[f(A)\in J]) =
    \hat E[f(A)\in J]   \label{Gi}
\end{equation}
which can be rewritten succinctly as
\begin{equation}
    {\bf G}(f_{\cal O})\circ i_{W_{f(A)}W_A} =
                    \mbox{id}_{W_{f(A)}}.
\end{equation}
This is expressed by saying that ${\bf G}(f_{\cal
O}):W_A\rightarrow W_{f(A)}$ is a {\em retraction\/}\footnote{In
general, a map $r:Y\rightarrow X$ is a {\em retraction\/} of a
subset embedding $i:X\subseteq Y$ if $r(x)=x$ for all $x\in
X\subseteq Y$; $X$ is then said to be a {\em retract\/} of $Y$.
Formally, we can write this as $r\circ i={\rm id}_X$.} map from
$W_A$ onto its embedded subalgebra $W_{f(A)}$.

\paragraph*{3. Composition Conditions:}

Since ${\bf G}$ is a contravariant functor from $\cal O$ to
$\mbox{Set}$, it follows that if $h_{\cal O}:\hat C\rightarrow
\hat B$ and $f_{\cal O}:\hat B\rightarrow\hat A$, then $f_{\cal
O}\circ h_{\cal O}:\hat C\rightarrow \hat A$, and
\begin{equation}
    {\bf G}(f_{\cal O}\circ h_{\cal O})=
    {\bf G}(h_{\cal O})\circ {\bf G}(f_{\cal O}).
\end{equation}
These can be thought of as the `composition conditions' that
must be satisfied by a coarse-graining operation.

\paragraph*{4. Monotonicity:}

If $\Delta_1\subseteq\Delta_2$, then $f(\Delta_1)\subseteq
f(\Delta_2)$, and hence $\hat E[f(A)\in f(\Delta_1)] \leq \hat
E[f(A)\in f(\Delta_2)]$.  From this we deduce the monotonicity
condition that is satisfied by the coarse-graining presheaf
${\bf G}$. Namely, if $\Delta_1\subseteq\Delta_2$, then
\begin{equation}
{\bf G}(f_{\cal O})(\hat E[A\in\Delta_1]) \leq {\bf G}(f_{\cal
O})(\hat E[A\in\Delta_2])\label{Gmono}
\end{equation}
Note that the partial-ordering operation `$\leq$' in Eq.\
(\ref{Gmono}) is taken in the Boolean algebra $W_{f(A)}$.

\subsection{The Definition of Coarse-Graining on $\cal W$}
\label{SubSec:defncoarse-graining}

\paragraph*{1. A Coarse-Graining Presheaf on $\cal W$:}

Motivated by the above we can now give our formal definition of
a `coarse-graining' operation on the category $\cal W$.

\begin{definition}
A {\em coarse-graining on $\cal W$} is an operation that
associates to each pair $W_2\subseteq W_1$, a `coarse-graining'
map $\theta_{W_1W_2}:W_1\rightarrow W_2$ with the following
properties:
\begin{enumerate}
\item {\em Coarse-graining:} For all $\hat\alpha\in W_1$,
\begin{equation}
    \hat\alpha\leq i_{W_2W_1}(\theta_{W_1W_2}(\hat\alpha)).
                            \label{Defn:CG-CG}
\end{equation}
If $W_2=W_1$, then $\theta_{W_1W_1}=\mbox{id}_{W_1}$.

\item{\em Monotonicity:} If $\hat\alpha,\hat\beta\in W_1$ are
such that $\alpha\leq\beta$, then
    \begin{equation}
        \theta_{W_1W_2}(\hat\alpha)\leq\
        \theta_{W_1W_2}(\hat\beta)\label{tha<b}
    \end{equation}

\item {\em Retraction:} For all $\hat\alpha\in W_2$,
\begin{equation}
\theta_{W_1W_2}(i_{W_2W_1}(\hat\alpha)) =\hat\alpha.
                                    \label{retraction}
\end{equation}
Thus $\theta_{W_1W_2}$ is a retraction of $W_2$ onto $W_1$; {\em
i.e.}, $\theta_{W_1W_2}\circ i_{W_2W_1}={\rm id}_{W_2}$.

\item{\em Composition conditions:} If
$W_3\subseteq W_2\subseteq W_1$ then
\begin{equation}
    \theta_{W_2W_3}\circ\theta_{W_1W_2}=\theta_{W_1W_3}.
                            \label{matching-conditions}
\end{equation}

\end{enumerate}

\end{definition}

From a topos perspective,  the composition conditions show that
$\theta$ defines a presheaf ${\bf\Theta}:{\cal W}^{\rm
op}\rightarrow{\rm Set}$ that is defined (i) on objects as
${\bf\Theta}(W):=W$; and (ii) on a morphism
$i_{W_2W_1}:W_2\rightarrow W_1$ as
${\bf\Theta}(i_{W_2W_1}):=\theta_{W_1W_2}$. Conversely, we could
{\em define\/} a `coarse-graining presheaf on $\cal W$' to be a
presheaf on $W$ that satisfies the remaining conditions, viz.
coarse-graining,  monotonicity, and retraction.

\paragraph*{2. The Canonical Coarse-Graining Presheaf:}

It is important to show that there exists at least one
coarse-graining presheaf. In the analogous case of
contextualizing over $\cal O$, there was a `canonical'
coarse-graining operation that came from considering the
implications of writing one operator $\hat B$ as a function
$f(\hat A)$ of another. The key to finding the analogue of this
construction for the category $\cal W$ is contained in Theorem
\ref{Theorem:inf-defn}. This result leads naturally to the
following definition:

\begin{definition}
\label{Defn:can-coarse-grain} The {\em canonical
coarse-graining\/} of $\cal W$ associates to each pair
$W_2\subseteq W_1$, the coarse-graining map
$\phi_{W_1W_2}:W_1\rightarrow W_2$ defined by
\begin{equation}
    \phi_{W_1W_2}(\hat\alpha):=
        \inf\{\hat \beta\in W_2\mid
            \hat\alpha\leq i_{W_2W_1}(\hat\beta)\}
                        \label{Def:cancoarse-grain}
\end{equation}
for all $\hat\alpha\in W_1$.
\end{definition}

    We shall leave as a straightforward exercise the task of
showing that the entity thus defined really does satisfy all the
requirements for a coarse-graining operation: coarse-graining,
monotonicity, retraction, and the composition condition.

\paragraph*{3. Generalized Valuations Associated with a
Coarse-Graining Presheaf:}

We shall now show that for any given coarse-graining presheaf,
there is an associated definition of a generalized valuation
that is constructed as a matching family of local valuations:

\begin{definition}
A {\em generalized valuation\/} on $\cal W$ associated with a
coarse-graining presheaf ${\bf\Theta}$ is a family of local
valuations $\phi_W:W\rightarrow{\bf\Omega}(W)$, $W\in\cal W$,
such that if $W_2\subseteq W_1$ then, for all $\hat\alpha\in
W_1$,
\begin{equation}
        \phi_{W_2}(\theta_{W_1W_2}(\hat\alpha))=i_{W_2W_1}^*(
\phi_{W_1}(\hat\alpha)).        \label{Def:genvalCG}
\end{equation}
\end{definition}

From a physical perspective, the interpretation of a generalized
valuation on $\cal W$ is closely analogous to that of
generalized valuation on $\cal O$ as given by the discussion
following Definition \ref{Defn:gen-val-O}. Specifically:
although a particular projector $\hat\alpha\in W_1$ may not be
assigned the value `totally true' at a stage of truth $W_1$, it
does have a partial truth-value that is given by the set of
coarser Boolean algebras $W_2$ that belong to the sieve
$\phi_{W_1}(\hat\alpha)$, where, on account of Eq.\
(\ref{Def:genvalCG}), each corresponding coarse-grained
projector $\theta_{W_1W_2}(\hat\alpha)$ is given the value
`totally true' at the corresponding stage of truth $W_2$. (This
should be compared with the discussion following Eq.\
(\ref{f*S}), and after Eq.\ (\ref{h*nuVA=1C}).)

\paragraph*{4. The Generalized Valuation Produced by a Density
Matrix:}

There is no difficulty in finding examples of generalized
valuations associated with any coarse-graining presheaf. In
particular, each density-matrix state $\rho$ produces one
according to the following definition.

\begin{definition}
The {\em generalized valuation\/} $\nu^\rho$ on $\cal W$
associated with a coarse-graining presheaf $\bf\Theta$ and a
density matrix $\rho$, is defined at each stage $W$ by
\begin{equation}
\nu^\rho_W(\hat\alpha):=\{W'\subseteq W\mid \mbox{tr}(\rho\,
                            \theta_{WW'}(\hat\alpha))=1\}
                            \label{Def:mu-rho}
\end{equation}
for all $\hat \alpha\in W$.
\end{definition}

To show that this is indeed a generalized valuation it is
necessary to show that (i) each $\nu^\rho_W:
W\rightarrow{\bf\Omega}(W)$ is a local valuation; and (ii) the
maps $\nu^\rho_W$ fit together in the way indicated by the
intertwining condition in Eq.\ (\ref{Def:genvalCG}). The proofs
are contained in the following theorem.

\begin{theorem}
The quantity $\nu^\rho$ defined in Eq.\ (\ref{Def:mu-rho})
satisfies all the conditions for a generalized valuation on
$\cal W$.
\end{theorem}

\smallskip\noindent{\bf Proof}\smallskip

\noindent {\bf A. For each stage $W\in\cal W$, $\nu^\rho_W$ is a
local valuation:}
\paragraph*{1. $\nu^\rho_W(\hat\alpha)$ is a sieve:}

The first step is to show that $\nu^\rho_W(\hat\alpha)$ is a
sieve on $W$ in $\cal W$. Thus suppose that
$W'\in\nu^\rho_W(\hat\alpha)$ and consider any subalgebra
$W{''}\subseteq W'$. The composition condition Eq.\
(\ref{matching-conditions}) applied to the chain $W''\subseteq
W'\subseteq W$ gives
\begin{equation}
        \theta_{WW''}(\hat\alpha)=\theta_{W'W''}(
            \theta_{WW'}(\hat\alpha))   \label{einstein}
\end{equation}
for all $\hat\alpha\in W$. Then applying the coarse-graining
condition Eq.\ (\ref{Defn:CG-CG}) to $\theta_{WW'}(\hat\alpha)$,
and using Eq.\ (\ref{einstein}), we get
\begin{equation}
        \theta_{WW'}(\hat\alpha)\leq i_{W''W'}\left(
        \theta_{W'W''}(\theta_{WW'}(\hat\alpha))\right)
                = i_{W''W'}(\theta_{WW''}(\hat\alpha)).
\end{equation}
Hence, in the Boolean algebra $W'$, we have
$\theta_{WW'}(\hat\alpha)\leq \theta_{WW''}(\hat\alpha)$. Thus,
in particular, ${\rm tr}(\rho\,\theta_{WW'}(\hat\alpha))=1$
implies ${\rm tr}(\rho\,\theta_{WW''}(\hat\alpha))=1$; and hence
$\nu^\rho_W(\hat\alpha)$ is a sieve on $W$ in $\cal W$.

\paragraph*{2. The null proposition condition:}

The equations Eq.\ (\ref{retraction}) and $i_{W'W}(0_{W'})=0_W$,
imply $\theta_{WW'}(\hat 0)=\hat 0$, from which the null
proposition condition follows at once. It is also trivial to
check that $\nu^\rho$ satisfies the unit proposition condition
$\nu^\rho_W(\hat 1)={\rm true}_W$.

\paragraph*{3. The monotonicity condition:}
To show monotonicity, suppose that $\hat\alpha,\hat\beta\in W$
satisfy $\hat\alpha\leq\hat\beta$, and that
$W'\in\nu_W^\rho(\hat\alpha)$, so that ${\rm
tr}(\rho\,\theta_{WW'}(\hat\alpha))=1$. Then the monotonicity
condition Eq.\ (\ref{tha<b}) obeyed by the coarse-graining
operation implies that $\theta_{WW'}(\hat\alpha)\leq
\theta_{WW'}(\hat\beta)$, and hence that ${\rm
tr}(\rho\,\theta_{WW'}(\hat\alpha)) \leq {\rm
tr}(\rho\,\theta_{WW'}(\hat\beta))$.  However, ${\rm
tr}(\rho\hat P)\leq 1$ for all projection operators $\hat P$,
and hence ${\rm tr}(\rho\,\theta_{WW'}(\hat\alpha))=1$ implies
${\rm tr}(\rho\,\theta_{WW'}(\hat\beta))=1$, which means that
$W'\in\nu_W^\rho(\beta)$; hence the monotonicity condition is
satisfied.

\paragraph*{4. The exclusivity condition:}
To show exclusivity, suppose that $\hat\alpha,\hat\beta\in W$
satisfy $\hat\alpha\land\hat\beta=0$, and that
$\nu_W^\rho(\hat\alpha):=1_W$.  The latter implies that
$W\in\nu^\rho_W(\hat\alpha)$, and hence, since $\theta_{WW}={\rm
id}_W$, we have ${\rm tr}(\rho\,\hat\alpha)=1$. However,
$\hat\alpha\land\hat\beta=0$ implies that
$\hat\beta\leq\neg\hat\alpha$ and, since $\neg\hat\alpha=\hat
1-\hat\alpha$, we get
\begin{equation}
    0\leq {\rm tr}(\rho\hat\beta)\leq
            {\rm tr}(\rho(\hat 1-\hat\alpha)) =0.
\end{equation}
Thus ${\rm tr}(\rho\hat\beta)=0$, and hence
$W\not\in\nu^\rho_W(\hat\beta)$. Therefore,
$\nu^\rho_W(\hat\beta)<1_W$, which proves exclusivity.

\medskip\noindent
{\bf B. For each stage $W\in\cal W$, $\nu^\rho$ satisfies the
intertwining condition Eq.\ (\ref{Def:genvalCG}):}

\noindent To see that Eq.\ (\ref{Def:genvalCG}) is satisfied,
let $W_2,W_1\in\cal W$ be such that $W_2\subseteq W_1$. Then,
for all $\hat\alpha\in W_1$,
\begin{eqnarray}
    \nu^\rho_{W_2}(\theta_{W_1W_2}(\hat\alpha))&:=&
    \{W'\subseteq W_2\mid {\rm tr}(\rho\,\theta_{W_2W'}
            (\theta_{W_1W_2}(\hat\alpha)))=1\}  \nonumber\\
        &\,=&\{W'\subseteq W_2\mid {\rm tr}(\rho\,\theta_{W_1W'}
            (\hat\alpha))=1\}   \label{fred}
\end{eqnarray}
where the last line follows from the composition conditions Eq.
(\ref{matching-conditions}). On the other hand,
\begin{eqnarray}
\{W'\subseteq W_2\mid {\rm tr}(\rho\,\theta_{W_1W'}
            (\hat \alpha))=1\} &=&\downarrow\!\!W_2\cap
\{W'\subseteq W_1\mid {\rm tr}(\rho\,\theta_{W_1W'}
            (\hat\alpha))=1\}       \nonumber\\
        &=&i^*_{W_2W_1}(\nu^\rho_{W_1}(\hat \alpha)),
\end{eqnarray}
so that $\nu_{W_2}(\theta_{W_1W_2}(\hat\alpha))=i_{W_2W_1}^*(
\nu_{W_1}(\hat\alpha))$, as required.   \hfill {\bf Q.E.D.}

\paragraph*{5. The Topos-Theoretic Perspective:}

From a topos-theoretic perspective we note that each generalized
valuation $\nu$ on $\cal W$ defines a natural transformation
$N^\nu$ between the coarse-graining presheaf $\Theta$ and the
subobject classifier ${\bf\Omega}$, in which, at each stage of
truth $W$, $N^\nu_W:{\bf\Theta}(W)\rightarrow{\bf\Omega}(W)$ is
defined by $N^\nu_W(\hat\alpha):=\nu_W(\hat\alpha)$. It is a
straightforward exercise in diagram chasing to show that $N^\nu$
really is a natural transformation.

Thus to each generalized valuation $\nu$ on $\cal W$ there
corresponds a morphism in the topos ${\rm Set}^{{\cal W}^{\rm
op}}$ between the coarse-graining presheaf ${\bf\Theta}$ and the
sub-object classifier. In particular, therefore, each
generalized valuation on $\cal W$ corresponds to a subobject of
${\bf\Theta}$. The overall implications of this are the same as
for the analogous result in the case of generalized valuations
defined on $\cal O$.

\section{Conclusion}
The Kochen-Specker theorem shows the non-existence of global
valuations on the self-adjoint operators in a quantum theory if
the dimension of the underlying Hilbert space $\cal H$ is
greater than two. We have shown that this theorem is equivalent
to the statement that a certain presheaf on the category of
bounded self-adjoint operators has no global sections. Then,
motivated by the underlying topos structure, we introduced a new
type of valuation which {\em is\/} globally defined, but whose
truth values (i) are contextual; and (ii) lie in a larger
Heyting algebra than the minimal $\{0,1\}$ Boolean algebra of
standard logic.

    Thus our construction shows clearly how contextual features
enter into a `neo-realist' interpretation of quantum theory. It
also shows that the use of multi-valued logic is perfectly
feasible. In particular, there is no ambiguity or uncertainty
about what the logical connectives are: the Heyting algebra of
the sieves at any particular stage of truth, or context, is
precisely fixed by the structure of the base category---in our
case $\cal O$ or $\cal W$---on which the relevant presheaves are
defined.

    As we generalize at the end of the Introduction, the main aim
of the present paper is to provide the main mathematical tools
and some of the general ideas involved in the application of
topos ideas in quantum theory. Much remains to be done to
develop both the mathematical and the conceptual implications of
these ideas; the latter in particular are discussed in a
forthcoming paper \cite{BI98a}.

    At the mathematical level, the work reported in this paper
suggests a number of topics for further research. Of particular
importance is the study of the space of all generalized
valuations which---as mentioned in Section
\ref{SubSec:topos-int-gen-val}---might carry an intuitionistic
logical structure by virtue of the identification of each
generalized valuation with a subobject of the coarse-graining
presheaf $\bf G$. An important part of any such study is likely
to involve a closer investigation of the negation operation in
the Heyting algebras, which we have not exploited in any
significant way so far.

A crucial question regarding the space of all generalized
valuations is to understand the mathematical status of the
valuations $\nu^\rho$ generated by the mixed states $\rho$ in
the quantum system. In particular, if we impose the `unit
proposition condition' of Eq.\ (\ref{unit-prop-cond}), is it
possible to find a set of extra conditions to be imposed on the
generalized valuations that will guarantee that {\em every\/}
subobject of $\bf G$ that satisfies these and the original
defining conditions Eqs.\ (\ref{Null-gen}---\ref{Excl-gen}), has
the form $\nu^\rho$ for some density matrix $\rho$? In effect,
we are asking for a contextualized, Heyting-algebra valued
analogue of the Gleason theorem. It seems likely that an
important role in such an analysis will be played by the
one-parameter family of generalized valuations $\nu^{r,\rho}$
defined in Eq.\ (\ref{Def:nurhor}).

    A number of other questions suggest themselves. For example,
is our theory of generalized values of physical quantities and
propositions related at all to existing ideas on `unsharp'
values of quantum quantities (as described, for example, in
\cite{BGL95})? Another important example is the relation of our
constructions to the standard probabilistic statements of
quantum theory.

Another important issue is to see how the phenomenon of quantum
entanglement is reflected in the truth-values assigned by our
generalized valuations. Thus we should study possible relations
between a generalized valuation $\nu^\psi$, where $\psi$ is an
entangled state in a tensor product ${\cal H}_1\otimes{\cal
H}_2$, and the generalized valuations associated with vectors in
the constituent Hilbert spaces ${\cal H}_1$ and ${\cal H}_2$.

The discussion in Section \ref{SubSec:defncoarse-graining} of
coarse-graining in the category $\cal W$ of Boolean subalgebras
implies that there might be coarse-graining functors other than
the canonical one given in Definition
\ref{Defn:can-coarse-grain}. It is clearly important to see if
this is indeed the case, since each such functor would give rise
to a whole new class of generalized valuations. In particular,
this is relevant to the problem mentioned above of classifying
generalized valuations. It would also be interesting to study
this question in a simple model quantum-logic situation in which
the orthoalgebra of propositions is not the projection lattice
of a Hilbert space.

Finally, there is the question of the Kochen-Specker theorem
itself: in particular, the possibility of finding a new proof
based on some theory of obstructions to the construction of
global sections of the spectral presheaf, rather as one studies
obstructions to the construction of global cross-sections of
non-trivial fibre bundles. This is an intriguing mathematical
challenge, and one whose solution could generate a deeper
insight into the ultimate significance of the Kochen-Specker
theorem. It could also suggest ways of using topos ideas in
quantum theory other than the coarse-graining scheme employed in
the present paper.

\section*{Acknowledgements}

Chris Isham is most grateful to the Mrs L.D.~Rope Third
Charitable Settlement for financial assistance during the course
of this work.

\appendix
\section{A Brief Account of the Relevant Parts of Topos Theory}
\label{Sec:mathprel}

\subsection{Presheaves on a Poset}
\label{SubSec:presheaves-poset} Topos theory is a remarkably
rich branch of mathematics which can be approached from a
variety of different viewpoints. The relevant general area of
mathematics is category theory; where, we recall, a category
consists of a collection of {\em objects\/} and a collection of
{\em morphisms\/} (or {\em arrows\/}). In the special case of
the category of sets, the objects are sets, and a morphism is a
function between a pair of sets. In general, each morphism $f$
in a category is associated with a pair of objects, known as its
`domain' and the `codomain', and is written in the form
$f:B\rightarrow A$ where $B$ and $A$ are the domain and codomain
respectively. Note that this arrow notation is used even if $f$
is not a function in the normal set-theoretic sense. A key
ingredient in the definition of a category is that if
$f:B\rightarrow A$ and $g:C\rightarrow B$ ({\em i.e.}, the
codomain of $g$ is equal to the domain of $f$) then $f$ and $g$
can be `composed' to give an arrow $f\circ g:C\rightarrow A$; in
the case of the category of sets, this is just the usual
composition of functions.

In many categories, the objects are sets equipped with some type
of additional structure, and the morphisms are functions that
preserve this structure; for example, in the category of groups,
an object is a group, and a morphism $f:G_1\rightarrow G_2$ is a
map from the group $G_1$ to the group $G_2$ that is also a
homomorphism. However, not all categories are of this type.  For
example, any partially-ordered set (`poset') $\cal C$ can be
regarded as a category in which (i) the objects are defined to
be the elements of $\cal C$; and (ii) if $p,q\in\cal C$, a
morphism from $p$ to $q$ is defined to exist if, and only if,
$p\leq q$ in the poset structure.  Thus, in a poset regarded as
a category, there is at most one morphism between any pair of
objects $p,q\in\cal C$; if it exists, we shall write this
morphism as $i_{pq}:p\rightarrow q$.

From our perspective, the most relevant feature of a topos is
that it is a category in which the subobjects of an object
behave in many ways like the subsets of a set in set theory
\cite{Gol84,MM92}. In particular, the subsets $K\subseteq X$ of
a set $X$ are in one-to-one correspondence with functions
$\chi^K:X\rightarrow\{0,1\}$, where $\chi^K(x)=1$ if $x\in K$,
and $\chi^K(x)=0$ otherwise. Thus the target space $\{0,1\}$ can
be regarded as the simplest `false-true' Boolean algebra, and
the proposition `$x\in K$' is true if $\chi^K(x)=1$, and false
otherwise.

In the case of a topos, the subobjects $K$ of an object $X$ in
the topos are in one-to-one correspondence with morphisms
$\chi^K:X\rightarrow \Omega$, where the special object $\Omega$
in the topos---called the `subobject classifier', or `object of
truth-values'---plays an analogous role to that of $\{0,1\}$ in
the category of sets.  In particular, we are interested in the
theory of presheaves where, as we shall see, a morphism
$\chi^K:X\rightarrow\Omega $ corresponds to a contextualized,
multi-valued truth assignment.

To illustrate the main ideas, we will first give a few
definitions from the theory of presheaves on a partially ordered
set (or `poset'); physically, this poset will represent the
space of `contexts' in which generalized truth-values are to be
assigned. We shall then use these ideas to motivate the
definition of a presheaf on a general category.  Only the
briefest of treatments is given here, and the reader is referred
to the standard literature for more information
\cite{Gol84,MM92}.

    A {\em presheaf\/} (also known as a {\em varying set\/}) $X$
on a poset $\cal C$ is a function that assigns to each $p\in\cal
C$, a set $X_p$; and to each pair $p\leq q$, a map
$X_{qp}:X_q\rightarrow X_p$ such that (i) $X_{pp}:X_p\rightarrow
X_p$ is the identity map ${\rm id}_{{X_p}}$ on $X_p$, and (ii)
whenever $p\leq q\leq r$, the composite map
$X_r\stackrel{X_{rq}}\longrightarrow
X_q\stackrel{X_{qp}}\longrightarrow X_p$ is equal to
$X_r\stackrel {X_{rp}}\longrightarrow X_p$, so that\footnote{A
matter of convention is involved here. Sometimes a presheaf is
defined as above except that, to each $p\leq q$, one associates
a function $X_{pq}:X_p\rightarrow X_q$ that maps $X_p$ to $X_q$,
rather than the function $X_{qp}$ that maps $X_q$ to $X_p$. To
reflect this, equation Eq.\ (\ref{Xrp=XqpXrq}) is replaced by
$X_{pr}=X_{qr}\circ X_{pq}$ for $p\leq q\leq r$. Presheaves in
the sense of the main text are in one-to-one correspondence with
presheaves in this alternative sense, in which the latter are
defined on the {\em opposite\/} poset $\cal C^{\rm
op}$---defined to be the same set as $\cal C$ but with all the
partial ordering relations reversed.}
\begin{equation}
        X_{rp}= X_{qp}\circ X_{rq}.     \label{Xrp=XqpXrq}
\end{equation}

    A {\em morphism\/} $\eta:X\rightarrow Y$ between two
presheaves $X,Y$ on $\cal C$ is a family of maps
$\eta_p:X_p\rightarrow Y_p$, $p\in\cal C$, that satisfy the
intertwining conditions
\begin{equation}
        \eta_p\circ X_{qp}=Y_{qp}\circ\eta_q
\end{equation}
whenever $p\leq q$. This is equivalent to the commutative
diagram
\begin{equation}
    \bundle{X_q}{\eta_q}{Y_q}\bundlemap{X_{qp}}{Y_{qp}}
    \bundle{X_p}{\eta_p}{Y_p}       \label{Def:eta}
\end{equation}

    A {\em subobject\/} of a presheaf $X$ is a presheaf $K$, with
a morphism $i:K\rightarrow X$ such that (i) $K_p\subseteq X_p$
for all $p\in\cal C$; and (ii) for all $p\leq q$, the map
$K_{qp}:K_q\rightarrow K_p$ is the restriction of
$X_{qp}:X_q\rightarrow X_p$ to the subset $K_q\subseteq X_q$.
This is shown in the commutative diagram
\begin{equation}
\bundle{K_q}{}{X_q}\bundlemap{K_{qp}}{X_{qp}}
                            \bundle{K_p}{}{X_p}\label{cd}
\end{equation}
where the vertical arrows are subset inclusions.

The collection of all presheaves on a poset $\cal C$ forms a
category, denoted ${\rm Set}^{{\cal C}^{\rm op}}$.  The
morphisms between presheaves in this category are defined as the
morphisms above.

\subsection{Presheaves on a General Category}
\label{SubSec:presheaves-gen-cat} The ideas sketched above admit
an immediate generalization to the theory of presheaves on an
arbitrary `small' category $\cal C$ (the qualification `small'
means that the collection of objects is a genuine set, as is the
collection of all morphisms between any pair of objects). To
make the necessary definition we first need the idea of a
`functor':

\paragraph*{1. The Idea of a Functor:}

A central concept is that of a `functor' between a pair of
categories $\cal C$ and $\cal D$. Broadly speaking, this is a
morphism-preserving function from one category to the other. The
precise definition is as follows.

\begin{definition}
\
\begin{enumerate}
\item {A {\em covariant functor\/} $\bf F$ from a category
$\cal C$ to a category $\cal D$ is a function that assigns
    \begin{enumerate}
        \item to each $\cal C$-object $A$, a $\cal D$-object
        ${\bf F}(A)$;

        \item {to each $\cal C$-morphism $f:B\rightarrow A$, a
$\cal D$-morphism ${\bf F}(f):{\bf F}(B)\rightarrow {\bf F}(A)$
such that ${\bf F}({\rm id}_A)={\rm id}_{{\bf F}(A)}$; and, if
$g:C\rightarrow B$, and $f:B\rightarrow A$ then
    \begin{equation}
        {\bf F}(f\circ g)={\bf F}(f)\circ
                {\bf F}(g).     \label{Def:covfunct}
    \end{equation}
        }
    \end{enumerate}
    }

\item {A {\em contravariant functor\/} $\bf X$ from a category
$\cal C$ to a category $\cal D$ is a function that assigns
\begin{enumerate} \item to each $\cal C$-object $A$, a $\cal
D$-object ${\bf X}(A)$;

    \item {to each $\cal C$-morphism $f:B\rightarrow A$, a $\cal
D$-morphism ${\bf X}(f):{\bf X}(A)\rightarrow {\bf X}(B)$ such
that ${\bf X}({\rm id}_A)={\rm id}_{{\bf X}(A)}$; and, if
$g:C\rightarrow B$, and $f:B\rightarrow A$ then
    \begin{equation}
        {\bf X}(f\circ g)={\bf X}(g)\circ{\bf X}(f).
                        \label{Def:confunct}
    \end{equation}
        }
    \end{enumerate}
    }
\end{enumerate}

\end{definition}

The connection with the idea of a presheaf on a poset is
straightforward. As mentioned above, a poset $\cal C$ can be
regarded as a category in its own right, and it is clear that a
presheaf on the poset $\cal C$ is the same thing as a
contravariant functor $\bf X$ from the category $\cal C$ to the
category `${\rm Set}$' of normal sets. Equivalently, it is a
covariant functor from the `opposite' category\footnote{The
`opposite' of a category $\cal C$ is a category, denoted ${\cal
C}^{\rm op}$, whose objects are the same as those of $\cal C$,
and whose morphisms are defined to be the opposite of those of
$\cal C$; {\em i.e.}, a morphism $f:A\rightarrow B$ in ${\cal
C}^{\rm op}$ is said to exist if, and only if, there is a
morphism $f:B\rightarrow A$ in $\cal C$.} ${\cal C}^{\rm op}$ to
${\rm Set}$.  More precisely, in terms of the notation used
earlier, the sets $X_p$, $p\in\cal C$, are defined as
\begin{equation}
    X_p:={\bf X}(p)
\end{equation}
and, if $p\leq q$ (so that $i_{pq}:p\rightarrow q$), the map
$X_{qp}:X_q\rightarrow X_p$ is defined as
\begin{equation}
    X_{qp}:={\bf X}(i_{pq}).
\end{equation}
Clearly, Eq.\ (\ref{Xrp=XqpXrq}) corresponds to the
contravariant condition Eq.\ (\ref{Def:confunct}).

\paragraph*{2. Presheaves on an Arbitrary Category $\cal C$:}

These remarks motivate the definition of a presheaf on an
arbitrary small category $\cal C$: namely, a {\em presheaf\/} on
$\cal C$ is a covariant functor ${\bf X}:{{\cal C}^{\rm
op}}\rightarrow {\rm Set}$ from ${\cal C}^{\rm op}$ to the
category of sets.  Equivalently, a presheaf is a contravariant
functor from $\cal C$ to the category of sets.

We want to make the collection of presheaves on $\cal C$ into a
category, and therefore we need to define what is meant by a
`morphism' between two presheaves $\bf X$ and $\bf Y$.  The
intuitive idea is that such a morphism from $\bf X$ to $\bf Y$
must give a `picture' of $\bf X$ within $\bf Y$. Formally, such
a morphism is defined to be a {\em natural transformation\/}
$N:{\bf X}\rightarrow{\bf Y}$, by which is meant a family of
maps (called the {\em components\/} of $N$) $N_A:{\bf
X}(A)\rightarrow{\bf Y}(A)$, $A$ in $\cal C$, such that if
$f:B\rightarrow A$ is a morphism in $\cal C$, then the composite
map ${\bf X}(A) \stackrel{N_A}\longrightarrow{\bf
Y}(A)\stackrel{{\bf Y}(f)} \longrightarrow{\bf Y}(B)$ is equal
to ${\bf X}(A) \stackrel{{\bf X}(f)}\longrightarrow{\bf
X}(B)\stackrel{N_B} \longrightarrow {\bf Y}(B)$. In other words,
we have the commutative diagram
\begin{equation}
    \bundle{{\bf X}(A)}{N_A}{{\bf Y}(A)}
    \bundlemap{{\bf X}(f)}{{\bf Y}(f)}
    \bundle{{\bf X}(B)}{N_B}{{\bf Y}(B)}    \label{cdNT}
\end{equation}
of which Eq.\ (\ref{Def:eta}) is clearly a special case. The
category of presheaves on $\cal C$ equipped with these morphisms
is denoted  ${\rm Set}^{{\cal C}^{\rm op}}$.

The idea of a subobject generalizes in an obvious way. Thus we
say that $\bf K$ is a {\em subobject\/} of $\bf X$ if there is a
morphism in the category of presheaves ({\em i.e.}, a natural
transformation) $i:{\bf K}\rightarrow{\bf X}$ with the property
that, for each $A$, the component map $i_A:{\bf
K}(A)\rightarrow{\bf X}(A)$ is a subset embedding, {\em i.e.},
${\bf K}(A)\subseteq {\bf X}(A)$.  Thus, if $f:B\rightarrow A$
is any morphism in $\cal C$, we get the analogue of the
commutative diagram Eq.\ (\ref{cd}):
\begin{equation}
\bundle{{\bf K}(A)}{}{{\bf X}(A)} \bundlemap{{\bf K}(f)}{{\bf
X}(f)} \bundle{{\bf K}(B)}{}{{\bf X}(B)}   \label{subobject}
\end{equation}
where, once again, the vertical arrows are subset inclusions.

The category of presheaves on $\cal C$, ${\rm Set}^{{\cal
O}^{\rm op}}$, forms a topos. We do not need the full definition
of a topos; but we do need the idea, mentioned in Section
\ref{SubSec:presheaves-poset}, that a topos has a subobject
classifier $\Omega$, to which we now turn.

\paragraph*{3. Sieves and The Subobject Classifier ${\bf\Omega}$:}

Among the key concepts in presheaf theory, and something of
particular importance for this paper, is that of a `sieve',
which plays a central role in the construction of the subobject
classifier in the topos of emphasized on a category $\cal C$.

A {\em sieve\/} on an object $A$ in $\cal C$ is defined to be a
collection $S$ of morphisms $f:B\rightarrow A$ in $\cal C$ with
the property that if $f:B\rightarrow A$ belongs to $S$, and if
$g:C\rightarrow B$ is any morphism, then $f\circ g:C\rightarrow
A$ also belongs to $S$. \footnote{A {\em cosieve\/} on $A$ is
defined to be a collection $S$ of morphisms $f:A\rightarrow B$
with the property that if $f:A\rightarrow B$ belongs to $S$, and
if $g:B\rightarrow C$ is any morphism, then $g\circ
f:A\rightarrow C$ also belongs to $S$.  However, another matter
of convention is involved here: some authors interchange our
usage of the words `sieve' and `cosieve'.  Note that, in any
event,  a sieve in $\cal C$ is the same thing as a cosieve in
${\cal C}^{\rm op}$, and vice versa.} In the simple case where
$\cal C$ is a poset, a sieve on $p\in\cal C$ is any subset $S$
of $\cal C$ such that if $r\in S$ then (i) $r\leq p$, and (ii)
$r'\in S$ for all $r'\leq r$; in other words, a sieve is nothing
but a {\em lower\/} set in the poset.

The presheaf ${\bf\Omega}:{\cal C}\rightarrow {\rm Set}$ is now
defined as follows. If $A$ is an object in $\cal C$, then
${\bf\Omega}(A)$ is defined to be the set of all sieves on $A$;
and if $f:B\rightarrow A$, then
${\bf\Omega}(f):{\bf\Omega}(A)\rightarrow{\bf\Omega}(B)$ is
defined as
\begin{equation}
{\bf\Omega}(f)(S):= \{h:C\rightarrow B\mid f\circ h\in S\}
                                \label{Def:Om(f)}
\end{equation}
for all $S\in{\bf\Omega}(A)$; the sieve ${\bf\Omega}(f)(S)$ is
often written as $f^*(S)$, and is known as the {\em pull-back\/}
to $B$ of the sieve $S$ on $A$ by the morphism $f:B\rightarrow
A$.

For our purposes in what follows, it is important to note that
if $S$ is a sieve on $A$, and if $f:B\rightarrow A$ belongs to
$S$, then from the defining property of a sieve we have
\begin{equation}
        f^*(S):=\{h:C\rightarrow B\mid f\circ h\in S\}=
\{h:C\rightarrow B\}=:\ \downarrow\!\!B     \label{f*S}
\end{equation}
where $\downarrow\!\!B$ denotes the {\em principal\/} sieve on
$B$, defined to be the set of all morphisms in $\cal C$ whose
codomain is $B$. In words: the pull-back of any sieve on $A$ by
a morphism from $B$ to $A$ that belongs to the sieve, is the
{\em principal\/} sieve on $B$.

If $\cal C$ is a poset, the pull-back operation corresponds to a
family of maps $\Omega_{qp}:\Omega_q\rightarrow\Omega_p$ (where
$\Omega_p$ denotes the set of all sieves on $p$ in the poset)
defined by $\Omega_{qp}={\bf\Omega}(i_{pq})$ if
$i_{pq}:p\rightarrow q$ ({\em i.e.}, $p\leq q$). It is
straightforward to check that if $S\in\Omega_q$, then
\begin{equation}
\Omega_{qp}(S):=\downarrow\!{p}\cap S \label{Def:Omqp}
\end{equation}
where $\downarrow\!{p}:=\{r\in{\cal C}\mid r\leq p\}$.

A crucial property of sieves is that the set ${\bf\Omega}(A)$ of
sieves on $A$ has the structure of a Heyting
algebra.\footnote{The paradigmatic example of a Heyting algebra
is the set of all open sets in a topological space $Z$.  The
algebraic operations are defined as $O_1\land O_2:=O_1\cap O_2$;
$O_1\lor O_2:=O_1\cup O_2$; and $\neg O:={\rm int}(Z-O)$.} This
is defined to be a distributive lattice, with null and unit
elements, that is {\em relatively complemented\/}, which means
that to any pair $S_1,S_2$ in ${\bf \Omega}(A)$, there exists an
element $S_1\Rightarrow S_2$ of ${\bf\Omega}(A)$ with the
property that, for all $S\in{\bf\Omega}(A)$,
\begin{equation}
        S\leq (S_1\Rightarrow S_2)\mbox{  if and only if
$S\land S_1\leq S_2$}.
\end{equation}
Specifically, ${\bf\Omega}(A)$ is a Heyting algebra where the
unit element $1_{{\bf\Omega}(A)}$ in ${\bf\Omega}(A)$ is the
principal sieve $\downarrow\!\!A$, and the null element
$0_{{\bf\Omega}(A)}$ is the empty sieve $\emptyset$.  The
partial ordering in ${\bf\Omega}(A)$ is defined by $S_1\leq S_2$
if, and only if, $S_1\subseteq S_2$; and the logical connectives
are defined as:
\begin{eqnarray}
    && S_1\land S_2:=S_1\cap S_2    \label{Def:S1landS2}\\
    && S_1\lor S_2:=S_1\cup S_2     \label{Def:S1lorS2} \\
    &&S_1\Rightarrow S_2:=\{f:B\rightarrow A\mid
    \mbox{ for all $g:C\rightarrow B$ if $f\circ g\in S_1$ then
                $f\circ g\in S_2$}\}.
\end{eqnarray}
As in any Heyting algebra, the negation of an element $S$
(called the {\em pseudo-complement\/} of $S$) is defined as
$\neg S:=S\Rightarrow 0$; so that
\begin{equation}
    \neg S:=\{f:B\rightarrow A\mid \mbox{for all
$g:C\rightarrow B$, $f\circ g\not\in S$} \}.    \label{Def:negS}
\end{equation}
The main distinction between a Heyting algebra and a Boolean
algebra is that, in the former, the negation operation does not
necessarily obey the law of excluded middle: instead, all that
be can said is that, for any element $S$,
\begin{equation}
        S\lor\neg S\leq 1.
\end{equation}

It can be shown that the presheaf ${\bf\Omega}$ is a subobject
classifier for the topos ${\rm Set}^{{\cal C}^{\rm op}}$. That
is to say, subobjects of any object $\bf X$ in this topos ({\em
i.e.}, any presheaf on $\cal C$) are in one-to-one
correspondence with morphisms $\chi:{\bf X}\rightarrow
{\bf\Omega}$. This works as follows.  First, let $\bf K$ be a
subobject of $\bf X$.  Then there is an associated {\em
characteristic\/} morphism $\chi^{{\bf K}}:{\bf
X}\rightarrow{\bf\Omega}$, whose `component' $\chi^{{\bf
K}}_A:{\bf X}(A)\rightarrow{\bf\Omega}(A)$ at each `stage of
truth' $A$ in $\cal C$ is defined as
\begin{equation}
    \chi^{{\bf K}}_A(x):=\{f:B\rightarrow A\mid {\bf X}(f)(x)\in
{\bf K}(B)\} \label{Def:chiKA}
\end{equation}
for all $x\in {\bf X}(A)$. That the right hand side of Eq.\
(\ref{Def:chiKA}) actually {\em is\/} a sieve on $A$ follows
from the defining properties of a subobject.

Thus, in each `branch' of the category $\cal C$ going `down'
from the stage $A$, $\chi^{{\bf K}}_A(x)$ picks out the first
member $B$ in that branch for which ${\bf X}(f)(x)$ lies in the
subset ${\bf K}(B)$, and the commutative diagram Eq.\
(\ref{subobject}) then guarantees that ${\bf X}(h\circ f)(x)$
will lie in ${\bf K}(C)$ for all $h:C\rightarrow B$.  Thus each
stage of truth $A$ in $\cal C$ serves as a possible context for
an assignment to each $x\in {\bf X}(A)$ of a generalized
truth-value: which is a sieve, belonging to the Heyting algebra
${\bf\Omega}(A)$, rather than an element of the Boolean algebra
$\{0,1\}$ of normal set theory. This is the sense in which
contextual, generalized truth-values arise naturally in a topos
of presheaves.

There is a converse to Eq.\ (\ref{Def:chiKA}): namely, each
morphism $\chi:{\bf X}\rightarrow{\bf\Omega}$ ({\em i.e.}, a
natural transformation between the presheaves ${\bf X}$ and
${\bf\Omega}$) defines a subobject ${\bf K}^\chi$ of $\bf X$ via
\begin{equation}
    {\bf K}^\chi(A):=\chi_A^{-1}\{1_{{\bf\Omega}(A)}\}.
                            \label{Def:KchiA}
\end{equation}
at each stage of truth $A$.

For this reason, the presheaf ${\bf\Omega}$ is known as {\em the
subobject classifier\/} in the category ${\rm Set}^{{\cal
C}^{\rm op}}$. As mentioned above, the existence of such an
object is one of the defining properties for a category to be a
topos, which ${\rm Set}^{{\cal C}^{\rm op}}$ is.

\paragraph*{3. Global Sections of a Presheaf:}

In any category, a {\em terminal object\/} is defined to be an
object $1$ with the property that, for any object $X$ in the
category, there is a unique morphism $X\rightarrow 1$; it is
easy to show that terminal objects are unique up to isomorphism.
A {\em global element\/} of an object $X$ is then defined to be
any morphism $1\rightarrow X$. The motivation for this
nomenclature is that, in the case of the category of sets, a
terminal object is any singleton set $\{*\}$; and then it is
true that there is a one-to-one correspondence between the
elements of a set $X$ and functions from $\{*\}$ to $X$.

For the category of presheaves on $\cal C$, a terminal object
${\bf 1}:{\cal C}\rightarrow {\rm Set}$ can be defined by ${\bf
1}(A):=\{*\}$ at all stages $A$ in $\cal C$; if $f:B\rightarrow
A$ is a morphism in $\cal C$ then ${\bf
1}(f):\{*\}\rightarrow\{*\}$ is defined to be the map $*\mapsto
*$. This is indeed a terminal object since, for any presheaf
$\bf X$, we can define a unique natural transformation $N:{\bf
X}\rightarrow{\bf 1}$ whose components $N_A:{\bf
X}(A)\rightarrow{\bf 1}(A)=\{*\}$ are the constant maps
$x\mapsto *$ for all $x\in{\bf X}(A)$.

A global element of a presheaf $\bf X$ is also called a {\em
global section\/}. As a morphism $\gamma:1\rightarrow{\bf X}$ in
the topos ${\rm Set}^{{\cal C}^{\rm op}}$, a global section
corresponds to a choice of an element $\gamma_A\in{\bf X}(A)$
for each stage of truth $A$ in $\cal C$, such that, if
$f:B\rightarrow A$, the `matching condition'
\begin{equation}
    {\bf X}(f)(\gamma_A)=\gamma_B \label{Def:global}
\end{equation}
is satisfied. As we shall see, the Kochen-Specker theorem can be
read as asserting the non-existence of any global sections of
certain presheaves that arises naturally in any quantum theory.

\paragraph*{4. Local Sections of a Presheaf:}

One of the important properties of a general topos category is
that an object may have `partial', or `local', elements even if
there are no global ones. In general, a {\em local element\/} of
an object $X$ in a category with a terminal object is defined to
be a morphism $U\rightarrow X$, where $U$ is a subobject of the
terminal object $1$. In the category of sets, there are
no-nontrivial subobjects of $1:=\{*\}$, but this is not the case
in a general topos.

In particular, in the case of presheaves on $\cal C$, a
subobject ${\bf U}$ of ${\bf 1}$ is a collection of subsets
${\bf U}(A)\subseteq\{*\}$, $A$ in $\cal C$, that satisfy the
appropriate form of the commutative diagram Eq.\
(\ref{subobject}) that describes a subobject.  However, the only
subsets of $\{*\}$ are $\{*\}$ itself, and the empty set
$\emptyset$.  Furthermore, there is a unique function
$\emptyset\rightarrow\{*\}$ (the `empty' function) but no
function $\{*\}\rightarrow\emptyset$. It follows, therefore,
that in assigning the sets $\emptyset$ or $\{*\}$ to each stage
$A$ for a subobject $\bf U$ of $\bf 1$, the assignments of the
singleton sets $\{*\}$ must be `closed downwards' in the sense
that if ${\bf U}(A)=\{*\}$ and if $f:B\rightarrow A$ is a
morphism in $\cal C$, then we must have ${\bf U}(B)=\{*\}$ also.

We deduce from this that a partial element of a presheaf $\bf X$
is an assignment $\gamma$ of an element $\gamma_A$ to a certain
{\em subset\/} of objects $A$ in $\cal C$---what we shall call
the {\em domain\/} ${\rm dom\,}\gamma $ of $\gamma$---with the
properties that (i) the domain is closed downwards in the sense
that if $A\in {\rm dom\,}\gamma$ and $f:B\rightarrow A$, then
$B\in {\rm dom\,}\gamma$; and (ii) for objects in this domain,
the matching condition Eq.\ (\ref{Def:global}) is satisfied.


\end{document}